\documentclass[CRPHYS,Unicode,manuscript,english]{cedram}

\usepackage{amssymb,amsmath,amsbsy}
\usepackage{caption}
\usepackage{comment}
\hypersetup{breaklinks=true}
\hypersetup{colorlinks=true}
\hypersetup{urlcolor=blue}
\hypersetup{citecolor=blue}
\hypersetup{linkcolor=blue}

\newcommand{\yvan}{\color{black}}
\DeclareMathOperator\re{Re}
\DeclareMathOperator\im{Im}
\DeclareMathOperator\argsh{argsh}
\DeclareMathOperator\argth{argth}
\DeclareMathOperator\sh{sh}
\DeclareMathOperator\ch{ch}
\DeclareMathOperator\thf{th}
\DeclareMathOperator\Ci{Ci}
\DeclareMathOperator\sgn{sgn}
\newcommand{\zero}{\mathbf{0}}
\newcommand{\kk}{\mathbf{k}}
\newcommand{\KK}{\mathbf{K}}
\newcommand{\qq}{\mathbf{q}}

\newcommand{\rr}{\mathbf{r}}

\newcommand{\g}[1]{"#1"}
\newcommand{\be}{\begin{equation}}
\newcommand{\ee}{\end{equation}}
\newcommand{\bea}{\begin{eqnarray}}
\newcommand{\eea}{\end{eqnarray}}
\newcommand{\ii}{\mathrm{i}}
\newcommand{\eee}{\mathrm{e}}
\newcommand{\dd}{\mathrm{d}}
\newcommand{\veps}{\varepsilon}
\newcommand{\eps}{\epsilon}

\title{Pair distribution functions of a superfluid spin{\yvan -}$1/2$ Fermi gas with contact interactions in the linearized time-dependent BCS theory}

\author{\firstname{Yvan} \lastname{Castin}}
\address{Laboratoire Kastler Brossel, ENS-Université PSL, CNRS, Université Sorbonne and Collège de France, 24 rue Lhomond, 75231 Paris, France}
\email[Y. Castin]{yvan.castin@lkb.ens.fr}
\keywords{Fermi gases, pair density, structure factor, fluctuation-dissipation theorem, superfluidity, unitary limit, contact interactions}

\begin{abstract}
We show that the minimal mean-field theory to use for calculating the pair distribution functions $g_{\sigma\sigma'}(\rr,\rr')$ of a spatially homogeneous, unpolarized spin-1/2 superfluid Fermi gas is not the ordinary static BCS theory, but the linearized time-dependent BCS theory implemented via the fluctuation-dissipation theorem. Indeed, the former completely ignores the acoustic excitation branch -- the phonons -- of the superfluid, while the latter explicitly takes it into account, as well as the quantum fluctuations induced by the broken-pair continuum. Unlike the first, the second theory (i) reflects the effect of these collective excitations on the system's equation of state, including at zero temperature, (ii) allows the function $g_{\uparrow\downarrow}(\rr,\rr')$ to go at sufficiently large distances strictly below its asymptotic value $(\rho/2)^2$ where $\rho$ is the gas density, as expected according to the quantum hydrodynamics of Landau and Khalatnikov at low temperatures, and (iii) predicts in the function $g_{\uparrow\uparrow}(\rr,\rr')$ at short distances subdominant contributions $|\rr-\rr'|^2\ln|\rr-\rr'|$ in 3D and $|\rr-\rr'|^2\ln(-\ln|\rr-\rr'|)$ in 2D, alongside the dominant contributions $|\rr-\rr'|$ in 3D and $|\rr-\rr'|^2\ln|\rr-\rr'|$ in 2D already present in static BCS theory but with a lower coefficient. This discussion is relevant to the recent theoretical work of Obeso-Jureidini and Romero-Roch\'{\i}n, and to the ongoing experiments on cold atomic gases at ENS and MIT. {\yvan A multilingual version is available in separate files on the open archive HAL at \url{https://hal.science/hal-05426463}.}
\end{abstract}

\begin{document}

\maketitle

\tableofcontents

\section{Introduction, motivations, inadequacy of static BCS theory}\label{sec1}

The pair distribution functions $g_{\sigma\sigma'}(\rr,\rr')$ of a Fermi gas with two internal states $\uparrow,\downarrow$ in contact interaction, in thermal equilibrium in the spatially homogeneous and unpolarized case (same density $\rho_\sigma=\rho/2$ in each component), defined as 
\begin{equation}
\label{eq:defg}
\boxed{g_{\sigma\sigma'}(\rr,\rr')\equiv \rho_\sigma \rho_{\sigma'}+\delta g_{\sigma\sigma'}(\rr,\rr') \equiv \langle\hat{\psi}_\sigma^\dagger(\rr)\hat{\psi}_{\sigma'}^\dagger(\rr')\hat{\psi}_{\sigma'}(\rr')\hat{\psi}_{\sigma}(\rr)\rangle}
\end{equation}
 where $\hat{\psi}_\sigma(\rr)$ is the fermionic field operator for spin $\sigma\in\{\uparrow,\downarrow\}$ and $\delta g_{\sigma\sigma'}(\rr,\rr')$ are the deviations from asymptotic values, have recently been the subject of in-depth studies \cite{Romero2D,Romero3D} in dimension $d=2$ or $d=3$ at zero temperature $T=0$ using ordinary BCS theory \cite{thBCS}, known as {\it static}, a variational theory reducing the ground state of the gas to a coherent state of bound pairs $\uparrow\downarrow$ (the equivalents for our neutral gas of the Cooper pairs of superconductors, here in the spin singlet state). However, in this article, we propose to highlight the shortcomings of this theory and to use a better one, known as {\it time-dependent linearized}, for the calculation of the functions $g_{\sigma\sigma'}(\rr,\rr')$.

Indeed, the static BCS theory has the genius of taking into account the phenomenon of pair condensation of the gas, giving a variational bound on energy and predicting an excitation spectrum $\eps_\kk$ of fermionic quasi-particles with a band gap of width $E_{\rm gap}>0$, the famous BCS spectrum. However, it is not suitable for all observables, particularly multimode observables such as $g_{\sigma\sigma'}(\rr,\rr')$. This can be clearly seen in the BEC (Bose-Einstein condensate) limit where the density $\rho$ tends towards zero in the case where two fermions $\uparrow$ and $\downarrow$ have a bound state in the vacuum: the static BCS theory is then reduced to a pure condensate of bosonic dimers; however, in a condensed Bose gas, a pure condensate ansatz leads to a pair distribution function $g_2^{\rm B}(\rr,\rr')$ that is trivial, equal everywhere—{\it after} taking the thermodynamic limit—to the density $\rho_{\rm B}$ of the bosons squared, thus to a deviation $\delta g_2^{\rm B}(\rr,\rr')$ from the asymptotic value $\rho_{\rm B}^2$ that is identically zero. Bogoliubov's multimode theory \cite{Bogo}, which correctly takes into account the non-condensed fraction and the modes of bosonic quasi-particles—known as Bogoliubov quasi-particles—shows, however, that this is not the case; in particular, at zero temperature, it predicts that $\delta g_2^{\rm B}(\rr,\rr')$ has a tail proportional to $-1/|\rr-\rr'|^{d+1}$, and that the integral of $\delta g_2^{\rm B}(\rr,\rr'=\zero)$ over the entire space $\mathbf{r}\in\mathbb{R}^d$ takes the negative value $-\rho_{\rm B}$, see section 8.7.2 of reference \cite{livre}. Remember the adage: zero is a poor approximation for any non-zero quantity...

\begin{figure}[t]
\includegraphics[width=6cm,clip=]{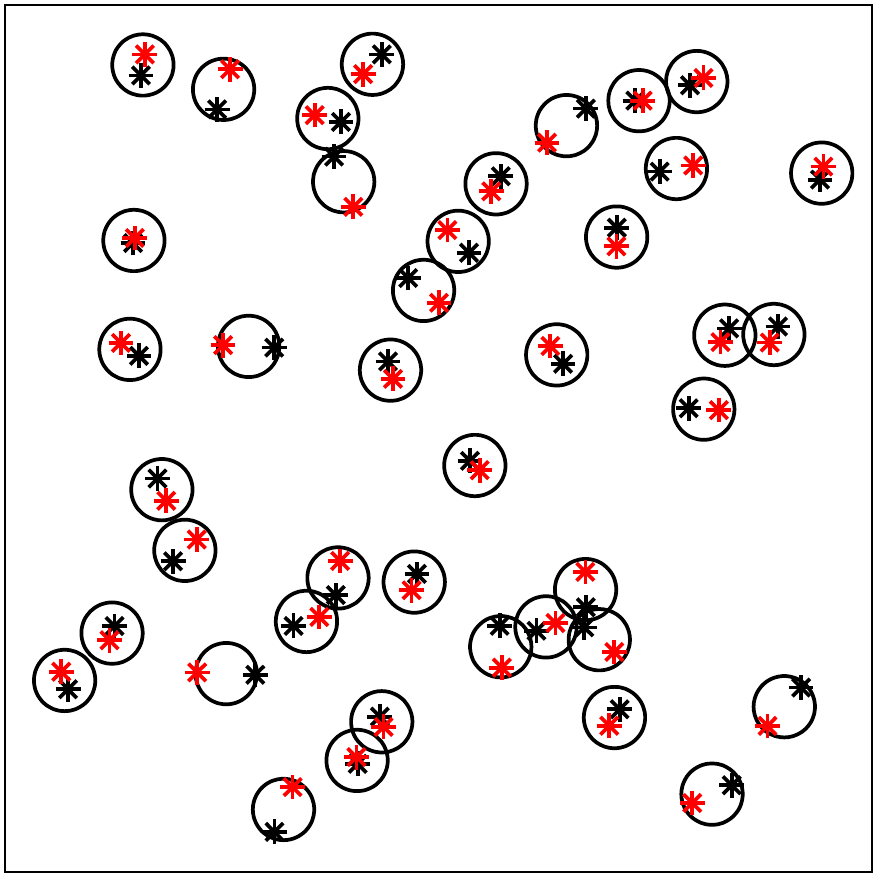}
\caption{Purely schematic representation of the spatial distribution of fermions $\uparrow$ and $\downarrow$ in a 2D pair-condensed Fermi gas, in an intermediate interaction regime—between the BEC limit and the BCS limit—where the size $\ell_{\rm pair}$ of a bound pair is smaller but of the same order of magnitude as the average distance $\rho^{-1/d}$ between particles. The fermions $\uparrow$ (black stars) and $\downarrow$ (red stars) in the same circle belong to the same bound pair (the circles are centered on the barycenters of the bound pairs). As shown in equation (\ref{eq:dgsta2}), the distribution function $g_{\uparrow\downarrow}^{\rm sta}(\rr,\rr')$ of static BCS theory takes into account the spatial correlations between two fermions of opposite spins within the same bound pair, but ignores those between two different bound pairs, as it neglects the acoustic modes of the superfluid and the quantum or thermal density fluctuations they induce.}
\label{fig:schema}
\end{figure}

But let us extend this objection to fermions in the general case. The inadequacy of static BCS theory, at zero temperature or otherwise, is clearly seen in the distribution of pairs of fermions with opposite spins: since the variational ansatz is a Gaussian state for the fermionic field, Wick's theorem applies and we have 
\begin{equation}
\label{eq:dgsta}
\delta g_{\uparrow\downarrow}^{\rm sta}(\rr,\rr')=|\langle\hat{\psi}_\downarrow(\rr')\hat{\psi}_\uparrow(\rr)\rangle_{\rm sta}|^2
\end{equation}
 where “sta” as a subscript or superscript means “static BCS.” However, applying the Penrose and Onsager criterion \cite{PenOnsa} to the two-body density matrix $\langle\hat{\psi}_{\sigma_1}^\dagger(\rr_1)\hat{\psi}_{\sigma_2}^\dagger(\rr_2)\hat{\psi}_{\sigma_2'}(\rr_2')\hat{\psi}_{\sigma_1'}(\rr_1')\rangle_{\rm sta}$ in the quantization volume $[0,L]^d$ as in references \cite{LivreLeggett,PhaseLong}, i.e., by identifying its macroscopically populated eigenvector—correctly normalized to unity in $[0,L]^{2d}$—with the wave function $\phi_0(\rr_1,\rr_2)$ of the bound pair $\uparrow\downarrow$ of the BCS condensate, we obtain\footnote{The ground state of the gas in static BCS theory is a coherent state of pairs $|\Psi_{\rm BCS}\rangle\propto \exp\left[\int\dd^dr_1\dd^dr_2\, \alpha\phi_{\rm BCS}(\rr_1,\rr_2)\hat{\psi}^\dagger_{\uparrow}(\rr_1)\hat{\psi}^\dagger_{\downarrow}(\rr_2)\right]|0\rangle$ where $\alpha$ is a complex amplitude, $\phi_{\rm BCS}(\rr_1,\rr_2)$ is a normalized pair wave function—we use the notation from references \cite{Romero3D,Romero2D}—and $|0\rangle$ is the fermion vacuum. However, it would be incorrect to equate $\phi_{\rm BCS}$ with $\phi_0$. These functions are not proportional; they are even quite distinct, except in the BEC limit where they reduce to the wave function of a $\uparrow\downarrow$ dimer with zero momentum. This pitfall to avoid illustrates the subtle effects of fermionic antisymmetrization.} 
\begin{equation}
\phi_0(\rr_1,\rr_2)=\langle\hat{\psi}_\downarrow(\rr_2)\hat{\psi}_\uparrow(\rr_1)\rangle_{\rm sta}/(\bar{N}_0^{\rm pairs})^{1/2} \ \ \mbox{where}\ \ \bar{N}_0^{\rm pairs}=\int_{[0,L]^d}\dd^dr_1 \int_{[0,L]^d}\dd^dr_2 |\langle\hat{\psi}_\downarrow(\rr_2)\hat{\psi}_\uparrow(\rr_1)\rangle_{\rm sta}|^2
\end{equation}
 is the average number of condensed pairs, so that 
\begin{equation}
\label{eq:dgsta2}
\delta g_{\uparrow\downarrow}^{\rm sta}(\rr,\rr') = \bar{N}_0^{\rm pairs} |\phi_0(\rr,\rr')|^2 
\end{equation}
 The prediction (\ref{eq:dgsta}) therefore only describes the internal structure of a bound pair $\uparrow\downarrow$ but corresponds to a flat distribution of bound pairs, i.e., to a complete decorrelation between the positions of the pairs, see Figure~\ref{fig:schema}, which is purely illustrative. It follows that at relative distances $|\rr-\rr'|\gg \ell_{\rm pair}$, where $\ell_{\rm pair}$ is the size of a bound pair, the function $g^{\rm sta}_{\uparrow\downarrow}(\rr,\rr')$ is sadly reduced to the product of densities $\rho_\uparrow^{\rm sta} \rho_\downarrow^{\rm sta}$ to an exponentially small degree, the anomalous mean $\langle\hat{\psi}_\downarrow(\rr')\hat{\psi}_\uparrow(\rr)\rangle_{\rm sta}$ decreasing rapidly for variable $|\rr-\rr'|$ \cite{Romero2D,Romero3D}.\footnote{\label{note:asymp} In dimension $d=3$, we find at $T=0$ that 
\begin{equation}
\langle\hat{\psi}_\downarrow(\rr')\hat{\psi}_\uparrow(\rr)\rangle_{\rm sta}\stackrel{d=3}{\underset{|\rr-\rr'|\to +\infty}{\sim}} -\frac{(m\Delta/\hbar^2)^{1/2}}{(2\pi |\rr-\rr'|)^{3/2}}\re\left[\mathrm{i} k_0^{1/2}\exp(-\mathrm{i} k_0 |\rr-\rr'|)\right]
\end{equation}
 with $k_0=[2 m(\mu-\mathrm{i}\Delta)/\hbar^2]^{1/2}$, $\mu$ the chemical potential of the Fermi gas, $m$ the mass of a fermion, $\Delta>0$ the order parameter written at order zero in $\exp(-\beta E_{\rm gap})$ (see section \ref{sec2_1}). To do this, we perform the angular integral in expression (\ref{eq:moya}) and use the formal parity in $k$ of the integrand to write the result in the form $\langle\hat{\psi}_\downarrow(\rr')\hat{\psi}_\uparrow(\rr)\rangle_{\rm sta}=[\Delta/(8\mathrm{i}\pi^2|\rr-\rr'|)]\int_{-\infty}^{+\infty}\mathrm{d} k\, k \exp(-\mathrm{i} k |\rr-\rr'|)/\eps_k$. The function $k\mapsto\eps_k$, with an analytic continuation to the complex values of $k$, has in each quadrant a branch cut starting from $k_0^*$, $-k_0$, $-k_0^*$, or $k_0$ and tending towards infinity along the bisector. Let us then deform the integration path as in \cite{Romero3D} by folding it in the lower half-plane onto the branch cuts originating from $k_0$ and $-k_0^*$, the contribution of the second being the complex conjugate of the first. It remains to describe parametrically the first as follows, $\eps\in[0,+\infty[\mapsto k(\eps)=\left[2m\left(\mu-\mathrm{i}\sqrt{\Delta^2+\eps^2}\right)/\hbar^2\right]^{1/2}$, and to note that $\eps_{k(\eps)}$ jumps from $-\mathrm{i}\eps$ to $+\mathrm{i}\eps$ if we cross the branch cut from top to bottom (indeed, $\hbar^2 k(\eps)\frac{\mathrm{d}}{\mathrm{d}\eps}k(\eps)/m=-\mathrm{i}\eps/\sqrt{\Delta^2+\eps^2}$, therefore $k(\eps+\mathrm{i}\eta)-k(\eps)\sim \mathrm{i}\eta\frac{\mathrm{d}}{\mathrm{d}\eps}k(\eps)\to 0^{\pm}$ and $\eps_{k(\eps+\mathrm{i}\eta)}=[-(\eps+\mathrm{i}\eta)^2]^{1/2}\to\mp\mathrm{i}\eps$ if $\eta\to 0^\pm$), to arrive at 
\begin{equation}
\langle\hat{\psi}_\downarrow(\rr')\hat{\psi}_\uparrow(\rr)\rangle_{\rm sta}\stackrel{d=3}{=}\frac{m\Delta}{2\pi^2\hbar^2|\rr-\rr'|}\im \int_0^{+\infty}\mathrm{d}\eps\, \frac{\exp[-\mathrm{i} k(\eps) |\rr-\rr'|]}{(\Delta^2+\eps^2)^{1/2}}
\end{equation}
 If $|\rr-\rr'|\to+\infty$, this integral is dominated by the interval $\eps=O(1/|\rr-\rr'|^{1/2})$, i.e. by a neighborhood of the branch point $k_0$; it remains to expand $k(\eps)$ to order $\eps^2$ under the exponential and to neglect $\eps$ under the square root in the denominator of the integrand, to obtain—after Gaussian integration—the announced equivalent. The same procedure gives for the normal mean (\ref{eq:moyn}): 
\begin{equation}
\langle\hat{\psi}^\dagger_\uparrow(\rr)\hat{\psi}_\uparrow(\rr')\rangle_{\rm sta}\stackrel{d=3}{=}\frac{-m}{2\pi^2\hbar^2|\rr-\rr'|} \re \int_0^{+\infty}\mathrm{d}\eps\, \eee^{-\mathrm{i} k(\eps) |\rr-\rr'|}\underset{|\rr-\rr'|\to+\infty}{\sim} -\frac{(m\Delta/\hbar^2)^{1/2}}{(2\pi |\rr-\rr'|)^{3/2}}\im\left[\mathrm{i} k_0^{1/2}\exp(-\mathrm{i} k_0 |\rr-\rr'|)\right]
\end{equation}
 In dimension $d=2$, still at $T=0$, we use the analytical expressions (\ref{eq:moya2d},\ref{eq:moyn2d}) of the normal and anomalous means established in reference \cite{Romero2D}, the known asymptotic behavior of Bessel functions and the specific property in 2D that $k_0\equiv[2 m(\mu-\mathrm{i}\Delta)/\hbar^2]^{1/2}$ can also be written as $k_0=k_{\rm F}^{\rm sta}-\mathrm{i} q_{\rm dim}$ with the notations and relations (\ref{eq:rel2d}) in section \ref{sec2_1}. Then, after cleverly grouping the two averages into real and imaginary parts of the same complex quantity (as could also be done in 3D), we find that 
\begin{equation}
\langle\hat{\psi}_\downarrow(\rr')\hat{\psi}_\uparrow(\rr)\rangle_{\rm sta}+\mathrm{i} \langle\hat{\psi}^\dagger_\uparrow(\rr)\hat{\psi}_\uparrow(\rr')\rangle_{\rm sta}\stackrel{d=2}{\underset{|\rr-\rr'|\to +\infty}{\sim}} -\frac{(m\Delta/\hbar^2)^{1/2}}{2\pi|\rr-\rr'|} \exp[\ii(\pi/4-k_0 |\rr-\rr'|)]
\end{equation}
which is compatible with the less precise equations (40,41) in reference \cite{Romero2D}. At non-zero temperature $T\neq 0$, a case not considered in references \cite{Romero3D,Romero2D}, the function $k\mapsto 1/\eps_\kk$ in the integrands is replaced by $k\mapsto (1-2f_\kk)/\eps_\kk=\thf(\beta\eps_\kk/2)/\eps_\kk$, which has neither (i) a branch cut for $k\in\mathbb{C}$ nor (ii) a pole for $k$ root of equation $\eps_\kk=0$. In fact, (i) $(\thf z)/z$ is an even function of $z$, which keeps the same value when crossing a branch cut of $k\mapsto\eps_\kk$—we can also say that its Taylor series expansion only includes even powers of $z$, and (ii) $(\thf z)/z$ tends towards $1$ when $z\to 0$, it is an analytic function of $z$ on a neighborhood of the origin. The only singularities remaining in the complex plane are therefore fermionic thermal poles, such as $\exp(\beta\eps_\kk)=-1$; those in the lower right quadrant are written as $k_n(T)=[2m(\mu-\ii|\Delta+\mathrm{i}\pi(2n+1)k_{\rm B}T|)/\hbar^2]^{1/2}$, $n\in\mathbb{N}$, the others can be deduced by global sign change and complex conjugation. The clever formulation by contour integral in $d=3$ still applies (the thermal factor $f_\kk$ does not change the formal parity in $k$ of the integrand and does not prevent the integration path from being closed by a semicircle at infinity in the lower half-plane). At large distances, according to the residue theorem, the anomalous and normal averages are dominated by the poles closest to the real axis, namely $k_0(T)$ and $-k_0^*(T)$; following section 8.7.3.2 of reference \cite{livre}, in particular for the more delicate case $d=2$, we obtain the asymptotic behaviors 
\begin{equation}
\langle\hat{\psi}_\downarrow(\rr')\hat{\psi}_\uparrow(\rr)\rangle_{\rm sta}\left|1+\mathrm{i}\pi\frac{k_{\rm B}T}{\Delta}\right|
+\mathrm{i} \langle\hat{\psi}^\dagger_\uparrow(\rr)\hat{\psi}_\uparrow(\rr')\rangle_{\rm sta}\stackrel{T\neq 0}{\underset{|\rr-\rr'|\to +\infty}{\sim}} -\frac{mk_{\rm B}T/\hbar^2}{(\pi|\rr-\rr'|)^{(d-1)/2}}\times\left\{\begin{array}{ll}
\displaystyle \mathrm{i}\, \eee^{-\mathrm{i} k_0(T)|\rr-\rr'|} & \quad \mathrm{in}\ d=3\\
& \\
\displaystyle \left[\frac{2}{k_0(T)}\right]^{1/2} \eee^{\mathrm{i}[\pi/4-k_0(T)|\rr-\rr'|]} & \quad \mathrm{in}\ d=2
\end{array}\right.
\end{equation}
 } In particular, it is obvious that $g_{\uparrow\downarrow}^{\rm sta}(\rr,\rr')$ is everywhere {\it greater} than its asymptotic value.

 In reality, in short-range interacting Fermi gases, there are collective bosonic excitations at low temperatures (here at $T=0$ for simplicity) that are completely ignored by static BCS theory. They correspond to an isotropic branch of acoustic excitation—by sound waves—
whose departure is linear in wave number, 
\begin{equation}
\label{eq:depart}
\omega_\mathbf{q} \underset{q\to 0^+}{=} cq + O(q^3)
\end{equation}
 where $c$ is the speed of sound in the ground state of the gas. The associated quanta are therefore phonons with wave vector $\qq$ and energy $\hbar\omega_\qq$ ($\omega_\qq$ is the natural angular frequency of the corresponding sound wave). In the case of Fermi superfluids, they play the role of Bogoliubov's quasi-particles in Bose gases. Bogoliubov—him again—and Anderson, contemporaries of Bardeen, Cooper, and Schrieffer, clearly saw the enormous gap in reference \cite{thBCS} and quickly remedied it \cite{Bogfer,RPA} using a rather cumbersome formalism, that of the Random Phase Approximation (RPA) for Anderson. Hence the name "Bogoliubov-Anderson" often associated with this acoustic branch and these phonons \cite{Minguzzi}.

It has since been understood that the same dispersion relation $q\mapsto \omega_{\mathbf{q}}$ can be obtained more simply by trivially generalizing the BCS theory to the time-dependent case (at $T=0$, the coherent state is then in a time-dependent spatial $\uparrow\downarrow$-pair mode) and then linearizing it around its steady state, i.e. around the static BCS theory \cite{brouifer}. A careful verification and comparison with RPA was carried out in reference \cite{theseHK}. The analogy with Bose gases becomes obvious: the time-dependent BCS theory is analogous to the time-dependent Gross-Pitaevskii equation; the linearization of the former reproduces the Bogoliubov-Anderson phonon excitation spectrum, just as the linearization of the latter reproduces that of Bogoliubov's quasiparticles. In both cases, a classical field theory reproduces the natural frequencies of a quantum field theory.

The key point, as the reader will have understood, is that the amplitudes of the phonon modes—sound waves in the gas—are subject to thermal and quantum fluctuations, the latter persisting at $T=0$ even if there are no phonons present (a statement made tangible and quantitative by Landau and Khalatnikov's quantum hydrodynamics in section \ref{sec3_1}). This inevitably leads to fluctuations in the density of bound pairs due to the movement of their center of mass, and therefore to the appearance of a non-trivial structure in $g_{\uparrow\downarrow}(\rr,\rr')$ at distances $|\rr-\rr'|\gg\ell_{\rm pair}$. It is this effect that leads to the existence of the negative parts of $\delta g_2^{\rm B}(\rr,\rr')$, mentioned above, in a sufficiently cold Bose gas. In the case of fermions, at sufficiently low temperatures,
 we should therefore expect to lose the property $\delta g_{\uparrow\downarrow}(\rr,\rr')\geq 0$ rashly deduced from Wick's theorem, since bound pairs $\uparrow\downarrow$ resemble bosons for phonons with long wavelengths $\lambda_{\phi}\gg\ell_{\rm pair}$. To drive the point home, let us note that at $T=0$, the difference $\delta g_{\uparrow\downarrow}(\rr,\rr')$, like $\delta g_2^{\rm B}(\rr,\rr')$, must obey in the thermodynamic limit to a simple but exact sum rule, in this case 
\begin{equation}
\int_{\mathbb{R}^d} \dd^dr\, \delta g_{\uparrow\downarrow}(\rr,\rr'=\mathbf{0})\stackrel{T=0}{=}0
\label{eq:relsomhb}
\end{equation}
which prohibits it—unlike (\ref{eq:dgsta})—from being positive everywhere. see Appendix \ref{sec:rel_som} for more details, which shows in particular the role played by phonons in equation (\ref{eq:relsomhb}) and in its generalization to $T>0$.

The moral of this story is that the minimum acceptable theory for the calculation of the pair distribution functions $g_{\sigma\sigma'}(\rr,\rr')$ in a Fermi gas of spin-$1/2$ fermions with contact interaction must take into account their acoustic excitation branch $\omega_\qq$. This cannot be the static BCS theory, which only contains the elementary excitations of fermionic quasi-particles $\eps_\kk$. We propose here as a natural candidate its direct improvement, the time-dependent BCS theory linearized around its stationary solution, which retains the simplicity of a mean-field theory. In what follows, we show how to obtain the functions $g_{\sigma\sigma'}(\rr,\rr')$ in this theory, linking response functions and density-density correlation functions through the fluctuation-dissipation theorem in section \ref{sec2}, then we analyze in detail the predictions obtained in section \ref{sec3}, also comparing them to numerical results beyond mean field and to experimental measurements in cold atomic gases. We conclude in section \ref{sec4}.

\section{Obtaining the $g_{\sigma\sigma'}(\rr,\rr')$ in the linearized time-dependent BCS theory}\label{sec2}

In this section, we implement in the grand canonical ensemble the linearized time-dependent BCS formalism and its density-density response functions. After defining the problem and providing targeted reminders on static BCS theory in section \ref{sec2_1}, we obtain, using the fluctuation-dissipation theorem in section \ref{sec2_2}, the deviations $\delta g_{\sigma\sigma'}(\rr,\rr')$ of the pair distribution functions $g_{\sigma\sigma'}(\rr,\rr')$ to their asymptotic values in the form of integrals of the density-density susceptibilities $\tilde{\chi}_{\sigma\sigma'}(\qq,\eps+\mathrm{i} 0^+)$ over the wave vector $\qq$ and the energy $\eps$. In section \ref{sec2_3}, we isolate the core of the theory, that is, a common correction $\tilde{\chi}_{\rm coll}(\qq,\eps+\mathrm{i} 0^+)$ of these susceptibilities to their static BCS approximations, thus a common correction on the deviations $\delta g_{\sigma\sigma'}(\rr,\rr')$. The integral of $\tilde{\chi}_{\rm coll}(\qq,\eps+\mathrm{i} 0^+)$ over $\eps$ has in section \ref{sec2_4} useful rewritings in terms of contour integrals in the complex plane, which
 facilitate numerical evaluation or allow the contributions of the Bogoliubov-Anderson phonon and the broken pair continuum to be separated. Finally, we show in section \ref{sec2_5} that, when we modify the density correlation functions with respect to static BCS, as we do here, we must also modify its mean value, i.e., the grand canonical equation of state of the gas, if we want to obtain sensible values of $g_{\uparrow\uparrow}(\rr,\rr')$ at short distances; this leads to a beautiful rewriting of the common correction of the pair distribution functions $g_{\sigma\sigma'}(\rr,\rr')$ to their static BCS approximations $g^{\rm sta}_{\sigma\sigma'}(\rr,\rr')$.

\subsection{Interaction, system state, and brief reminders on static BCS theory}\label{sec2_1}

Our fermions of spin $1/2$ and mass $m$ live in dimension $d\in\{2;3\}$ without a trapping potential but with periodic boundary conditions in the quantization box $[0,L]^d$, whose volume $L^d$ we let tend to infinity at the end of the calculations.\footnote{Case $d=1$ is treated in mean field and beyond, for example in reference \cite{Iac1D}, and in any case falls under the theory of integrable systems by Bethe ansatz \cite{Gaudin}.} They do not interact when they are in the same internal state $\sigma\in\{\uparrow,\downarrow\}$ (no scattering resonance in the $p$ wave between $\uparrow$ and $\uparrow$, nor between $\downarrow$ and $\downarrow$); otherwise, they interact through a negligible range potential with scattering length $a_{d\mathrm{D}}$ ($s$-wave scattering resonance between $\uparrow$ and $\downarrow$). We follow the convention that the stationary two-body $\uparrow\downarrow$ scattering state with zero energy is written (outside the interaction potential) as 
\begin{equation}
\label{eq:conva}
\psi_0(\rr,\rr')\stackrel{d=3}{=}1/{a_{\rm 3D}-1}/|\rr-\rr'| \quad\mbox{and}\quad \psi_0(\rr,\rr')\stackrel{d=2}{=}\ln(|\rr-\rr'|/a_{\rm 2D})
\end{equation}
 In 2D, the two-body problem always has a bound state with binding energy 
\begin{equation}
\label{eq:edim2d}
\eps_{\rm dim}=\hbar^2 q^2_{\rm dim}/m \quad \mbox{with}\quad q_{\rm dim}=2\exp(-\gamma)/a_{\rm 2D}
\end{equation}
 where $\gamma=0.577\, 215\ldots$ is the Euler-Mascheroni constant. In 3D, it has only one, with binding energy $\hbar^2/m a_{\rm 3D}^2$, as for $a_{\rm 3D}>0$.

The gas is taken at thermal equilibrium in the grand canonical ensemble at a temperature $T$ that is non-zero but lower than the superfluid transition temperature, $0<T<T_{\rm c}$. As it is unpolarized, it has the same chemical potential in both internal states, $\mu_\uparrow=\mu_\downarrow=\mu$. The BCS theory approximates its state by the most general Gaussian many-body density operator, $\hat{\sigma}_{\rm sta}\propto \exp(-\beta \hat{H}_{\rm quad})$, where $\beta=1/k_{\rm B}T$ and where $\hat{H}_{\rm quad}$, a quadratic form in the fermionic field operators $\hat{\psi}_\sigma(\rr)$ and their Hermitian conjugates, contains not only the normal terms $\hat{\psi}^\dagger_\sigma\hat{\psi}_\sigma$ but also the anomalous terms $\hat{\psi}_\uparrow\hat{\psi}_\downarrow$ and $\hat{\psi}^\dagger_\downarrow\hat{\psi}^\dagger_\uparrow$ describing pair condensation $\uparrow\downarrow$.\footnote{In the absence of (i) coherent Rabi coupling between $\uparrow$ and $\downarrow$ and (ii) pairing in the $p$ wave between the same spin states, it is unnecessary to include terms (i) as $\hat{\psi}^\dagger_\sigma\hat{\psi}_{-\sigma}$ and (ii) as $\hat{\psi}_\sigma\hat{\psi}_\sigma$.}${}^{,}$\footnote{In 2D, there is in fact no condensation of bound pairs at the thermodynamic limit (except if $T=0$), but this limitation of static BCS theory is of little importance here because the pair distribution functions $g_{\sigma\sigma'}$ are local observables—that is, insensitive to whether the non-diagonal order has infinite range or not. See also reference \cite{Iac1D}, where the same problem arises in 1D (this time including $T=0$).}
 After optimization, we find that $\hat{H}_{\rm quad}=\hat{H}_{\rm sta}$ where the BCS Hamiltonian $\hat{H}_{\rm sta}$ is deduced from the complete grand canonical Hamiltonian $\hat{H}_{\rm GC}$ by replacing the quartic interaction term with quadratic terms by means of incomplete Wick contractions (see the Hartree-Fock-Bogoliubov section in reference \cite{Blaizot}; the limiting case $T=0$ is discussed in detail in section 3.2 of reference \cite{Varenna}). 
This quadratic Hamiltonian is diagonalized by the Bogoliubov transformation, here an expansion of the field operators or their Hermitian conjugates over the eigenmodes of the BCS quasi-particles, plane waves in the present spatially homogeneous case: 
\begin{equation}
\label{eq:hsta}
\binom{\hat{\psi}_\uparrow(\rr)}{\hat{\psi}^\dagger_\downarrow(\rr)}=\frac{1}{L^{d/2}} \sum_\mathbf{k} \hat{\gamma}_{\mathbf{k}\uparrow} \binom{U_\mathbf{k}}{V_\mathbf{k}} \exp(\mathrm{i}\mathbf{k}\cdot\rr) + \gamma^\dagger_{\mathbf{k}\downarrow} \binom{-V_{\mathbf{k}}}{U_{\mathbf{k}}} \exp(-\mathrm{i}\mathbf{k}\cdot\rr)\quad\Longrightarrow\quad \hat{H}_{\rm sta}=E_0+\sum_{\kk,\sigma} \eps_\mathbf{k} \hat{\gamma}^\dagger_{\mathbf{k}\sigma}\hat{\gamma}_{\mathbf{k}\sigma}
\end{equation}
 The creation $\hat{\gamma}^\dagger_{\mathbf{k}\sigma}$ and annihilation $\hat{\gamma}_{\mathbf{k}\sigma}$ operators of a BCS quasi-particle with wave vector $\kk$, spin $\sigma$, and eigenenergy $\eps_\kk$, which are the coefficients of this expansion, obey the usual discrete fermionic anticommutation relations, for example $\{\hat{\gamma}_{\mathbf{k}\sigma},\hat{\gamma}^\dagger_{\kk'\sigma'}\}=\delta_{\sigma\sigma'}\delta_{\mathbf{k}\kk'}$ with $\delta$ the Kronecker delta. The expression for the energy $E_0$ of the quasi-particle vacuum is not required here. We find for the modal amplitudes and eigenenergies 
\begin{equation}
\label{eq:ampmodener}
U_\kk=\frac{1}{\sqrt{2}}\left(1+\frac{\xi_\mathbf{k}}{\eps_\mathbf{k}}\right)^{1/2} \quad ; \quad V_\kk=\frac{1}{\sqrt{2}}\left(1-\frac{\xi_\mathbf{k}}{\eps_\mathbf{k}}\right)^{1/2} \quad ; \quad \eps_\kk=(\xi_\kk^2+\Delta^2)^{1/2}
\end{equation}
 with the usual notation shortcuts 
\begin{equation}
\label{eq:rac}
\xi_\kk=E_\kk-\mu \quad\mbox{and}\quad E_\kk=\frac{\hbar^2k^2}{2m}
\end{equation}
 As announced, the excitation spectrum $\eps_\kk$ has a forbidden energy band $[0,E_{\rm gap}]$ with $E_{\rm gap}=\Delta$ if $\mu\geq 0$, $E_{\rm gap}=(\mu^2+\Delta^2)^{1/2}$ if $\mu<0$. The order parameter $\Delta$, taken here as positive real, satisfies according to the dimensionality of the space at 
\begin{equation}
\label{eq:delta}
-\frac{m}{4\pi\hbar^2 a_{\rm 3D}}\stackrel{d=3}{=} \int_{\mathbb{R}^3} \frac{\dd^3k}{(2\pi)^3} \left(\frac{1-2f_\mathbf{k}}{2\eps_\mathbf{k}}-\frac{1}{2 E_\mathbf{k}}\right)\quad ; \quad 0\stackrel{d=2}{=} \int_{\mathbb{R}^2} \frac{\dd^2k}{(2\pi)^2} \left(\frac{1-2f_\mathbf{k}}{2\eps_\mathbf{k}}-\frac{1}{2E_\kk+\eps_{\rm dim}}\right)
\end{equation}
 where the Fermi-Dirac law $f_\kk=1/[\exp(\beta\eps_\kk)+1]$ gives the average thermal number $\langle\hat{\gamma}^\dagger_{\mathbf{k}\sigma}\hat{\gamma}_{\mathbf{k}\sigma}\rangle_{\rm sta}$ of fermionic quasi-particles $\mathbf{k}\sigma$ and we have used the identity $U_\mathbf{k} V_\kk=\Delta/2\eps_\kk$.\footnote{\label{note:micro} We do not specify the microscopic model of the interaction; in practice, it has a non-zero range $b$ that we make tend to zero at the end of the calculations. One possibility is the lattice model $\mathbf{r}\in b\mathbb{Z}^d$ from reference \cite{MoraQC}, with on-site interaction $V(\rr_i,\rr_j)=g_0\delta_{\rr_i\rr_j}/b^d$ and free dispersion relation $\mathbf{k}\in\mathcal{D}\mapsto E_\kk$ ($\mathcal{D}=[-\pi/b,\pi/b[^d$ is the first Brillouin zone). In this case, we simply have $\Delta=g_0\langle\hat{\psi}_\downarrow(\rr)\hat{\psi}_\uparrow(\rr)\rangle_{\rm sta}$, i.e., $-1/g_0=\int_{\mathcal{D}}[\dd^dk/(2\pi)^d](1-2f_\kk)/2\eps_\kk$; by adjusting the bare coupling constant $g_0$ to reproduce the scattering length $a_{d{\rm D}}$ (or, more conveniently in 2D, the dimer binding energy $\eps_{\rm dim}$), then taking the limit $b\to 0^+$, we arrive at equation (\ref{eq:delta}). }

Since the variational density operator $\hat{\sigma}_{\rm sta}$ is Gaussian, Wick's theorem applies and the averages of observables in static BCS theory—marked by the index “sta”—are deduced from the following bilinear averages, called anomalous or normal depending on whether they break $U(1)$ symmetry or not: 
\begin{eqnarray}
\label{eq:moyad}
\langle\hat{\psi}_\downarrow(\rr',t')\psi_\uparrow(\rr,t)\rangle_{\rm sta} &=& \int_{\mathbb{R}^d} \frac{\dd^dk}{(2\pi)^d} U_\mathbf{k} V_\mathbf{k} \left[f_\mathbf{k} \exp(-\mathrm{i}\phi_\kk)-(1-f_\kk)\exp(\mathrm{i}\phi_\kk)\right]\exp[\mathrm{i}\mathbf{k}\cdot(\rr-\rr')]\\
\label{eq:moyai}
\langle\psi_\uparrow(\rr,t)\hat{\psi}_\downarrow(\rr',t')\rangle_{\rm sta} &=& \int_{\mathbb{R}^d} \frac{\dd^dk}{(2\pi)^d} U_\mathbf{k} V_\mathbf{k} \left[(1-f_\kk)\exp(-\mathrm{i}\phi_\kk)-f_\mathbf{k}\exp(\mathrm{i}\phi_\kk)\right]\exp[\mathrm{i}\mathbf{k}\cdot(\rr-\rr')]\\
\label{eq:moynd}
\langle\hat{\psi}^\dagger_\uparrow(\rr,t)\hat{\psi}_\uparrow(\rr',t')\rangle_{\rm sta}&=&\int_{\mathbb{R}^d} \frac{\dd^dk}{(2\pi)^d} [U_\kk^2 f_\mathbf{k} \exp(\mathrm{i}\phi_\kk)+V_\kk^2 (1-f_\kk) \exp(-\mathrm{i}\phi_\kk)] \exp[\mathrm{i}\mathbf{k}\cdot(\rr-\rr')] \\
\label{eq:moyni}
\langle\hat{\psi}_\uparrow(\rr,t)\hat{\psi}^\dagger_\uparrow(\rr',t')\rangle_{\rm sta}&=&\int_{\mathbb{R}^d} \frac{\dd^dk}{(2\pi)^d} \left[U_\kk^2(1-f_\kk)\exp(-\mathrm{i}\phi_\kk)+V_\kk^2f_\mathbf{k}\exp(\mathrm{i}\phi_\kk)\right] \exp[\mathrm{i}\mathbf{k}\cdot(\rr-\rr')]
\end{eqnarray}
 where the time dependence of the field operators is that of Heisenberg's picture for the Hamiltonian $\hat{H}_{\rm sta}$, and we have set $\phi_\kk=\eps_\kk(t-t')/\hbar$ for brevity. Needless to say, for our unpolarized gas without Rabi coupling, the normal averages (\ref{eq:moynd},\ref{eq:moyni}) do not depend on the internal state $\sigma=\uparrow,\downarrow$ chosen, and the cross-normal averages (between two different spin states) are zero. 

By specializing relation (\ref{eq:moynd}) to equal positions and times, we obtain the equation of state of the static BCS theory, relating
 the total average density of the gas $\rho^{\rm sta}=\rho^{\rm sta}_\uparrow+\rho^{\rm sta}_\downarrow$ to the chemical potential $\mu$ and the temperature $T$, 
\begin{equation}
\label{eq:etat}
\rho^{\rm sta}=2\rho^{\rm sta}_\sigma=\int_{\mathbb{R}^d} \frac{\dd^dk}{(2\pi)^d} \left[1-(1-2 f_\kk)\frac{\xi_\mathbf{k}}{\eps_\mathbf{k}}\right]
\end{equation}
 At the same time, we introduce the Fermi wave number $k_{\rm F}^{\rm sta}$ of the ideal gas that is unpolarized at zero temperature but has the same density as the interacting gas: 
\begin{equation}
\label{eq:kf}
k_{\rm F}\stackrel{d=3}{\equiv}(3\pi^2\rho)^{1/3} \quad ; \quad k_{\rm F}\stackrel{d=2}{\equiv} (2\pi\rho)^{1/2}
\end{equation}
 which allows us to define the Fermi energy and temperature, $E_{\rm F}=k_{\rm B}T_{\rm F}=\hbar^2k_{\rm F}^2/2m$. Finally, starting from the definitions (\ref{eq:defg}), we give the pair distribution functions and their deviations from their asymptotic values in the static BCS theory, whose shortcomings we highlighted in section \ref{sec1}: 
\begin{eqnarray}
\label{eq:wickhb}
\delta g^{\rm sta}_{\uparrow\downarrow}(\rr,\rr') &\equiv& g^{\rm sta}_{\uparrow\downarrow}(\rr,\rr')-\rho^{\rm sta}_\uparrow\rho^{\rm sta}_{\downarrow}=+|\langle\hat{\psi}_\downarrow(\rr')\hat{\psi}_\uparrow(\rr)\rangle_{\rm sta}|^2 \\
\label{eq:wickhh}
\delta g^{\rm sta}_{\uparrow\uparrow}(\rr,\rr') &\equiv& g^{\rm sta}_{\uparrow\uparrow}(\rr,\rr')-\rho^{\rm sta}_\uparrow\rho^{\rm sta}_{\uparrow}=-|\langle\hat{\psi}^\dagger_\uparrow(\rr)\hat{\psi}_\uparrow(\rr')\rangle_{\rm sta}|^2
\end{eqnarray}
 with the anomalous and normal averages from equations (\ref{eq:moyad},\ref{eq:moynd}) taken at equal times: 
\begin{eqnarray}
\label{eq:moya}
\langle\hat{\psi}_\downarrow(\rr')\hat{\psi}_\uparrow(\rr)\rangle_{\rm sta}&=&\int_{\mathbb{R}^d} \frac{\dd^dk}{(2\pi)^d} \frac{(-\Delta)}{2\eps_\mathbf{k}}(1-2 f_\kk)\exp[\mathrm{i}\mathbf{k}\cdot(\rr-\rr')] \\
\label{eq:moyn}
\langle\hat{\psi}^\dagger_\uparrow(\rr)\hat{\psi}_\uparrow(\rr')\rangle_{\rm sta}&=&\int_{\mathbb{R}^d} \frac{\dd^dk}{(2\pi)^d} \frac{1}{2}\left[1-(1-2 f_\kk)\frac{\xi_\mathbf{k}}{\eps_\mathbf{k}}\right]\exp[\mathrm{i}\mathbf{k}\cdot(\rr-\rr')]
\end{eqnarray}

 Note that at a temperature $T\ll E_{\rm gap}/k_{\rm B}$, we can neglect the thermal dissociation of bound pairs $\uparrow\downarrow$, thus placing ourselves at order {\it zero} in the small parameter $\exp(-\beta E_{\rm gap})$ without this being very physically restrictive given its exponential decay, and omit the thermal occupation numbers $f_\kk$. We will see in section \ref{sec3_2} that this approximation is justified in cold atomic gases at the minimum accessible temperatures. Analytical forms can then be obtained for the equation on the order parameter (\ref{eq:delta}) and the equation of state (\ref{eq:etat}). In dimension $d=3$, they involve elliptic integrals \cite{Strinati}; inspired by reference \cite{Higgslong}, we simplify them somewhat by means of hyperbolic parameterization $\tau=\argsh (\mu/\Delta)$ and identities $K(\mathrm{i} x)=K(x/\sqrt{x^2+1})/\sqrt{x^2+1}$ and $E(\mathrm{i} x)=E(x/\sqrt{1+x^2})\sqrt{1+x^2}$ ($\forall x\in\mathbb{R}$) on complete elliptic integrals of the first kind $K(k)=F(\pi/2,k)$ and of the second kind $E(k)=E(\pi/2,k)$ \cite{GR}: 
\begin{eqnarray}
\label{eq:delta3d}
-\frac{\pi}{4a_{\rm 3D}}\left(\frac{\hbar^2}{m\Delta}\right)^{1/2} &\stackrel{d=3}{=}& \eee^{+\tau/2} \left[\ch\tau\, K\left(\mathrm{i}\eee^{\tau}\right)-\eee^{-\tau} E\left(\mathrm{i}\eee^{\tau}\right)\right] \\
\label{eq:rho3d}
\frac{3}{2}\pi^2\rho^{\rm sta} \left(\frac{\hbar^2}{m\Delta}\right)^{3/2} &\stackrel{d=3}{=}&\eee^{-\tau/2} \left[\ch\tau\, K\left(\mathrm{i}\eee^{\tau}\right)+\sh\tau\, E\left(\mathrm{i}\eee^{\tau}\right)\right]
\end{eqnarray}
 In dimension $d=2$, on the other hand, they are explicit \cite{Randeria2D} (it suffices to use $\xi_\kk$ as the radial integration variable): 
\begin{multline}
\label{eq:rel2d}
0\stackrel{d=2}{=}\argsh\frac{\mu}{\Delta}+\ln\frac{\eps_{\rm dim}}{\Delta} \quad \mbox{and} \quad \rho^{\rm sta}\stackrel{d=2}{=}\frac{m}{2\pi\hbar^2}\left(\mu+\sqrt{\Delta^2+\mu^2}\right) \\ \Longrightarrow \quad \Delta\stackrel{d=2}{=}[\eps_{\rm dim}(2\mu+\eps_{\rm dim})]^{1/2}=\frac{\hbar^2}{m} k_{\rm F}^{\rm sta} q_{\rm dim} \quad \mbox{and} \quad \rho^{\rm sta} \stackrel{d=2}{=} \frac{m}{2\pi\hbar^2} (2\mu+\eps_{\rm dim})
\end{multline}
 In this case, reference \cite{Romero2D} manages to express the averages (\ref{eq:moya},\ref{eq:moyn}) in terms of Bessel functions, which is remarkable: 
\begin{eqnarray}
\label{eq:moya2d}
\langle\hat{\psi}_\downarrow(\rr')\hat{\psi}_\uparrow(\rr)\rangle_{\rm sta}&\stackrel{d=2}{=}&-\frac{m\Delta}{2\pi\hbar^2} \, J_0(k_{\rm F}^{\rm sta}|\rr-\rr'|) \, K_0(q_{\rm dim}|\rr-\rr'|) \\
\label{eq:moyn2d}
\langle\hat{\psi}^\dagger_\uparrow(\rr)\hat{\psi}_\uparrow(\rr')\rangle_{\rm sta}&\stackrel{d=2}{=}&\frac{m\Delta}{2\pi\hbar^2} \, J_1(k_{\rm F}^{\rm sta}|\rr-\rr'|) \, K_1(q_{\rm dim}|\rr-\rr'|)\end{eqnarray}
 Let us mention a particular interest of the equation of state at order zero in $\exp(-\beta E_{\rm gap})$ for our problem: in the linearized time-dependent BCS theory, the slope $c$ (speed of sound at zero temperature) of the dispersion relation (\ref{eq:depart}) of phonons, whose importance was highlighted in section \ref{sec1}, agrees with the exact relation {\sl} of hydrodynamics of superfluids 
\begin{equation}
\label{eq:hydro}
m c^2(T=0)=\rho \frac{\mathrm{d} \mu}{\mathrm{d}\rho}
\end{equation}
 for the {\it approximated} equation of state $\rho^{\rm sta}(\mu)$ of static BCS theory, due to the non-perturbative variational nature of this theory \cite{CKS}. Hence the expressions in reference \cite{Strinati} in 3D (also simplified by us) and references \cite{Strinati,VanLoon} in 2D: 
\begin{eqnarray}
\label{eq:c3d}
mc^2&\stackrel{d=3}{=}& \frac{2}{3}\Delta \frac{\left[E\left(\mathrm{i}\eee^{\tau}\right)-K\left(\mathrm{i}\eee^{\tau}\right)\right]\,\left[\ch\tau\, K\left(\mathrm{i}\eee^{\tau}\right)+\sh\tau\, E\left(\mathrm{i}\eee^{\tau}\right)\right]}{\left|E\left(\mathrm{i}\eee^{\tau}\right)+\left(\mathrm{i}\eee^{\tau}-1\right)\, K\left(\mathrm{i}\eee^{\tau}\right)\right|^2} = \frac{2}{3}\mu \frac{1+\frac{2\Delta^2}{3\mu\rho^{\rm sta} g_{\rm 3D}}}{1+\left(\frac{2\Delta}{3\rho^{\rm sta} g_{\rm 3D}}\right)^2}\\
\label{eq:c2d}
mc^2&\stackrel{d=2}{=}& \frac{1}{2}\left(\mu+\sqrt{\Delta^2+\mu^2}\right)=\mu+\frac{1}{2}\eps_{\rm dim}= \frac{\pi\hbar^2}{m}\rho^{\rm sta}=E_{\rm F}^{\rm sta}
\end{eqnarray}
 The expression in the right-hand side of the second equality in (\ref{eq:c3d}), which involves the effective coupling constant $g_{\rm 3D}=4\pi\hbar^2 a_{\rm 3D}/m$, is absent from reference \cite{Strinati}; to obtain it, we eliminated the elliptic integrals by solving equations (\ref{eq:delta3d}, \ref{eq:rho3d}) considered formally as a linear system on the unknowns $E(\mathrm{i}\eee^\tau)$ and $K(\mathrm{i}\eee^\tau)$. In the BEC limit where, in 3D, $a_{\rm 3D}>0$ is fixed, $\mu\to -\hbar^2/2 m a^2_{\rm 3D}$, $\rho^{\rm sta}\to 0^+$, and $\Delta^2/\rho^{\rm sta} g_{\rm 3D}\to \hbar^2/m a^2_{\rm 3D}$ \cite{RanderiaBEC}, it allows us to easily verify that $m c^2\sim \rho^{\rm sta} g_{\rm 3D}/4$, which is the expected form $m_{\rm d} c^2=\rho_{\rm d} g^{\rm dd}_{\rm 3D}$ for a weakly interacting Bose gas of dimers $\uparrow\downarrow$ of mass $m_{\rm d}$, density $\rho_{\rm d}$ and effective coupling constant $g^{\rm dd}_{\rm 3D}$; however, the predicted value of $g^{\rm dd}_{\rm 3D}$ is rather poor because the static BCS theory ignores the acoustic excitation branch of the superfluid.\footnote{By definition, we have $g_{\rm 3D}^{\rm dd}=4\pi\hbar^2 a_{\rm 3D}^{\rm dd}/m_{\rm d}$ with $a_{\rm 3D}^{\rm dd}$ the three-dimensional dimer-dimer scattering length. The static BCS theory therefore predicts $a_{\rm 3D}^{\rm dd}=2 a_{\rm 3D}$, which is quite far from the exact value $a_{\rm 3D}^{\rm dd}\simeq 0.60 a_{\rm 3D}$ obtained by direct solution of the Schrödinger equation for four fermions $\uparrow\uparrow\downarrow\downarrow$ with contact interaction \cite{Petrov} or by a diagrammatic method of quantum field theory \cite{Leyronas}. The improvement of the static BCS theory by means of the Gaussian fluctuation approximation, which takes phonon modes into account, see our note \ref{note:afg}, gives the much better result $a_{\rm 3D}^{\rm dd}\simeq 0.55a_{\rm 3D}$ \cite{Randeria3D}.} In 2D, on the other hand, expression (\ref{eq:c2d}) and therefore the equation of state (\ref{eq:rel2d}) are catastrophically wrong in this limit, because they do not reproduce the correct scaling law $m_{\rm d} c^2\sim 2\pi\hbar^2\rho_{\rm d}/[m_{\rm d}|\ln(\rho_{\rm d}^{1/2} a^{\rm dd}_{\rm 2D})|]$ \cite{Schick,Popov}, with $a^{\rm dd}_{\rm 2D}$ the dimer-dimer scattering length. This shortcoming is again due to the omission of phonons by static BCS theory; if we take them into account, we find the expected logarithm of the density \cite{Lianyi2D}.\footnote{The Gaussian fluctuation approximation, see our note \ref{note:afg}, reproduces the logarithmic behavior with a two-dimensional dimer-dimer scattering length $a_{\rm 2D}^{\rm dd}\simeq 0.56 a_{\rm 2D}$.}

Finally, it should be noted that the transition temperature $T_{\rm c}^{\rm sta}$ predicted in 3D by static BCS theory, the solution of equation (\ref{eq:delta}) for $\Delta=0$ and therefore for $\eps_\kk=|\xi_\kk|$, is approximately correct $\propto \Delta(T=0)/k_{\rm B}$ in the BCS limit, of the correct order of magnitude $\propto T_{\rm F}/k_{\rm B}$ at the unitary limit, but catastrophically wrong in the BEC limit where it reproduces—up to a factor—the dissociation temperature $\propto\hbar^2/[m a_{\rm 3D}^2 k_{\rm B}|\ln(k_{\rm F}a_{\rm 3D})|]$ of dimers in the gas (the logarithm is an entropic contribution favoring pair breaking)
 rather than the transition temperature $\propto T_{\rm F}$ of the ideal gas of bosonic dimers \cite{RanderiaBEC}.\footnote{Let us introduce the binding energy of the three-dimensional dimer $\eps_{\rm d}\equiv\hbar^2/m a_{\rm 3D}^2$ for brevity. Then, as stated in reference \cite{RanderiaBEC}, equation (\ref{eq:delta}) shows that $x_{\rm c}\equiv \eps_{\rm d}/k_{\rm B}T_{\rm c}^{\rm sta}\to +\infty$ in the BEC limit and connects it to $\delta\mu\equiv \mu+\eps_{\rm d}/2\to 0^+$; the equation of state (\ref{eq:etat}) written for $\Delta=0$ then connects $x_{\rm c}$ to the density. We find that $\int_{\mathbb{R}^3}[\dd^3k/(2\pi)^3][2\eps_{\rm d}\delta\mu/(2E_\kk+\eps_{\rm d})^2]\sim \int_{\mathbb{R}^3}[\dd^3k/(2\pi)^3] 2 f_k|_{\Delta=0}=\rho^{\rm sta}|_{\Delta=0}$, and therefore ultimately that $(\pi/4)\delta\mu/\eps_{\rm d}\sim (\pi/2)^{1/2}\exp(-x_{\rm c}/2)/x_{\rm c}^{3/2}\sim \pi^2\rho^{\rm sta} a_{\rm 3D}^3$, hence the scaling law announced on $T_{\rm c}^{\rm sta}$ (more precisely, $x_{\rm c}\sim 6|\ln(k^{\rm sta}_{\rm F}a_{\rm 3D})|$, in agreement with \cite{RanderiaBEC}). The only simplifying trick is to eliminate $g_{\rm 3D}$ in (\ref{eq:delta}) by means of the relation $g_{\rm 3D}^{-1}=\int_{\mathbb{R}^3}[\dd^3k/(2\pi)^3][(2 E_\kk)^{-1}-(2 E_\kk+\eps_{\rm d})^{-1}]$, valid for $g_{\rm 3D}>0$.} This will not be very relevant here because we will limit ourselves in practice to temperatures $T\ll T_{\rm F}, \Delta/k_{\rm B}$ and we will set ourselves to order zero in $\exp(-\beta E_{\rm gap})$ in our explicit applications and results, see section \ref{sec3}. Reference \cite{SdMTc} shows how to go beyond static BCS theory in the calculation of $T_{\rm c}$; in particular, it finds exactly the expected value in the BEC limit. The key point is still the consideration of the acoustic excitation branch $\hbar\omega_\qq$ ignored by static BCS: in the BEC limit at fixed wave number $q>0$, the branch trivially reduces to the setting in motion of the center of mass of a bound pair with wave vector $\qq$ and purely kinetic energy $\hbar^2q^2/4m$, thus reproducing the excitation spectrum and the partition function of a condensate of non-interacting bosons of mass $2m$.

\subsection{Obtaining $\delta g_{\sigma\sigma'}^{\rm ltd}(\rr,\rr')$ itself}\label{sec2_2}

It is obtained from a basic sequence of known results, initially based on no approximations.

First, the dynamic structure factor $S_{\sigma\sigma'}(\qq,\omega)$ of the gas in thermal equilibrium, here a two-index tensor since our fermions have non-zero spin, is by definition the Fourier transform (at wave vector $\qq$ and angular frequency $\omega$) of the space-time correlation function of the density, so that 
\begin{equation}
\label{eq:sqom}
\langle\hat{\rho}_\sigma(\rr,t)\hat{\rho}_{\sigma'}(\rr',t')\rangle-\langle\hat{\rho}_\sigma(\rr,t)\rangle\,\langle\hat{\rho}_{\sigma'}(\rr',t')\rangle = \int_{\mathbb{R}^d} \frac{\dd^dq}{(2\pi)^d} \int_\mathbb{R} \frac{\mathrm{d}\omega}{2\pi} \exp\{\mathrm{i}[\mathbf{q}\cdot(\rr-\rr')-\omega(t-t')]\} S_{\sigma\sigma'}(\qq,\omega)
\end{equation}
 where the average is taken in the exact thermal state, Heisenberg's evolution corresponds to the exact Hamiltonian and we have, of course, $\hat{\rho}_\sigma(\rr,t)=\hat{\psi}^\dagger_\sigma(\rr,t)\hat{\psi}_\sigma(\rr,t)$ in second quantization. By specializing equation (\ref{eq:sqom}) at equal times and using the fermionic anticommutation relations on the field operators to put them in normal order as in the definition (\ref{eq:defg}) of pair distribution functions, we obtain 
\begin{equation}
\delta g_{\sigma\sigma'}(\rr,\rr')=-\rho_\sigma{\delta_{\sigma\sigma'}}\delta(\rr-\rr') + \int_{\mathbb{R}^d} \frac{\dd^dq}{(2\pi)^d} \int_\mathbb{R} \frac{\mathrm{d}\omega}{2\pi} \exp[\mathrm{i}\mathbf{q}\cdot(\rr-\rr')] S_{\sigma\sigma'}(\qq,\omega)
\end{equation}
 where $\delta$ is the Dirac distribution in $\mathbb{R}^d$.

Second, the fluctuation-dissipation theorem (see section \S124 of reference \cite{Landaustat}), also known as the Callen-Welton theorem \cite{Callen}, relates the dynamic structure factor to the susceptibility $\tilde{\chi}_{\sigma\sigma'}(\qq,\hbar\omega+\mathrm{i} 0^+)$, itself the Fourier transform of the linear density-density response function,\footnote{The application of an infinitesimal time-dependent external potential $U$ in the internal state $\sigma'$ induces in the internal state $\sigma$, at time $t$ and at point $\rr$, a change in average density $\delta\rho_{\sigma}(\rr,t)=\int \dd^dr'\mathrm{d} t' \chi_{\sigma\sigma'}(\rr,t;\rr',t') U(\rr',t')$, where the response function is given by Kubo's formula, 
\begin{equation}
\label{eq:Kubo}
\chi_{\sigma\sigma'}(\rr,t;\rr',t')=\langle{\frac{1}{\mathrm{i}\hbar}}[\hat{\rho}_\sigma(\rr,t),\hat{\rho}_{\sigma'}(\rr',t')]\rangle \Theta(t-t')
\end{equation}
 In this formula, the average and Heisenberg evolution are those of the unperturbed system, and $\Theta$ is the Heaviside function. Since the gas is initially spatially homogeneous and in equilibrium, $\chi_{\sigma\sigma'}(\rr,t;\rr',t')$ is a function only of the relative variables $\rr-\rr'$ and $t-t'$, whose susceptibility is the Fourier transform regularized as follows: 
\begin{equation}
\label{eq:gen}
\tilde{\chi}_{\sigma\sigma'}(\qq,z)=\int_{\mathbb{R}^d}\dd^dr\int_{\mathbb{R}}\mathrm{d} t\, \chi_{\sigma\sigma'}(\rr,t;\rr',t')\eee^{\mathrm{i}[z (t-t')/\hbar-\mathbf{q}\cdot(\rr-\rr')]}
\end{equation}
 with $z=\hbar\omega+\mathrm{i}\eta$, $\eta\to 0^+$. The factor $\exp[-\eta(t-t')/\hbar]$ thus introduced ensures the convergence of the integral over time. Expression (\ref{eq:gen}) allows $\tilde{\chi}_{\sigma\sigma'}(\qq,z)$ to be extended to the upper complex half-plane $\im z>0$, and property $[{\tilde{\chi}}_{\sigma\sigma'}(\qq,z)]^*={\tilde{\chi}}_{\sigma\sigma'}(\qq,z^*)$ to the lower half-plane. The function obtained is analytic on $\mathbb{C}\setminus\mathbb{R}$ but has singularities (poles, branch cuts) on the real axis.} as follows: 
\begin{equation}
\label{eq:theo}
\im \left[\tilde{\chi}_{\sigma\sigma'}(\qq,\hbar\omega+\mathrm{i} 0^+)\right]= -\frac{1}{2\hbar} S_{\sigma\sigma'}(\qq,\omega) [1-\exp(-\beta\hbar\omega)]
\end{equation}
 It follows that\footnote{For a finite-size system, it would generally be necessary to take into account, in the dynamic structure factor, a contribution proportional to $\delta(\omega)$ that is not constrained by (\ref{eq:theo}), due in particular to a difference between the thermodynamic susceptibility (which reflects the change in density at thermal equilibrium in the presence of a reservoir at temperature $T$) and the zero-frequency limit of the dynamic susceptibility $\tilde{\chi}_{\sigma\sigma'}(\qq,\hbar\omega+i 0^+)$ (which reflects the response of a purely Hamiltonian isolated system), see Chapter 8 of reference \cite{Levy}, equation (66) in reference \cite{Iac1D} and its appendix C.}, after the change of variable $\omega=\eps/\hbar$: 
\begin{equation}
\label{eq:expdg}
\delta g_{\sigma\sigma'}(\rr,\rr')=-\rho_\sigma{\delta_{\sigma\sigma'}}\delta(\rr-\rr') + \int_{\mathbb{R}^d} \frac{\dd^dq}{(2\pi)^d} \exp[\mathrm{i}\mathbf{q}\cdot(\rr-\rr')] \int_\mathbb{R} \frac{\mathrm{d}\eps}{(-\pi)} \frac{\im[\tilde{\chi}_{\sigma\sigma'}(\qq,\eps+\mathrm{i} 0^+)]}{1-\exp(-\beta\eps)}
\end{equation}

 Thirdly, density-density susceptibilities have already been calculated approximately using the time-dependent BCS theory linearized around its steady state.\footnote{\label{note:plutot} The literature tends to give susceptibilities on total density and differential density. Those of interest to us can be deduced by bilinearity and $\uparrow$-$\downarrow$ exchange invariance: $\chi_{\hat{\rho}_\uparrow+\varepsilon\hat{\rho}_{\downarrow},\hat{\rho}_\uparrow+\varepsilon\hat{\rho}_{\downarrow}}=\chi_{\uparrow\uparrow}+\varepsilon\chi_{\uparrow\downarrow}+\varepsilon\chi_{\downarrow\uparrow}+\chi_{\uparrow\downarrow}=2(\chi_{\uparrow\uparrow}+\varepsilon\chi_{\uparrow\downarrow})$ with $\varepsilon\in\{-1;+1\}$.}
 By generalizing the results of references \cite{LiHe,Higgslong,YvanSpectro} to non-zero temperatures (as in reference \cite{HK}\footnote{The susceptibilities in reference \cite{HK}, see equations (43,46) in this reference, are extensive rather than intensive; they differ from ours by a factor of $L^d/2$ for the total density-total density response (noted $\tilde{\chi}_{33}$ in \cite{HK}), and by a factor of $-L^d/2$ for the differential density-differential density response (noted $\tilde{\chi}_{44}$ in \cite{HK} and $2\Sigma_{44}$ here). Note that references \cite{Higgslong,YvanSpectro} omitted the differential density.}), we find at wave vector $\qq$ and complex energy $z$ that 
\begin{equation}
\label{eq:chihb}
\tilde{\chi}^{\rm ltd}_{\uparrow\downarrow}(\qq,z)=\frac{1}{2}\left(\Sigma_{33}-\Sigma_{44}\right)+\frac{2\Sigma_{13}\Sigma_{23}\Sigma_{12}-\Sigma_{13}^2\Sigma_{22}-\Sigma_{23}^2\Sigma_{11}}{2\left(\Sigma_{11}\Sigma_{22}-\Sigma_{12}^2\right)}
\end{equation}
 \begin{equation}
\label{eq:chihh}
\tilde{\chi}^{\rm ltd}_{\uparrow\uparrow}(\qq,z)=\frac{1}{2}\left(\Sigma_{33}+\Sigma_{44}\right)+\frac{2\Sigma_{13}\Sigma_{23}\Sigma_{12}-\Sigma_{13}^2\Sigma_{22}-\Sigma_{23}^2\Sigma_{11}}{2\left(\Sigma_{11}\Sigma_{22}-\Sigma_{12}^2\right)}
\end{equation}
 where the exponent “ltd” stands for "linearized time-dependent BCS theory". We have introduced the following functions in the right-hand side, implying their dependence on the variables $(\qq,z)\in\mathbb{R}^d\times(\mathbb{C}\setminus\mathbb{R})$ for simplicity:\footnote{In the lattice model $\mathbf{r}\in b\mathbb{Z}^d$ in note \ref{note:micro}, the counterterm of equations (\ref{eq:s11},\ref{eq:s22}) is simply $-1/g_0$; it was replaced by its expression given in the note before taking the limit $b\to 0^+$. We verify that $\Sigma_{11}(\qq,z)\to 0$ when $(\qq,z)\to (\mathbf{0},0)$, a key property for the existence of an acoustic excitation branch at all orders in $\exp(-\beta E_{\rm gap})$ (in this case, equation (\ref{eq:valpro}) for $\hbar\omega_\qq$ must be generalized by analytic continuation to the lower complex half-plane; we retain the linear-start property (\ref{eq:depart}) but with a complex speed of sound $c(T)$, see references \cite{Kulik,Tempere}; however, this complex speed ignores Landau damping on the convex acoustic branch because the linearized time-dependent BCS theory ignores the interaction between phonons, see note 64 of reference \cite{qouv}).} 
\begin{eqnarray}
\nonumber
\Sigma_{11}&\hspace{-1mm}=\hspace{-1mm}&\int_{\mathbb{R}^d}\!\!\frac{\dd^dk}{(2\pi)^d} \left\{\frac{(\eps_+\!\!+\!\!\eps_-)(\eps_+\eps_-\!\!+\!\!\xi_+\xi_-\!\!+\!\!\Delta^2)(1\!\!-\!\!f_+\!\!-\!\!f_-)}{2\eps_+\eps_-[z^2\!\!-\!\!(\eps_+\!\!+\!\!\eps_-)^2]}\!\!+\!\!\frac{(\eps_+\!\!-\!\!\eps_-)(-\eps_+\eps_-\!\!+\!\!\xi_+\xi_-\!\!+\!\!\Delta^2)(f_+\!\!-\!\!f_-)}{2\eps_+\eps_-[z^2\!\!-\!\!(\eps_+\!\!-\!\!\eps_-)^2]}\!\!+\!\!\frac{1{\!-2\!f_\mathbf{k}}}{2\eps_\mathbf{k}}\right\}\\
\label{eq:s11}&&\\
\nonumber
\Sigma_{22}&\hspace{-1mm}=\hspace{-1mm}&\int_{\mathbb{R}^d}\!\!\frac{\dd^dk}{(2\pi)^d}\left\{\frac{(\eps_+\!\!+\!\!\eps_-)(\eps_+\eps_-\!\!+\!\!\xi_+\xi_-\!\!-\!\!\Delta^2)(1\!\!-\!\!f_+\!\!-\!\!f_-)}{2\eps_+\eps_-[z^2\!\!-\!\!(\eps_+\!\!+\!\!\eps_-)^2]}\!\!+\!\!\frac{(\eps_+\!\!-\!\!\eps_-)(-\eps_+\eps_-\!\!+\!\!\xi_+\xi_-\!\!-\!\!\Delta^2)(f_+\!\!-\!\!f_-)}{2\eps_+\eps_-[z^2\!\!-\!\!(\eps_+\!\!-\!\!\eps_-)^2]}\!\! +\!\! \frac{1{\!-2\!f_\mathbf{k}}}{2\eps_\mathbf{k}}\right\}\\
\label{eq:s22}&&\\
\label{eq:s33}
\Sigma_{33}&\hspace{-1mm}=\hspace{-1mm}&\int_{\mathbb{R}^d}\!\!\frac{\dd^dk}{(2\pi)^d} \left\{\frac{(\eps_+\!\!+\!\!\eps_-)(\eps_+\eps_-\!\!-\!\!\xi_+\xi_-\!\!+\!\!\Delta^2)(1\!\!-\!\!f_+\!\!-\!\!f_-)}{2\eps_+\eps_-[z^2\!\!-\!\!(\eps_+\!\!+\!\!\eps_-)^2]}\!\!+\!\!\frac{(\eps_+\!\!-\!\!\eps_-)(-\eps_+\eps_-\!\!-\!\!\xi_+\xi_-\!\!+\!\!\Delta^2)(f_+\!\!-\!\!f_-)}{2\eps_+\eps_-[z^2\!\!-\!\!(\eps_+\!\!-\!\!\eps_-)^2]} \right\}\\
\label{eq:s44}
\Sigma_{44}&\hspace{-1mm}=\hspace{-1mm}&\int_{\mathbb{R}^d}\!\!\frac{\dd^dk}{(2\pi)^d} \left\{\frac{(\eps_+\!\!+\!\!\eps_-)(\eps_+\eps_-\!\!-\!\!\xi_+\xi_-\!\!-\!\!\Delta^2)(1\!\!-\!\!f_+\!\!-\!\!f_-)}{2\eps_+\eps_-[z^2\!\!-\!\!(\eps_+\!\!+\!\!\eps_-)^2]}\!\!+\!\!\frac{(\eps_+\!\!-\!\!\eps_-)(-\eps_+\eps_-\!\!-\!\!\xi_+\xi_-\!\!-\!\!\Delta^2)(f_+\!\!-\!\!f_-)}{2\eps_+\eps_-[z^2\!\!-\!\!(\eps_+\!\!-\!\!\eps_-)^2]} \right\}\\
\label{eq:s12}
\Sigma_{12}&\hspace{-1mm}=\hspace{-1mm}&\int_{\mathbb{R}^d}\!\!\frac{\dd^dk}{(2\pi)^d} \left\{\frac{z(\eps_+\xi_-\!\!+\!\!\eps_-\xi_+)(1\!\!-\!\!f_+\!\!-\!\!f_-)}{2\eps_+\eps_-[z^2\!\!-\!\!(\eps_+\!\!+\!\!\eps_-)^2]}\!\!+\!\!\frac{z(\eps_+\xi_-\!\!-\!\!\eps_-\xi_+)(f_+\!\!-\!\!f_-)}{2\eps_+\eps_-[z^2\!\!-\!\!(\eps_+\!\!-\!\!\eps_-)^2]} \right\}\\
\label{eq:s13}
\Sigma_{13}&\hspace{-1mm}=\hspace{-1mm}&\int_{\mathbb{R}^d}\!\!\frac{\dd^dk}{(2\pi)^d} \left\{\frac{z\Delta(\eps_+\!\!+\!\!\eps_-)(1\!\!-\!\!f_+\!\!-\!\!f_-)}{2\eps_+\eps_-[z^2\!\!-\!\!(\eps_+\!\!+\!\!\eps_-)^2]}\!\!+\!\!\frac{z\Delta(\eps_+\!\!-\!\!\eps_-)(f_+\!\!-\!\!f_-)}{2\eps_+\eps_-[z^2\!\!-\!\!(\eps_+\!\!-\!\!\eps_-)^2]} \right\}\\
\label{eq:s23} 
\Sigma_{23}&\hspace{-1mm}=\hspace{-1mm}&\int_{\mathbb{R}^d}\!\!\frac{\dd^dk}{(2\pi)^d} \left\{\frac{\Delta(\eps_+\!+\!\eps_-)(\xi_+\!+\!\xi_-)(1\!-\!f_+\!-\!f_-)}{2\eps_+\eps_-[z^2\!-\!(\eps_+\!+\!\eps_-)^2]}\!+\!\frac{\Delta(\eps_+\!-\!\eps_-)(\xi_+\!+\!\xi_-)(f_+\!-\!f_-)}{2\eps_+\eps_-[z^2\!-\!(\eps_+\!-\!\eps_-)^2]} \right\}
\end{eqnarray}
 with the very compact notations\footnote{Starting from definitions (\ref{eq:ampmodener}), we eliminated modal amplitudes $U_\pm=U_{\kk_\pm}$ and $V_\pm=V_{\kk_\pm}$ using relations $(U_+V_-+s U_-V_+)^2=(\eps_+\eps_--\xi_+\xi_-+s\Delta^2)/2\eps_+\eps_-$, $(U_+U_-+s V_+V_-)^2=(\eps_+\eps_-+\xi_+\xi_-+s\Delta^2)/2\eps_+\eps_-$, $(U_+V_-+s U_-V_+) (U_+U_-+s V_+V_-)=\Delta(\eps_++s\eps_-)/2\eps_+\eps_-$, and $(U_+V_-+s U_-V_+) (U_+U_--s V_+V_-)=\Delta(\xi_++s\xi_-)/2\eps_+\eps_-$ ($s\in\{-1;+1\}$). Note that, in functions $\Sigma_{ij}$, the second term can be deduced from the first by making the formal change $\eps_-\to-\eps_-$ (which has the effect of changing $f_-$ to $1-f_-$) without affecting $\eps_+$ or $\xi_\pm$.} 
\begin{equation}
\label{eq:defkpm}
\kk_\pm=\mathbf{k}\pm\qq/2 \quad;\quad \xi_\pm=\xi_{\kk_\pm} \quad ; \quad \eps_\pm=\eps_{\kk_\pm} \quad ; \quad f_\pm=f_{\kk_\pm}
\end{equation}
 Note that the functions $\Sigma_{ij}$ have well-defined parity for the variable $z$ (they are even, except for $\Sigma_{12}$ and $\Sigma_{13}$, which are odd), and that they are not independent, in the sense that 
\begin{equation}
\label{eq:lien}
\Sigma_{11}(\qq,z)-\Sigma_{22}(\qq,z)=\Sigma_{33}(\qq,z)-\Sigma_{44}(\qq,z) \quad\mbox{and}\quad \Sigma_{13}(\qq,z)=\frac{z}{2\Delta} \left[\Sigma_{11}(\qq,z)-\Sigma_{22}(\qq,z)\right]\quad\forall (\qq,z)
\end{equation}
 We have of course $\tilde{\chi}^{\rm ltd}_{\downarrow\uparrow}\equiv\tilde{\chi}^{\rm ltd}_{\uparrow\downarrow}$ and $\tilde{\chi}^{\rm ltd}_{\downarrow\downarrow}\equiv\tilde{\chi}^{\rm ltd}_{\uparrow\uparrow}$ by symmetry between the two spin states. All these susceptibilities are even functions of $z$.

Carrying over expressions (\ref{eq:chihb},\ref{eq:chihh}) into equation (\ref{eq:expdg}) leads to the functions $\delta g^{\rm ltd}_{\sigma\sigma'}(\rr,\rr')$ of the linearized time-dependent BCS theory. We will study them in detail below.

\subsection{Isolating the collective part in $\delta g^{\rm ltd}_{\sigma\sigma'}(\rr,\rr')$}\label{sec2_3}

It turns out that the first contributions to the right-hand side of (\ref{eq:chihb},\ref{eq:chihh}), to the left of the large fractions, are simply the density-density susceptibilities of the static BCS theory. To see this, we apply Kubo's formula (\ref{eq:Kubo}) to this theory and then directly use Wick's theorem and the two-time correlation functions (\ref{eq:moyad},\ref{eq:moyai},\ref{eq:moynd},\ref{eq:moyni}), without trying to perform the calculation for the commutator. Elementary manipulations\footnote{Parity invariance allows us to change the sign of the wave vectors under the Fourier exponentials at will, so in particular—as in (\ref{eq:chistahh})—symmetrize the integrand in integrals over two wave vectors using the identities $\int\dd^dk_1\int\dd^dk_2 f(\kk_1,\kk_2)\exp[\ii(\kk_1-\kk_2)\cdot(\rr-\rr')]=\int\dd^dk_1\int\dd^dk_2 f(\kk_2,\kk_1)\exp[\ii(\kk_2-\kk_1)\cdot(\rr-\rr')]=\int\dd^dk_1\int\dd^dk_2 (1/2)[f(\kk_1,\kk_2)+f(-\kk_2,-\kk_1)]\exp[\ii(\kk_1-\kk_2)\cdot(\rr-\rr')]$. In our case, the modal amplitudes and eigenvalues, i.e., the functions $f$ are even and depend only on the moduli of the wave vectors.} lead to the response functions for $t>t'$ (they are zero for $t<t'$): 
\begin{eqnarray}
\nonumber
\chi^{\rm sta}_{\uparrow\downarrow}(\rr,t;\rr',t')&\hspace{-1mm}=\hspace{-1mm}&\frac{1}{\mathrm{i}\hbar}\int_{\mathbb{R}^d}\hspace{-1mm}\frac{\dd^dk_1}{(2\pi)^d} \int_{\mathbb{R}^d}\hspace{-1mm}\frac{\dd^dk_2}{(2\pi)^d} U_{\kk_1} V_{\kk_1} U_{\kk_2} V_{\kk_2} \left\{\left[\eee^{\ii(\phi_{\kk_1}+\phi_{\kk_2})}-\eee^{-\ii(\phi_{\kk_1}+\phi_{\kk_2})}\right](f_{\kk_1}+f_{\kk_2}-1)\right. \\
&& \left.+\left[\eee^{\ii(\phi_{\kk_1}-\phi_{\kk_2})}-\eee^{-\ii(\phi_{\kk_1}-\phi_{\kk_2})}\right](f_{\kk_2}-f_{\kk_1})\right\}\exp[\ii(\kk_1-\kk_2)\cdot(\rr-\rr')] 
\label{eq:chistahb} \\
\nonumber
\chi^{\rm sta}_{\uparrow\uparrow}(\rr,t;\rr',t')&\hspace{-1mm}=\hspace{-1mm}&\frac{1}{\mathrm{i}\hbar}\int_{\mathbb{R}^d}\hspace{-1mm}\frac{\dd^dk_1}{(2\pi)^d} \int_{\mathbb{R}^d}\hspace{-1mm}\frac{\dd^dk_2}{(2\pi)^d} \left\{\left[\eee^{\ii(\phi_{\kk_1}+\phi_{\kk_2})}-\eee^{-\ii(\phi_{\kk_1}+\phi_{\kk_2})}\right]\frac{1}{2}(U^2_{\kk_1}V^2_{\kk_2}+U^2_{\kk_2}V^2_{\kk_1})(f_{\kk_1}+f_{\kk_2}-1)\right. \\
\nonumber
&& \left.+\left[\eee^{\ii(\phi_{\kk_1}-\phi_{\kk_2})}-\eee^{-\ii(\phi_{\kk_1}-\phi_{\kk_2})}\right](f_{\kk_1}-f_{\kk_2})\frac{1}{2}(U^2_{\kk_1}U^2_{\kk_2}+V^2_{\kk_1}V^2_{\kk_2})\right\} \exp[\ii(\kk_1-\kk_2)\cdot(\rr-\rr')]\\
&& \label{eq:chistahh}
\end{eqnarray}
 It remains to take the spatio-temporal Fourier transform as in equation (\ref{eq:gen}) after the change of variables of unit Jacobian $\kk_{1,2}=\mathbf{k}\pm\qq/2$ to find the announced result: 
\begin{equation}
\label{eq:chitsta}
\tilde{\chi}^{\rm sta}_{\uparrow\downarrow}(\qq,z)=\frac{1}{2}\left[\Sigma_{33}(\qq,z)-\Sigma_{44}(\qq,z)\right] \quad ; \quad \tilde{\chi}^{\rm sta}_{\uparrow\uparrow}(\qq,z)=\frac{1}{2}\left[\Sigma_{33}(\qq,z)+\Sigma_{44}(\qq,z)\right]
\end{equation}

 Thus, the large fraction in (\ref{eq:chihb},\ref{eq:chihh})—it is exactly the same in both cases—takes into account the collective excitations (Bogoliubov-Anderson phonons) ignored by the static BCS theory and by references \cite{Romero3D,Romero2D}. In view of the studies that will follow, let us isolate its contribution to the pair distribution functions by setting (an exponent “ltd” would be redundant on the collective contribution): 
\begin{equation}
\label{eq:defcoll}
\tilde{\chi}_{\rm coll}(\qq,z)=\frac{2\Sigma_{13}\Sigma_{23}\Sigma_{12}-\Sigma_{13}^2\Sigma_{22}-\Sigma_{23}^2\Sigma_{11}}{2\left(\Sigma_{11}\Sigma_{22}-\Sigma_{12}^2\right)} \quad\mbox{and}\quad 
\delta\tilde{g}_{\rm coll}(\qq) \equiv \int_\mathbb{R} \frac{\mathrm{d}\eps}{(-\pi)} \frac{\im[\tilde{\chi}_{\rm coll}(\qq,\eps+\mathrm{i} 0^+)]}{1-\exp(-\beta\eps)}
\end{equation}
 It follows that 
\begin{equation}
\label{eq:sepa}
\boxed{\delta g^{\rm ltd}_{\sigma\sigma'}(\rr,\rr')=\delta g^{\rm sta}_{\sigma\sigma'}(\rr,\rr')+\delta g_{\rm coll}(\rr,\rr')}
\end{equation}
 with the common collective contribution 
\begin{equation}
\label{eq:tfdgcoll}
\boxed{\delta g_{\rm coll}(\rr,\rr')=\int_{\mathbb{R}^d} \frac{\dd^dq}{(2\pi)^d} \delta\tilde{g}_{\rm coll}(\qq) \exp[\mathrm{i}\mathbf{q}\cdot(\rr-\rr')]}
\end{equation}
 We deduce in particular the beautiful naturally equivalent relations 
\begin{eqnarray}
\label{eq:belle}
\delta g^{\rm ltd}_{\uparrow\uparrow}(\rr,\rr')-\delta g^{\rm ltd}_{\uparrow\downarrow}(\rr,\rr') &=&
\delta g^{\rm sta}_{\uparrow\uparrow}(\rr,\rr')-\delta g^{\rm sta}_{\uparrow\downarrow}(\rr,\rr')\\
\label{eq:belle2}
\delta g^{\rm ltd}_{\uparrow\downarrow}(\rr,\rr')-\delta g^{\rm sta}_{\uparrow\downarrow}(\rr,\rr')&=&
\delta g^{\rm ltd}_{\uparrow\uparrow}(\rr,\rr') - \delta g^{\rm sta}_{\uparrow\uparrow}(\rr,\rr')
\end{eqnarray}
the second of which we will return to in equation (\ref{eq:belle3}). For completeness, let us perform the calculation of the Fourier transform of $\delta g^{\rm sta}_{\sigma\sigma'}(\rr,\rr')$ in the sense of equation (\ref{eq:tfdgcoll}) by applying the fluctuation-dissipation theorem—more precisely, the resulting equation (\ref{eq:expdg})—to the susceptibilities (\ref{eq:chitsta}) of the static BCS theory:\footnote{Basically, it suffices to use property $\im\{\eps_0/[(\eps+\mathrm{i} 0^+)^2-\eps_0^2]\}=(\pi/2)[\delta(\eps+\eps_0)-\delta(\eps-\eps_0)]$ and not to overlook remarkable simplifications such as $(1-f_+-f_-)\{\exp[\beta(\eps_++\eps_-)]+1\}/\{\exp[\beta(\eps_++\eps_-)]-1\}-(f_+-f_-)\{\exp[\beta(\eps_+-\eps_-)]+1\}/\{\exp[\beta(\eps_+-\eps_-)]-1\}=1$. To eliminate the Dirac distribution in (\ref{eq:expdg}), we recall the expression (\ref{eq:etat}) from $\rho^{\rm sta}_{{\sigma}}$ and realize that $\int\dd^dq\,\exp[\mathrm{i}\mathbf{q}\cdot(\rr-\rr')]\int\dd^dk\, f(\kk_\pm)=(2\pi)^d \delta(\rr-\rr')\int\dd^dk \, f(\kk)$ for any function $f$ of integral convergent on $\mathbb{R}^d$.} 
\begin{eqnarray}
\label{eq:fourhb}
\delta\tilde{g}^{\rm sta}_{\uparrow\downarrow}(\qq)&=&+\int_{\mathbb{R}^d} \frac{\dd^dk}{(2\pi)^d} \frac{\Delta^2}{4\eps_+\eps_-} (1-2 f_+)(1-2 f_-)\\
\label{eq:fourhh}
\delta\tilde{g}^{\rm sta}_{\uparrow\uparrow}(\qq)&=&-\int_{\mathbb{R}^d} \frac{\dd^dk}{(2\pi)^d} \frac{1}{4}\left[1-\frac{\xi_+}{\eps_+}(1-2 f_+)\right] \left[1-\frac{\xi_-}{\eps_-}(1-2 f_-)\right]\end{eqnarray}
 The factorized forms of the integrand—as a function of $\eps_+$ times the same function of $\eps_-$—herald the Wick expressions (\ref{eq:wickhb},\ref{eq:wickhh}) of $\delta g^{\rm sta}_{\sigma\sigma'}(\rr,\rr')$, which are perfect squares, and which are found by the change of variables of unit Jacobian $(\kk,\qq)\mapsto (\kk_+,\kk_-)$ in the inverse Fourier transform of the $\delta\tilde{g}^{\rm sta}_{\sigma\sigma'}(\qq)$.

\subsection{Different formulations in terms of curvilinear integrals and applications}\label{sec2_4}

The writing of the collective contribution $\delta\tilde{g}_{\rm coll}(\qq)$ in (\ref{eq:defcoll}), directly reflecting the original one of the fluctuation-dissipation theorem (\ref{eq:theo}), is inconvenient to use, analytically and numerically, because its integrand is very singular.

Indeed, as can be seen from their definitions (\ref{eq:s11}--\ref{eq:s23}), the functions $z\mapsto\Sigma_{ij}(\qq,z)$ have branch cuts on the real axis, where the denominators vanish for certain values of the integration vector $\kk$, in practice of its modulus $k$ and its angle $\theta$ with the wave vector $\qq$ (such as $\mathbf{k}\cdot\mathbf{q} = k q\cos \theta$). The denominator of the second contributions between braces vanishes if $z=\pm(\eps_+-\eps_-)$; this is the resonance condition for the transfer process of a pre-existing — thermal — fermionic quasi-particle from wave vector $\kk_-=\kk-\qq/2$ to wave vector $\kk_+=\kk+\qq/2$ or the opposite by an external excitation potential of wave vector $\qq$ and angular frequency $z/\hbar=\omega$ (a Bragg excitation in cold atom experiments \cite{Bragg}). However, when $\kk$ describes $\mathbb{R}^d$, the energy difference $\eps_+-\eps_-$ describes $\mathbb{R}$;\footnote{To see this, for a fixed nonzero $\qq$, we take the limits $k\to +\infty$ and $\cos\theta\to 0$ while keeping the product $k\cos\theta$ constant; then $\eps_+-\eps_-\sim \xi_+-\xi_-=\hbar q k (\cos\theta)/m$ tends towards any desired value.} the functions $\Sigma_{ij}$ therefore have the entire real axis as a branch cut.

Fortunately, the situation simplifies somewhat at order zero in $\exp(-\beta E_{\rm gap})$, where we will place ourselves—unless otherwise stated—in section \ref{sec3}: the fermionic occupation numbers $f_\pm$ are then set to zero, and the functions $\Sigma_{ij}$ are reduced to the first contributions between curly brackets. In this case, the functions $\Sigma_{ij}$ have a branch cut only on the real half-lines $]-\infty,-E_{\rm bord}(\qq)]$ and $[E_{\rm bord}(\qq),+\infty[$, where $E_{\rm bord}(\qq)=\inf_\mathbf{k} {(}\eps_++\eps_-{)}$ is the lower edge of the broken pair continuum with a fixed total wave vector $\qq$ (under the effect of an excitation potential transmitting a well-defined momentum $\hbar\qq$, a bound pair $\uparrow\downarrow$ of the condensate can break into two fermionic quasi-particles with a fixed total momentum $\hbar\qq$ but with a free relative wave vector $\kk$ in $\mathbb{R}^d$; so the corresponding energy variation describes an interval of $\mathbb{R}$). We have explicitly \cite{CKS,vcsuper}:\footnote{It would be too naive to believe that the branch cuts have only their endpoints $\pm E_{\rm bord}(\qq)$ as branch points. Depending on the values of $q$ and the strength of the interactions, there may be one or two other branch points in $]E_{\rm bord}(\qq),+\infty[$ and in $]-\infty,-E_{\rm bord}(\qq)[$, which we identify as angular points in spectral densities $\rho_{ij}(\qq,\eps)$ (such as $\Sigma_{ij}(\qq,z)\propto\int\mathrm{d}\eps \rho_{ij}(\qq,\eps)/(z-\eps)$ or, if you prefer, $\im\Sigma_{ij}(\qq,\eps+\mathrm{i} 0^+)\propto \rho_{ij}(\qq,\eps)$) \cite{Higgslong}. The same phenomenon occurs beyond order zero in $\exp(-\beta E_{\rm gap})$, with the possibility of additional, purely thermal branch points \cite{Klimin}. References \cite{Higgslong,Klimin} are in 3D, but their considerations apply to 2D, except, of course, for the final expression of spectral densities in terms of elliptic integrals in \cite{Higgslong}, which results from the form $\int\dd(\cos\theta)$ of the polar integral in three dimensions. } 
\begin{equation}
\label{eq:ebord}
E_{\rm bord}(\qq)=\left\{\begin{array}{cl} 2\Delta & \quad\mbox{if}\quad \mu\geq 0 \ \mbox{and}\ q<2(2m\mu/\hbar^2)^{1/2} \\
2\eps_{\qq/2} & \quad\mbox{otherwise} \end{array} \right.
\end{equation}
 where, in the case $\mu>0$, $k_{\rm min}=(2m\mu/\hbar^2)^{1/2}$ is the position of the minimum $E_{\rm gap}=\Delta$ of the dispersion relation $k\mapsto\eps_\kk$ of the fermionic quasi-particles (for $\mu<0$, the minimum $E_{\rm gap}=(\mu^2+\Delta^2)^{1/2}$ is reached in $k=0$). Furthermore, if the wave number $q$ belongs to the domain of existence $q\in\mathcal{E}_{\rm son}$ of the acoustic excitation branch in the linearized time-dependent BCS theory,\footnote{\label{note:dom} In 2D, a phonon mode exists at any wave number $q>0$, $\mathcal{E}_{\rm son}^{d=2}=]0,+\infty[$ \cite{VanLoon}. In 3D, the domain of existence is bounded and connected, $\mathcal{E}_{\rm son}^{d=3}=]0,q_{\rm sup}[$, for $a_{\rm 3D}<0$, i.e. for $\Delta/\mu\in ]0\,;1.162 [$; it has two connected components, $\mathcal{E}_{\rm son}^{d=3}=]0,q_{\rm sup}[\cup]q_{\rm inf},+\infty[$ if $\Delta/\mu\in ]1.162\,; 1.729[$; finally, it is equal to half entire real axis, $\mathcal{E}_{\rm son}^{d=3}=]0,+\infty[$, if $\Delta/\mu>1.729$ or $\mu<0$ \cite{CKS,concav}.} the collective susceptibility (\ref{eq:defcoll}) has a pole at the eigenenergy $\hbar\omega_\mathbf{q}\in ]0,E_{\rm bord}(\qq)[$ of the corresponding Bogoliubov-Anderson phonon, and at its opposite $-\hbar\omega_\qq$, since by definition\footnote{\label{note:afg} We obtain the same eigenvalue equation in the Gaussian fluctuation approximation, that is in a path integral formulation where, after introducing an auxiliary field in the Hubbard-Stratonovich transformation and then summing over the fermionic degrees of freedom, we quadratize the action obtained around its static BCS stationary point \cite{SdMTc,RanderiaBEC,Strinati,Drummond,Randeria3D} (this gives us a Gaussian path integral that we know how to calculate, but that's another story ). To convince the reader, let us compare explicitly with \cite{Randeria3D}: in the notation of this reference, the eigenvalue equation is written as $M_{11}(\qq,z)M_{22}(\qq,z)=M_{12}(\qq,z)M_{21}(\qq,z)$; however, we find that $M_{11}=(\Sigma_{11}+\Sigma_{22})/2+\Sigma_{12}$, $M_{22}=(\Sigma_{11}+\Sigma_{22})/2-\Sigma_{12}$, and $M_{12}=M_{21}=(\Sigma_{22}-\Sigma_{11})/2$, again omitting the variables $(\qq,z)$. We simply need to expand and simplify to find (\ref{eq:valpro}).}${}^{,}$\footnote{If we leave order zero in $\exp(-\beta E_{\rm gap})$, $z\mapsto\tilde{\chi}_{\rm coll}(\qq,z)$ no longer has a pole; on the other hand, its analytic continuation through its branch cut from the upper half-plane to the lower half-plane does, and the corresponding eigenenergies of the sound are complex—the phonons become resonances \cite{Klimin}.} 
\begin{equation}
\label{eq:valpro}
\left[\Sigma_{11}(\qq,z)\Sigma_{22}(\qq,z)-\Sigma^2_{12}(\qq,z)\right]\Big|_{z=\pm\hbar\omega_\mathbf{q}}=0
\end{equation}

\begin{figure}[t]
\begin{center}
\includegraphics[width=6cm,clip=]{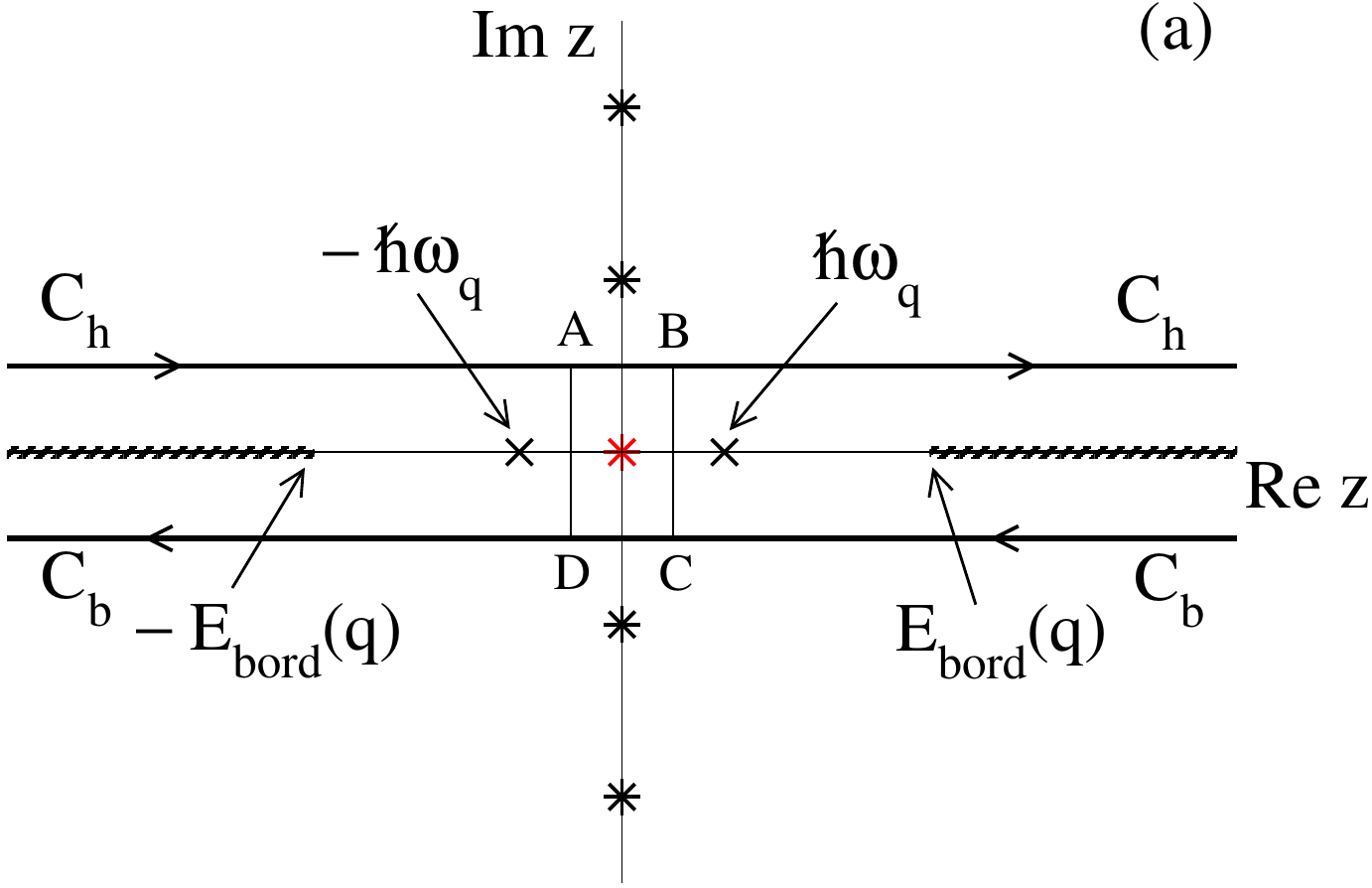}\hspace{1cm}\includegraphics[width=6cm,clip=]{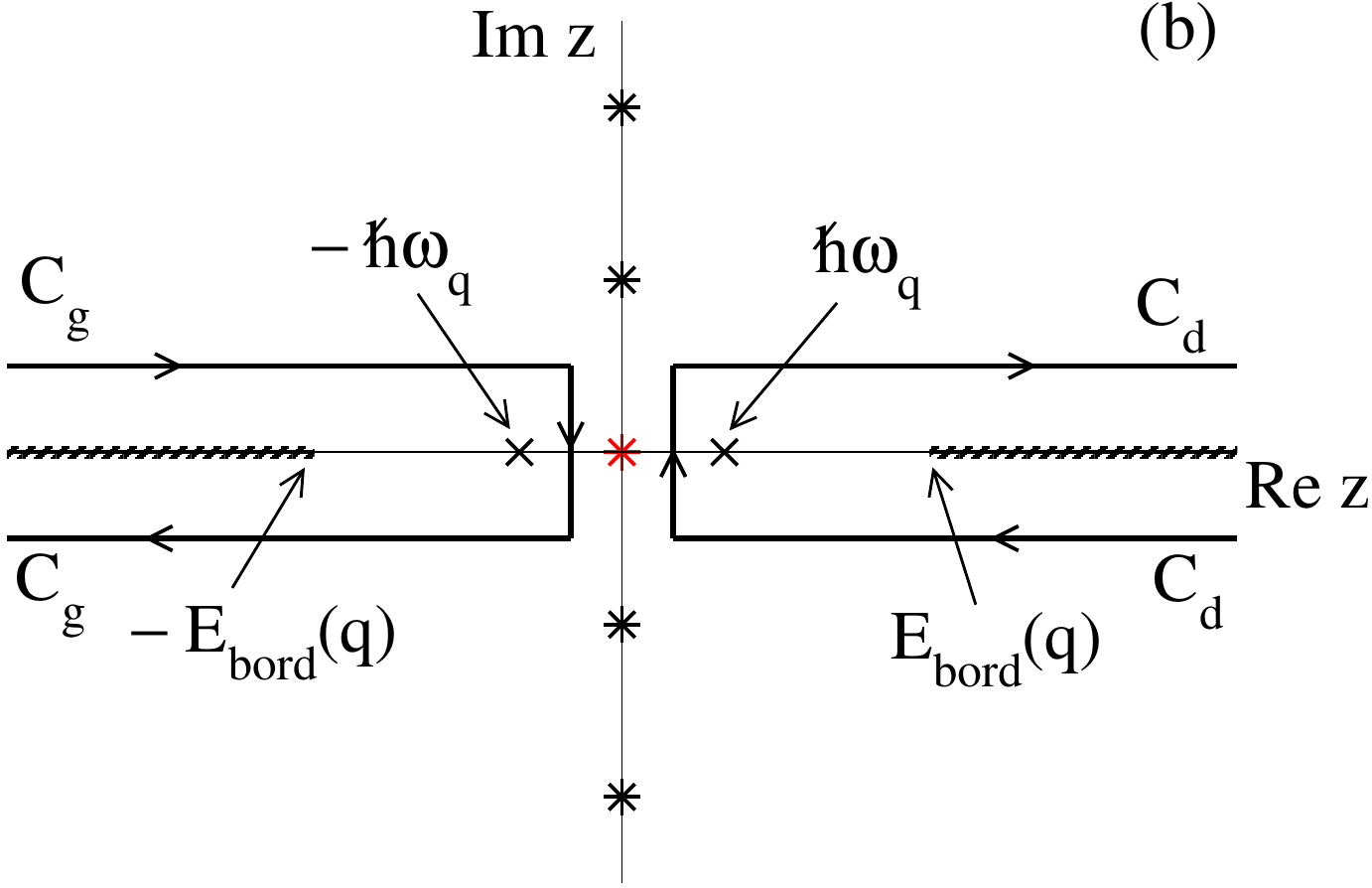} \\
\includegraphics[width=6cm,clip=]{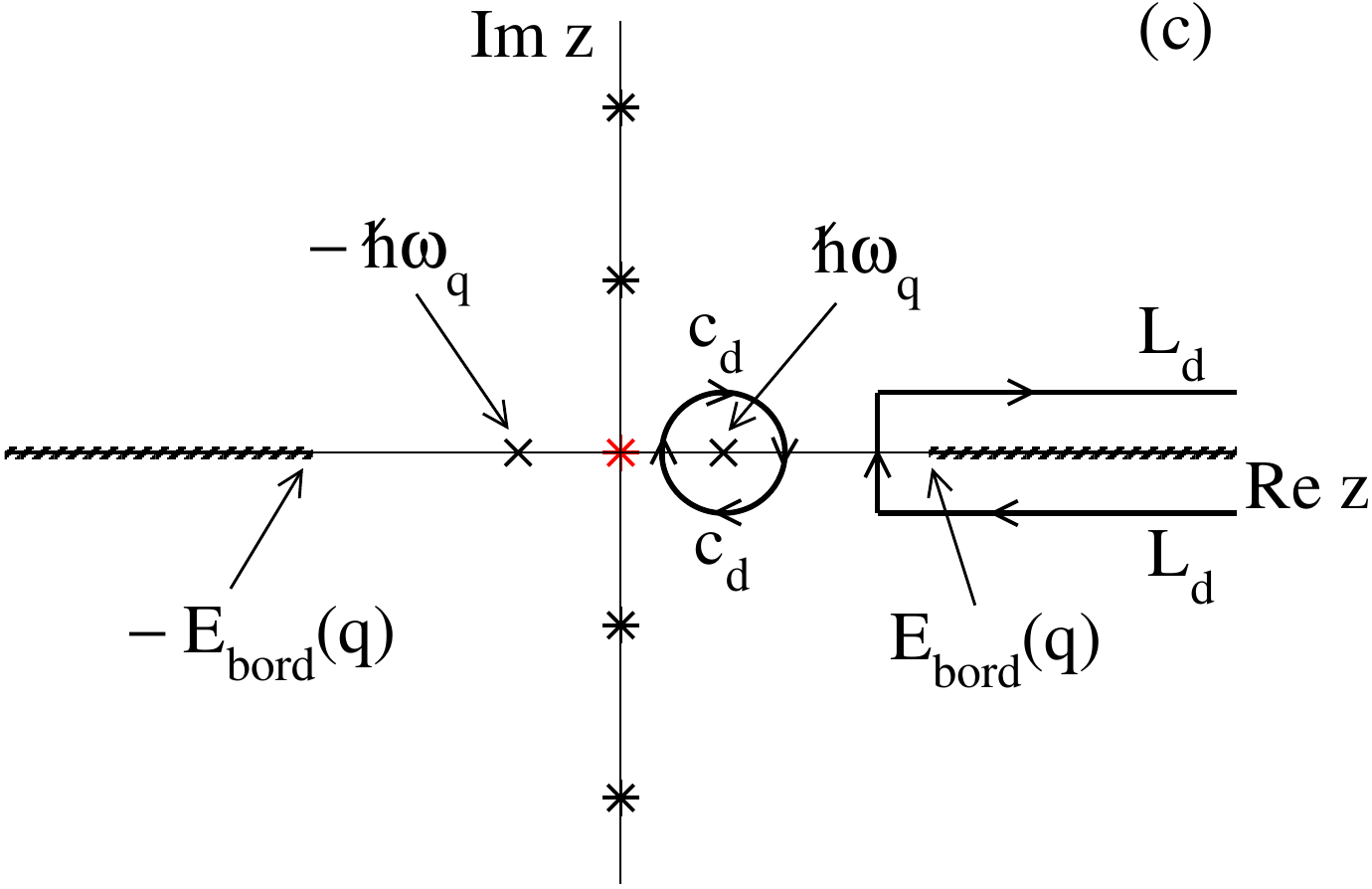}
\end{center}
\caption{Different integration paths used at temperature $T>0$ in the rewriting of quantity $\delta\tilde{g}_{\rm coll}(\qq)$ of equation (\ref{eq:defcoll}) in the form of curvilinear integrals in the complex plane, having the advantage—both numerically and analytically—of remaining at a non-infinitesimal distance from the singularities of the integrand. (a) In the general case, at all orders in $\exp(-\beta E_{\rm gap})$, see equation (\ref{eq:exp1}). (b) At order zero in $\exp(-\beta E_{\rm gap})$ in the susceptibility $\tilde{\chi}_{\rm coll}(\qq,z)$ but not in the denominator of the integrand of $\delta\tilde{g}_{\rm coll}(\qq)$, see equation (\ref{eq:exp2}). (c) At order zero in $\exp(-\beta E_{\rm gap})$ everywhere, see equation (\ref{eq:exp3}). Hatched lines: branch cuts (for the readability of the figure, we have also limited ourselves in (a) to order zero in $\exp(-\beta E_{\rm gap})$; in reality, the branch cut should cover the entire real axis there). Symbols $\times$: poles $\pm\hbar\omega_\qq$ of susceptibility at zero order in $\exp(-\beta E_{\rm gap})$ ($\omega_\qq>0$ is the natural frequency of the Bogoliubov-Anderson phonon, solution of equation (\ref{eq:valpro}), if it exists at the wave number $q$ considered). Stars: Bose poles $z=2\mathrm{i}\pi n k_{\rm B} T$, $n$ integer 
 (roots of the thermal denominator) in (\ref{eq:exp1}); the non-physical pole $n=0$ is in red. The edge $E_{\rm bord}(\qq)$ of the broken pair continuum with fixed total wave vector $\qq$ is given by equation (\ref{eq:ebord}). In the path names, the indices h and b mean high and low, the indices g and d mean left and right, $C$ means path, $c$ means contour, $L$ means loop. The arrows mark the orientation of the paths. Rectangle $ABCD$ is involved in the reasoning that leads from form (\ref{eq:exp1}) to form (\ref{eq:exp2}). In (a), paths $C_{\rm h}$ and $C_{\rm b}$ must pass between the real axis and the Bose poles $n=\pm 1$ that are closest to it. This constraint is of course inoperative in (b) and (c).}
\label{fig:contour}
\end{figure}

 Let us draw the conclusions from this discussion and replace in the expression (\ref{eq:defcoll}) of $\delta\tilde{g}_{\rm coll}(\qq)$ the real axis with useful integration paths remaining at a non-zero distance from the singularities in the complex plane. 
The first choice, applicable to all orders in $\exp(-\beta E_{\rm gap})$ and used in our numerical calculations in section \ref{sec3_2}, takes advantage of the analyticity of $\tilde{\chi}_{\rm coll}(\qq,z)$ on $\mathbb{C}\setminus\mathbb{R}$ and property $[\tilde{\chi}_{\rm coll}(\qq,z)]^*=\tilde{\chi}_{\rm coll}(\qq,z^*)$ (valid in reality beyond the linearized time-dependent BCS approximation) to write 
\begin{equation}
\im[\tilde{\chi}_{\rm coll}(\qq,\eps+\mathrm{i} 0^+)]=\left[\tilde{\chi}_{\rm coll}(\qq,\eps+\mathrm{i} 0^+)-\tilde{\chi}_{\rm coll}(\qq,\eps-\mathrm{i} 0^+)\right]/(2\ii)
\end{equation}
 and to separate the two domains of integration that appeared $\mathbb{R}+\mathrm{i} 0^+$ and $\mathbb{R}-\mathrm{i} 0^+$ into two paths $-\infty+\mathrm{i}\eta \to +\infty+\mathrm{i} \eta$ and $+\infty-\mathrm{i}\eta\to {-}\infty-\mathrm{i}\eta$, traveled in opposite directions and at a non-infinitesimal distance $\eta>0$ from the real axis, called respectively $C_{\rm h}$ (upper path) and $C_{{\rm b}}$
 (lower path) in Figure~\ref{fig:contour}a. This leads to 
\begin{multline}
\label{eq:exp1}
\boxed{\delta\tilde{g}_{\rm coll}(\qq)=}
-k_{\rm B} T \tilde{\chi}_{\rm coll}(\qq,0) +
\int_{C_{\rm h}\cup C_{\rm b}}\frac{\mathrm{d} z}{(-2\mathrm{i}\pi)} \frac{\tilde{\chi}_{\rm coll}(\qq,z)}{1-\exp(-\beta z)} 
\\
=\boxed{-k_{\rm B} T \tilde{\chi}_{\rm coll}(\qq,0)+\im \int_{-\infty}^{+\infty} \frac{\mathrm{d}\eps}{(-\pi)} \frac{\tilde{\chi}_{\rm coll}(\qq,\eps+\mathrm{i}\eta)}{1-\exp[-\beta(\eps+\mathrm{i}\eta)]}}
\end{multline}
Two observations are in order: (i) the separation into two paths has revealed a non-physical pole at $z=0$, whose counterterm $\propto T$ in the right-hand side of the first equality in (\ref{eq:exp1}) compensates for the contribution deduced from the residue formula;\footnote{We find this result again with a minus-plus trick and distribution theory: so that the integrand in (\ref{eq:defcoll}) becomes a function only of $\eps+\mathrm{i} 0^+$ (and not of $\eps$), we write $\im[\tilde\chi(\qq,\eps+\mathrm{i} 0^+)]/[1-\exp(-\beta\eps)]=\im\{\tilde\chi(\qq,\eps+\mathrm{i} 0^+)/[1-\exp(-\beta\eps)]\}=\im\{\tilde\chi(\qq,\eps+\mathrm{i} 0^+)/[1-\exp(-\beta(\eps+\mathrm{i} 0^+))]\}+\im[\tilde\chi(\qq,\eps+\mathrm{i} 0^+)\phi(\eps)]$, omitting the “coll” index for simplicity. The distribution introduced, $\phi(\eps)=1/[1-\exp(-\beta\eps)]-1/[1-\exp(-\beta(\eps+\mathrm{i} 0^+))]$, has a non-zero action only on a neighborhood of the origin; we can therefore linearize the exponential and write $\phi(\eps)=k_{\rm B}[1/\eps-1/(\eps+\mathrm{i} 0^+)]=k_{\rm B} T [1/\eps-\mathrm{v.p.}\,(1/\eps)+\mathrm{i}\pi \delta(\eps)]$ with $\mathrm{v.p.}$ as the principal value distribution and $\delta$ as the Dirac distribution. As $\im[\tilde\chi(\qq,\eps+\mathrm{i} 0^+)]$ tends towards zero at $\eps=0$, the distribution $1/\eps-\mathrm{v.p.}\,(1/\eps)$ has no effect, only the Dirac contribution is non-zero, and we reproduce the counter term of (\ref{eq:exp1}).} this pole did not pose any problem in the original formulation (\ref{eq:defcoll}) because $\tilde{\chi}_{\rm coll}(\qq,z=0)$ has a real value, and therefore a zero imaginary part; (ii) the denominator $1-\exp(-\beta z)$ reveals the thermal poles $z = 2\mathrm{i}\pi n k_{\rm B} T$, $n\in \mathbb{Z}^*$; those of indices $n=\pm 1$ limit the vertical excursion of $C_{\rm h}$ and $C_{\rm b}$; in the numerics, to avoid getting too close to it, we take $\eta=\pi k_{\rm B}T$ according to the “golden mean” doctrine, and in practice perform only the calculation of the contribution of $C_{\rm h}$, that of $C_{\rm b}$ being deduced by sign change and complex conjugation, as in the right-hand side of the second equality in (\ref{eq:exp1}).\footnote{\label{note:matsu} In this right-hand side, let us close the integration path $-\infty+\mathrm{i}\eta\to+\infty+\mathrm{i}\eta$ with a semicircle at infinity in the upper complex half-plane (assuming that the integral is zero on the semicircle) and then apply the residue theorem: since the susceptibility is analytic for $\im z>0$, the only enclosed singularities are the thermal poles, and we obtain 
\begin{equation}
\label{eq:matsu}
\delta\tilde{g}_{\rm coll}(\qq)=-k_{\rm B} T \tilde{\chi}_{\rm coll}(\qq,0)-2 k_{\rm B}T \re\sum_{n=1}^{+\infty}\tilde{\chi}_{\rm coll}(\qq,2\mathrm{i} \pi n k_{\rm B}T)=-k_{\rm B}T \sum_{n=-\infty}^{+\infty} \tilde{\chi}_{\rm coll}(\qq,2\mathrm{i} \pi n k_{\rm B}T)
\end{equation}
 after using property $[\tilde{\chi}_{\rm coll}(\qq,z)]^*=\tilde{\chi}_{\rm coll}(\qq,z^*)$ in the second equality. We thus arrive at a bosonic Matsubara series \cite{FW} (the $2\mathrm{i} \pi n k_{\rm B}T$ are precisely the complex Matsubara energies for bosons) without ever having introduced a many-body Green's function or periodic boundary conditions in imaginary time! For our numerical calculation in section \ref{sec3_2}, which sets a cutoff $k<k_{\rm max}$ on the internal wave vector of the pairs in the $\Sigma_{ij}$ functions, representation (\ref{eq:matsu}) would have, compared to the integration over $\eps$ in (\ref{eq:exp1}), the advantage of a larger step $\delta\eps$ but the disadvantage of much slower asymptotic convergence: the error made by truncating the Matsubara series at $n=n_{\rm max}$ tends towards zero as $1/n_{\rm max}$, instead of being exactly zero beyond a certain $\eps_{\rm max}$ in (\ref{eq:exp1}), see note \ref{note:num}; there would be a workaround (we could analytically estimate the neglected end $n>n_{\rm max}$ using an asymptotic expansion of the summand), but it would be laborious; therefore, we did not use it (\ref{eq:matsu}). Finally, note that, in the limit of zero temperature $T\to 0^+$, the Matsubara series—viewed as a Riemann sum—is replaced by an integral, that of expression (\ref{eq:chem0}) with $\eta=+\infty$.}

 At order zero in $\exp(-\beta E_{\rm gap})$, a second choice of integration path becomes natural. Since there is no longer a branch cut between $-E_{\rm bord}(\qq)$ and $E_{\rm bord}(\qq)$, in $C_{\rm h}\cup C_{\rm b}$, a central portion surrounding the non-physical pole $z=0$ as the only singularity can be removed, at the cost of adding vertical segments $AD$ and $BC$ connecting $C_{\rm h}$ to $C_{\rm b}$, see Figure~\ref{fig:contour}a; we therefore write $C_{\rm h}\cup C_{\rm b}=C_{\rm g}\cup ABCD \cup C_{\rm d}$ where rectangle $ABCD$ is traversed clockwise, contour $C_{\rm g}$ is on the left, and contour $C_{\rm d}$ is on the right, see Figure~\ref{fig:contour}b. The added vertical segments have zero contribution as a whole, as they are traversed in opposite directions; the curvilinear integral over $ABCD$ compensates for the counterterm $\propto T$ according to the residue theorem; it remains
\begin{equation}
\label{eq:exp2}
\boxed{\delta\tilde{g}_{\rm coll}(\qq)=}\int_{C_{\rm g}\cup C_{\rm d}}\frac{\mathrm{d} z}{(-2\mathrm{i}\pi)} \frac{\tilde{\chi}_{\rm coll}(\qq,z)}{1-\exp(-\beta z)}=\boxed{\int_{C_{\rm d}} \frac{\mathrm{d} z}{(-\mathrm{i}\pi)} \tilde{\chi}_{\rm coll}(\qq,z) \left(\frac{1}{2}+\frac{1}{\exp(\beta z)-1}\right)}
\end{equation}
We have further simplified the expression in the right-hand side of the second equality, using the fact that $C_{\rm g}$ can be deduced from $C_{\rm d}$ by changing $z$ to $-z$, including orientation; this results in $\int_{C_{\rm g}}\mathrm{d} z\, f(z)=\int_{C_{\rm d}}\mathrm{d} z\, [-f(-z)]$ for any function $f$ of the complex variable, and therefore only the odd part of the integrand remains, which reveals a Bose-Einstein distribution.\footnote{The susceptibility $\tilde{\chi}_{\rm coll}(\qq,z)$ is an even function of $z$—see the remarks below (\ref{eq:defkpm},\ref{eq:lien})—and we have $1/[1-\exp(-\beta z)]-1/[1-\exp(\beta z)]=1+2/[\exp(\beta z)-1]$.} However, the zero order in $\exp(-\beta E_{\rm gap})$ is not implemented in a completely consistent manner, since the fermionic occupation numbers $f_\pm$ in $\tilde{\chi}_{\rm coll}(\qq,z)$ have been omitted while retaining the thermal Bose law for complex energies $z$ of arbitrarily large real part in $C_{\rm d}$.

We therefore make a final change to the integration path, taking advantage of the absence of singularity between the phonon pole $z=\hbar\omega_\qq$ and the branch point $z=E_{\rm bord}(\qq)$ of the continuum, to replace $C_{\rm d}$ with $c_{\rm d}\cup L_{\rm d}$ where the circular contour $c_{\rm d}$ surrounds the phonon pole and the loop $L_d$ surrounds the branch cut—the broken pair continuum at $\qq$ fixed—clockwise as in Figure~\ref{fig:contour}c. Since the points of the loop have a real part $\re z\gtrsim E_{\rm bord}(\qq)\geq E_{\rm gap}$, we must neglect the thermal Bose law at order zero in $\exp(-\beta E_{\rm gap})$. The same approximation cannot be made on $c_{\rm d}$ because the phonon eigenenergy $\hbar\omega_\qq$ becomes arbitrarily smaller than $k_{\rm B}T>0$ at low wave numbers $q$. We ultimately separate the Fourier transform of $\delta g_{\rm coll}(\rr,\rr')$ into a contribution from the Bogoliubov-Anderson phonon $\phi$ (if it exists at the wave vector $\qq$ considered, see note \ref{note:dom} on the domain of existence $\mathcal{E}_{\rm son}$ of the acoustic branch, otherwise $\delta\tilde{g}^{\phi}_{\rm coll}(\qq)=0$) and a collective contribution from the broken pair continuum, 
\begin{equation}
\label{eq:exp3}
\delta\tilde{g}_{\rm coll}(\qq)=\delta\tilde{g}^{\phi}_{\rm coll}(\qq)+\delta\tilde{g}^{C^0}_{\rm coll}(\qq) \ \mbox{with}\ 
\left\{\begin{array}{lcl}
\delta\tilde{g}^{\phi}_{\rm coll}(\qq) &\!\!=\!\!&\displaystyle\int_{c_{\rm d}}\!\!\frac{\mathrm{d} z}{(-\mathrm{i}\pi)}\tilde{\chi}_{\rm coll}(\qq,z)\left(\frac{1}{2}\!+\!\frac{1}{\exp(\beta z)\!-\!1}\right)\!=\!\left(\bar{n}_\mathbf{q}\!+\!\frac{1}{2}\right) \Phi(\qq) \\
& & \\
\delta\tilde{g}^{C^0}_{\rm coll}(\qq) &\!\!=\!\!& \displaystyle\int_{L_{\rm d}}\!\!\frac{\mathrm{d} z}{(-2\mathrm{i}\pi)}\tilde{\chi}_{\rm coll}(\qq,z) 
\end{array}\right.
\end{equation}
 Here, by virtue of the residue theorem applied to the integral over the contour $c_{\rm d}$, $\bar{n}_\qq$ is the thermal occupation number of the phonon mode $\qq$ and $\Phi(\qq)$ is twice the residue of the collective susceptibility at $z=\hbar\omega_\qq$: 
\begin{equation}
\label{eq:nqPhi}
\bar{n}_\qq=\frac{1}{\exp(\beta\hbar\omega_\qq)-1}\quad\mbox{and}\quad 
\Phi(\qq)=\frac{-s_{11}\left(|\Sigma_{11}|^{1/2}\Sigma_{23}-s_{11}s_{12}|\Sigma_{22}|^{1/2}\Sigma_{13}\right)^2}{\frac{\mathrm{d}}{\mathrm{d} z}\left(\Sigma_{11}\Sigma_{22}-\Sigma^2_{12}\right)}\Big|_{(\qq,z=\hbar\omega_\qq)}
\end{equation}
 with $s_{ij}$ the sign of $\Sigma_{ij}(\qq,\hbar\omega_\qq)$. The expression of $\Phi(\qq)$ is verified by expanding the square in the numerator and recognizing the definition (\ref{eq:defcoll}) by means of equation (\ref{eq:valpro}), according to which $s_{11}=s_{22}$ and $\Sigma_{12}=s_{12}|\Sigma_{11}\Sigma_{22}|^{1/2}$ at the considered point.

In formulations (\ref{eq:exp2},\ref{eq:exp3}), the distances to the real axis on paths $C_{\rm d}$, $c_{\rm d}$, and $L_{\rm d}$ are no longer constrained by temperature (they do not have to be less than $2\pi k_{\rm B} T$), but the radius of $c_{\rm d}$ must remain less than $\hbar\omega_\qq$ and $E_{\rm bord}(\qq)-\hbar\omega_\qq$. If we want the integration path to remain at a non-infinitesimal distance from the singularities of the integrand, (i) path $c_{\rm d}\cup L_{\rm d}$ is not suitable near point $q_*$ where the acoustic branch joins the broken pair continuum ($E_{\rm bord}(\qq)-\hbar\omega_\mathbf{q}\to 0$ if $q\to q_*$) and it is better to return to path $C_{\rm d}$ and expression (\ref{eq:exp2}), (ii) even $C_{\rm d}$ is no longer suitable in the limit $q\to 0$ where $\hbar\omega_\mathbf{q}\to 0$ and it is necessary to return to $C_{\rm h}$ and expression (\ref{eq:exp1}). To conclude, let us explain the peculiarity of case $T=0$, which led us to avoid it in our numerical calculations: the definition (\ref{eq:defcoll}) of $\delta\tilde{g}_{\rm coll}(\qq)$ is reduced to 
\begin{equation}
\label{eq:chem0}
\delta\tilde{g}_{\rm coll}(\qq) \stackrel{T=0}{=} \int_0^{+\infty}\frac{\mathrm{d}\eps}{(-\pi)} \im[\tilde{\chi}_{\rm coll}(\qq,\eps+\mathrm{i} 0^+)]=\im\left[\left(\int_0^{\mathrm{i}\eta}+\int_{\mathrm{i}\eta}^{\mathrm{i}\eta+\infty}\right)\frac{\mathrm{d} z}{(-\pi)} \tilde{\chi}_{\rm coll}(\qq,z)\right],
\end{equation}
there is no longer any trace of a non-physical pole at $z=0$, and the vertical deviation $\eta>0$ from the branch cut can be chosen freely in the right-hand side of the second equality, but the integration path, necessarily starting from the origin, cannot remain at a non-infinitesimal distance from the phonon pole $z=\hbar\omega_\qq$ if $q\to 0$.

\subsection{Which equation of state in our linearized time-dependent BCS theory?}\label{sec2_5}

Essentially, the fluctuation-dissipation theorem gives access to the deviations $\delta g_{\sigma\sigma'}(\rr,\rr')$ of the pair distribution functions $g_{\sigma\sigma'}(\rr,\rr')$ from their asymptotic values $\rho_\sigma\rho_{\sigma'}$. However, in cold atomic gases, it is precisely these functions $g_{\sigma\sigma'}$ that interest numericians and experimentalists, see our section \ref{sec3_2}. We therefore need to know the density of the gas, which is unpolarized here ($\rho_\uparrow=\rho_\downarrow=\rho/2$), in order to convert the $\delta g_{\sigma\sigma'}(\rr,\rr')$ into the $g_{\sigma\sigma'}(\rr,\rr')$.

Thanks to the linearized time-dependent BCS theory, we have improved the calculation of the $\delta g_{\sigma\sigma'}(\rr,\rr')$ using the density-density response functions $\chi_{\sigma\sigma'}(\rr,t;\rr',t')$, but we have not touched on the rest of the static BCS theory, in particular the one-body observables. It would therefore seem natural that the grand canonical equation of state $\rho(\mu,T)$ remains the same, and that we have 
\begin{equation}
\label{eq:pas_bon}
g^{\rm ltd}_{\sigma\sigma'}(\rr,\rr') \stackrel{?}{=} \rho^{\rm sta}_\sigma\rho^{\rm sta}_{\sigma'}+\delta g^{\rm ltd}_{\sigma\sigma'}(\rr,\rr')
\end{equation}
 where $\rho^{\rm sta}_\sigma$ is given by (\ref{eq:etat}). In reality, this is not the case, because the function $\delta g_{\uparrow\uparrow}(\rr,\rr')$ contains some hidden information about the average density. In fact, the particles in the internal state $\uparrow$ are identical fermions with no direct interaction between them (no resonance in the $p$ wave), so that their pair density must be zero at zero distance: 
\begin{equation}
\label{eq:propri}
\lim_{|\rr-\rr'|\to 0^+} g_{\uparrow\uparrow}(\rr,\rr')=0 \quad \mbox{so}\quad \lim_{|\rr-\rr'|\to 0^+} \delta g_{\uparrow\uparrow}(\rr,\rr')=-\rho^{{2}}_\uparrow
\end{equation}
 Returning to the separations (\ref{eq:sepa},\ref{eq:tfdgcoll}) of $\delta g^{\rm ltd}_{\sigma\sigma'}(\rr,\rr')$ into static and collective parts, we deduce the correct value of the density of the unpolarized gas in our linearized time-dependent BCS theory and the correct expressions for the pair distributions:\footnote{We have used in (\ref{eq:sepa})—written for $\sigma=\sigma'=\uparrow$—the property (\ref{eq:propri}) for static BCS and linearized time-dependent BCS; the integral over $\qq$ in (\ref{eq:bon}) is simply the limit of $\delta g_{\rm coll}(\rr,\rr')$ when $|\rr-\rr'|\to 0^+$, a limit whose existence and expression are established in section \ref{sec3_3} for the zero-order theory in $\exp(-\beta E_{\rm gap})$.} 
\begin{equation}
\label{eq:bon}
\boxed{\frac{1}{4}[\rho^{\rm ltd}(\mu,T)]^2 = \frac{1}{4}[\rho^{\rm sta}(\mu,T)]^2-\int_{\mathbb{R}^d}\frac{\dd^d q}{(2\pi)^d} \delta\tilde{g}_{\rm coll}(\qq)} \quad\mbox{and}\quad \boxed{g^{\rm ltd}_{\sigma\sigma'}(\rr,\rr')=\rho^{\rm ltd}_\sigma\rho^{\rm ltd}_{\sigma'}+\delta g^{\rm ltd}_{\sigma\sigma'}(\rr,\rr')}
\end{equation}
 In the numerical examples in section \ref{sec3_2}, at very low temperature $k_{\rm B}T\ll m c^2, E_{\rm gap}$, i.e., practically in the ground state, we find that the function $\delta\tilde{g}_{\rm coll}(\qq)$, defined in equation (\ref{eq:defcoll}), is negative everywhere, see equation (\ref{eq:estneg}); we then have $\rho^{\rm ltd}>\rho^{\rm sta}$ with fixed chemical potential $\mu$ or, if you prefer, $\mu^{\rm ltd}<\mu^{\rm sta}$ with fixed density $\rho$ (the thermodynamic stability of the gas in a spatially homogeneous phase requires that $\partial_\mu\rho>0$, see equation (\ref{eq:hydro})). This is a step in the right direction, towards an improvement of the variational BCS theory.

Thanks to the equation of state (\ref{eq:bon}), we can now specify the common value of the two differences in the beautiful relation (\ref{eq:belle2}) and write nicely: 
\begin{equation}
\label{eq:belle3}
\boxed{
g_{\sigma\sigma'}^{\rm ltd}(\rr,\rr')-g_{\sigma\sigma'}^{\rm sta}(\rr,\rr')=\int_{\mathbb{R}^d} \frac{\dd^d q}{(2\pi)^d} \delta\tilde{g}_{\rm coll}(\qq)\left[\eee^{\mathrm{i}\mathbf{q}\cdot(\rr-\rr')}-1\right]}
\end{equation}
 Remember, the two theories (static or not) must be taken in (\ref{eq:belle3}) at the same chemical potential, not at the same density; this is obvious in the present section \ref{sec2} since we are in a grand canonical viewpoint, but it will not always be so in the following, see section \ref{sec3_2}. We have arrived at the result (\ref{eq:belle3}) — which will be used in particular in section \ref{sec3_3} — by means of a chain of obvious equalities based on equations (\ref{eq:defg}, \ref{eq:sepa}, \ref{eq:tfdgcoll}, \ref{eq:bon}): 
\begin{multline}
\label{eq:chaine}
g_{\sigma\sigma'}^{\rm ltd}(\rr,\rr')-g_{\sigma\sigma'}^{\rm sta}(\rr,\rr')=\rho_\sigma^{\rm ltd}\rho_{\sigma'}^{\rm ltd}-\rho_\sigma^{\rm sta}\rho_{\sigma'}^{\rm sta}+\left[\delta g_{\sigma\sigma'}^{\rm ltd}(\rr,\rr')-\delta g_{\sigma\sigma'}^{\rm sta}(\rr,\rr')\right]=\frac{1}{4}[\rho^{\rm ltd}(\mu,T)]^2 \\ 
-\frac{1}{4}[\rho^{\rm sta}(\mu,T)]^2+\delta g_{\rm coll}(\rr,\rr')=-\int_{\mathbb{R}^d} \frac{\dd^d q}{(2\pi)^d} \delta\tilde{g}_{\rm coll}(\qq) + \int_{\mathbb{R}^d} \frac{\dd^d q}{(2\pi)^d} \delta\tilde{g}_{\rm coll}(\qq) \eee^{\mathrm{i}\mathbf{q}\cdot(\rr-\rr')}
\end{multline}

\section{Explicit results on $g^{\rm ltd}_{\sigma\sigma'}(\rr,\rr')$, numerical and analytical}\label{sec3}

In this section, we present all of our results on pair distribution functions in linearized time-dependent BCS theory. First at large distances in section \ref{sec3_1}, where we analytically obtain at low temperature a beautiful law connecting the quantum asymptotic regime $|\rr-\rr'|\to +\infty$ but $|\rr-\rr'|\ll \lambda_\phi$ to the thermal asymptotic regime $|\rr-\rr'|\gg\lambda_\phi\to +\infty$ where $\lambda_\phi=\hbar c/{k_{\rm B}T}$ is the thermal wavelength of the phonons of the superfluid. Then numerically at intermediate distances in section \ref{sec3_2}, from a fraction of $k_{\rm F}^{-1}$ to a few $k_{\rm F}^{-1}$ ($k_{\rm F}$ is the Fermi wave number) in the strongly interacting regime, the domain of choice for quantum Monte Carlo simulations and experimental measurements in cold atomic gases. Finally, at short distances, i.e., at large wave numbers $q$ in section \ref{sec3_3}, where simple laws describe the divergence of $g^{\rm ltd}_{\uparrow\downarrow}(\rr,\rr')$ and the departure of $g^{\rm ltd}_{\uparrow\uparrow}(\rr,\rr')$ from zero. All these results are obtained at order zero in $\exp(-\beta E_{\rm gap})$, i.e., at a sufficiently low temperature that we can omit the occupation numbers $f_\kk$ of fermionic quasiparticles in static BCS theory and in the functions $\Sigma_{ij}$ giving the density-density susceptibilities $\tilde{\chi}^{\rm ltd}_{\sigma\sigma'}(\qq,z)$.

\subsection{Behavior of $\delta g^{\rm ltd}_{\sigma\sigma'}(\rr,\rr')$ at large distances and low temperatures}\label{sec3_1}

Given the separation of the deviations $\delta g_{\sigma\sigma'}^{\rm ltd}(\rr,\rr')$ into static and collective parts in (\ref{eq:sepa}), and the known asymptotic behavior—rapidly decreasing—of the static parts, see references \cite{Romero2D,Romero3D} supplemented by our note \ref{note:asymp}, work remains to be done on the collective part $\delta g_{\rm coll}(\rr,\rr')$ common to pair distribution functions. An elementary technique in section \ref{sec3_1_1} gives an asymptotic equivalent of $\delta g_{\rm coll}(\rr,\rr')$ at zero temperature, in the form of a power law and with a negative sign, but provides no other indication than the rapid decay at temperature $T>0$. A more sophisticated technique in section \ref{sec3_1_2} fills this gap and finds that the $\delta g^{\rm ltd}_{\sigma\sigma'}(\rr,\rr')$ are negative and approximately equal to $\delta g_{\rm coll}(\rr,\rr')$ at sufficiently large distance $|\rr-\rr'|\to +\infty$ and low temperature $T\to 0^+$. These results invalidate the static BCS theory, in particular its flagship prediction (\ref{eq:dgsta}) that it is positive everywhere, but are confirmed in section \ref{sec3_1_3} by a low-energy effective theory that is believed to be exact.

\subsubsection{What we deduce from integration by parts}\label{sec3_1_1}

A first technique for determining the long-range behavior of an isotropic Fourier integral such as (\ref{eq:tfdgcoll}) giving $\delta g_{{\rm coll}}(\rr,\rr')$ consists, after angular averaging $\langle \exp[\mathrm{i}\mathbf{q}\cdot(\rr-\rr')]\rangle_{\rm ang}=\sin(q|\rr-\rr'|)/(q|\rr-\rr'|)$ in dimension $d=3$ or $\langle \exp[\mathrm{i}\mathbf{q}\cdot(\rr-\rr')]\rangle_{\rm ang}=J_0(q|\rr-\rr'|)$ in dimension $d=2$, in integrating by parts repeatedly \cite{Holzmann}. We reproduce here equations (8.300,8.307) from reference \cite{livre}: 
\begin{eqnarray}
\label{eq:ipp3d}
\int_0^{+\infty} \mathrm{d} q \, \phi(q)\sin (q|\rr-\rr'|) &\underset{|\rr-\rr'|\to +\infty}{=}& \sum_{s=0}^S (-1)^s \frac{\phi^{(2s)}(0)}{|\rr-\rr'|^{2s+1}} + O\left(\frac{1}{|\rr-\rr'|^{2S+3}}\right) \\
\label{eq:ipp2d}
\int_0^{+\infty} \mathrm{d} q\, J_0(q|\rr-\rr'|) \phi(q) &\underset{|\rr-\rr'|\to+\infty}{=}& \sum_{s=0}^S (-1)^s \frac{\phi^{(2s)}(0)\Gamma(s+1/2)}{\Gamma(1/2)s!\, |\rr-\rr'|^{2s+1}} + O\left(\frac{1}{|\rr-\rr'|^{2S+3}}\right)
\end{eqnarray}
 where $S$ is any natural number, $\phi(q)=q\, \delta\tilde{g}_{\rm coll}(\qq)/2\pi^2|\rr-\rr'|$ in dimension $d=3$ and $\phi(q)=q\, \delta\tilde{g}_{\rm coll}(\qq)/2\pi$ in dimension $d=2$. 

Now, for $\im z\neq 0$, the functions $\Sigma_{ij}(\qq,z)$ in section \ref{sec2_2} are formally even functions of the wave number $q$, in the sense that they are integrals on an internal wave vector $\kk$ of smooth functions of $\kk$ invariant by change of $\qq$ in $-\qq$; this property extends to the susceptibility $\tilde{\chi}_{\rm coll}(\qq,z)$ giving $\delta\tilde{g}_{\rm coll}(\qq)$ in (\ref{eq:defcoll}) and, after integration in the complex plane on a path remaining {\it always} at a non-infinitesimal distance from the singularities of the integrand, to the function $\delta\tilde{g}_{\rm coll}(\qq)$ itself. At non-zero temperature, this constraint is satisfied by the representation (\ref{eq:exp1}), so that $\delta\tilde{g}_{\rm coll}(\qq)$ only has even powers in its Taylor expansion at $q=0$: 
\begin{equation}
\label{eq:devgen}
\delta\tilde{g}_{\rm coll}(\qq)\stackrel{T>0}{\underset{q\to 0}{=}}  \sum_{s=0}^{S} c_{2s} q^{2s} + O(q^{2S+2})
\end{equation}
 with coefficients that are unimportant for the moment. At order zero in $\exp(-\beta E_{\rm gap})$,
 the same reasoning on the loop $L_{\rm d}$ in (\ref{eq:exp3}) shows that the contribution of the continuum—independent of temperature—enjoys the same property, 
\begin{equation}
\delta\tilde{g}^{C^0}_{\rm coll}(\qq)\underset{q\to 0}{=}\sum_{s=0}^{S} c^{C^0}_{2s} q^{2s} + O(q^{2S+2})
\end{equation}
 with the first two coefficients calculated analytically in Appendix \ref{ann:deccont}, see equation (\ref{eq:repq}). Although the eigenvalue equation (\ref{eq:valpro}) is formally an even function of $q$ and $\omega_\qq$, the angular eigenfrequency $\omega_\qq$ of the phonon $\qq$ in the linearized time-dependent BCS theory is linear at the outset as in (\ref{eq:depart}) and therefore only has odd powers,\footnote{The angular frequency squared certainly only has even powers, $\omega_\qq^2=\omega_0^2+c^2 q^2+O(q^4)$, but taking the square root changes the situation since $\omega_0=0$.} 
\begin{equation}
\hbar\omega_\mathbf{q} \underset{q\to 0}{=} \hbar c q \left[1+\sum_{s=1}^{S} \zeta_s \left(\frac{\hbar q}{m c}\right)^{2s}+O(q^{2S+2})\right],
\end{equation}
 whose first coefficients $\zeta_s$ are known \cite{concav,VanLoon}.\footnote{In 3D, the quantum fluctuations of the bosonic field of phonons, ignored in the linearized time-dependent BCS classical field theory, appear in $\omega_\qq$ as a term proportional to $q^5\ln q$, and even as an imaginary term scaling as $q^5$ — Belyaev damping effect — if the acoustic branch is initially convex ($\zeta_2>0$) \cite{espagnols,insuffisance}. }${}^{,}$\footnote{We also know in any dimension that, in the BEC limit where the acoustic excitation spectrum is reduced to that of Bogoliubov for a condensate or quasi-condensate of bosonic dimers, $\hbar\omega_\qq^{\rm Bog}=[(\hbar^2 q^2/2 m_{\rm d})({2}m_{\rm d}c^2+\hbar^2 q^2/2 m_{\rm d})]^{1/2}$ with $m_{\rm d}=2m$ the mass of a dimer, we must have $\zeta_s\to (-1)^s\Gamma(s-1/2)/[\Gamma(-1/2) s!\, 4^{2s}]$ by virtue of equation (8.303) in reference \cite{livre} and relation $\Gamma(1/2)=(-1/2)\Gamma(-1/2)$.} Let us nevertheless write the coefficient $\zeta_2$ in 3D, because the writing trick (\ref{eq:c3d}) allows us to simplify it greatly:\footnote{There is a pitfall to avoid: when comparing equation (12) in reference \cite{concav}, $m c^2=(2\mu/3) (1+xy)/(1+y^2)$ where $y=\partial_\mu\Delta|_{a_{\rm 3D}}$, to our equation (\ref{eq:c3d}), one might think that $y=z$. In reality — as can be clearly seen at the unitary limit — we must take as $y$ the non-obvious root of equation $(1+xy)/(1+y^2)=(1+xz)/(1+z^2)$, namely $y=(x-z)/(1+xz)$; its transfer to equation (16) of \cite{concav} gives our expression (\ref{eq:zeta2_3D}).} 
\begin{equation}
\label{eq:zeta2_3D}
\zeta_2^{\rm ltd}\stackrel{d=3}{=} \frac{32+64z^2+52z^4+4xz(24+53z^2+39z^4)+x^2(-35-56z^2+13z^4+54z^6)}{-8\times 135 x^2(1+z^2)^3}\Big|_{x=\frac{\Delta}{\mu},z=\frac{2\Delta}{3\rho^{\rm sta}g_{\rm 3D}}}
\end{equation}
 Similarly, as its expression (\ref{eq:nqPhi}) is formally even in $q$ but odd in $\omega_\qq$ (because of the derivative $\dd/\mathrm{d} z$ in the denominator), the function $\Phi(\qq)$ in the contribution to $\delta\tilde{g}_{\rm coll}(\qq)$ of the Bogoliubov-Anderson phonon only has odd powers:\footnote{This is consistent with (\ref{eq:devgen}) because, by virtue of the term $1/2$, $\bar{n}_\qq+1/2=1/2\thf(\beta\hbar\omega_\qq/2)$ as a factor of $\Phi(\qq)$ in $\delta\tilde{g}^{\phi}_{\rm coll}(\qq)$ only has odd powers in its expansion at temperature $T>0$.} 
\begin{equation}
\Phi(\qq) \underset{q\to 0}{=}\sum_{s=0}^{S} \frac{\Phi^{(2s+1)}(0)}{{(2s+1)!}} q^{2s+1} + O(q^{2S+3})
\end{equation}
 The first derivative at zero is calculated in Appendix \ref{ann:deccont}, see equation (\ref{eq:apressimp}), and is written in any dimension: 
\begin{equation}
\label{eq:dphi0}
\Phi'(0) = \frac{\hbar \rho^{\rm sta}}{4 m c}
\end{equation}

 At non-zero temperature, the functions $\phi(q)$ in (\ref{eq:ipp3d},\ref{eq:ipp2d}) therefore only have odd powers of $q$ in their expansion. We can deduce that all their even-order derivatives $\phi^{(2s)}(q)$ are zero at $q=0$, and that the deviation $\delta g_{\rm coll}(\rr,\rr')$ decreases rapidly (more rapidly than any power law $1/|\rr-\rr'|^{2S+{1}}$). As was already the case for the deviations $\delta g^{\rm sta}_{\sigma\sigma'}(\rr,\rr')$ of the static BCS theory, see references \cite{Romero3D,Romero2D} and our note \ref{note:asymp},\footnote{References \cite{Romero3D,Romero2D} are of order zero in $\exp(-\beta E_{\rm gap})$, but the property remains true beyond; it suffices to apply (\ref{eq:ipp3d},\ref{eq:ipp2d}) to the functions $\delta\tilde{g}^{{\rm sta}}_{\sigma\sigma'}(\qq)$ of equations (\ref{eq:fourhb},\ref{eq:fourhh}). Moreover, our note \ref{note:asymp}, which applies to all orders in $\exp(-\beta E_{\rm gap})$ in its second part, explicitly confirms this. } this conclusion extends to the deviations $\delta g^{\rm ltd}_{\sigma\sigma'}(\rr,\rr')$ by virtue of relations (\ref{eq:sepa}) and we write symbolically: 
\begin{equation}
\delta g^{\rm ltd}_{\sigma\sigma'}(\rr,\rr')\stackrel{T>0}{\underset{|\rr-\rr'|\to +\infty}{=}} O(1/|\rr-\rr'|^{\infty})
\end{equation}
 without anything being said about the sign at this stage.
On the other hand, at zero temperature, Bose's law $\bar{n}_\qq$ disappears in (\ref{eq:exp3}) and $\delta\tilde{g}_{\rm coll}(\qq)$ acquires odd powers of $q$, those of the function $\Phi(\qq)$; there is no contradiction with the reasoning leading to (\ref{eq:devgen}) because the condition of non-infinitesimal distance to singularities
 cannot be satisfied in the ground state of the system when $q\to 0$, see equation (\ref{eq:chem0}). The dominant order is sufficient here ($S=1$ in (\ref{eq:ipp3d},\ref{eq:ipp2d})): 
\begin{equation}
\label{eq:asympzero}
\delta g^{\rm ltd}_{\sigma\sigma'}(\rr,\rr')\stackrel{T=0}{\underset{|\rr-\rr'|\to +\infty}{\sim}} \delta g_{\rm coll}(\rr,\rr') \stackrel{T=0}{\underset{|\rr-\rr'|\to +\infty}{\sim}} 
\left\{\begin{array}{cl}\displaystyle-\frac{\Phi'(0)}{2\pi^2|\rr-\rr'|^4} & \quad\mbox{in}\quad d=3\\
& \\
\displaystyle-\frac{\Phi'(0)}{4\pi |\rr-\rr'|^3} & \quad\mbox{in}\quad d=2\end{array}
\right.
\end{equation}
 Since $\Phi'(0)>0$, the distribution function of pairs of fermions with opposite spins tends towards its asymptotic value from below, in contradiction with the static BCS theory (\ref{eq:dgsta}) but in agreement with the physical discussion in section \ref{sec1}.

\subsubsection{What we deduce from a formulation by contour integral }\label{sec3_1_2}

Let us now move on to a second technique for studying isotropic Fourier integrals at large distances, which is useful when the result decays rapidly, i.e., at non-zero temperature. See, for example, reference \cite{Holzmann} and section 8.7.3.2 of reference \cite{livre}.

Let's start with the easier case $d=3$. The function $\phi(q)$ in (\ref{eq:ipp3d}) only has odd powers of $q$ in its expansion; we assume that it can be formally extended to a smooth and odd function of $q$ over the entire $\mathbb{R}$. Since the sine function is odd, the integrand is even and we can write 
\begin{equation}
\delta g_{\rm coll}(\rr,\rr') \stackrel{{d=3}}{=} \int_{-\infty}^{+\infty}\frac{\mathrm{d} q}{2\mathrm{i}} \phi(q) \exp(\mathrm{i} q |\rr-\rr'|)
\label{eq:avecexp}
\end{equation}
Assuming then that $\phi(q)$ has an analytic continuation to $q\in\mathbb{C}$ being $O(1/q^\alpha)$ at infinity, with $\alpha>1$,\footnote{ Appendix \ref{ann:grandq} shows that ${\delta\tilde{g}_{\rm coll}(\qq)}=O(1/q^4)$ and therefore $\phi(q)=O(1/q^3)$ on the real axis. } we transform (\ref{eq:avecexp}) into a contour integral by closing the integration path with a large semicircle in the upper half-plane, bypassing any branch cuts. Cauchy's theorem finally allows us to shift the horizontal part of the integration path upwards until we reach the first singularity $k_0$ of the integrand, whether it is a pole or a branch point. We then expect $\delta g_{\rm coll}(\rr,\rr')$ to decrease at infinity like $\exp(-\kappa_0|\rr-\rr'|)$ with $\kappa_0=\im k_0>0$, up to a factor of power law.\footnote{ In particular, we have on the integration contour the upper bound $|\exp(\mathrm{i} q |\rr-\rr'|)|\leq \exp(-\kappa_0|\rr-\rr'|)$.} 

Unfortunately, unlike note \ref{note:asymp}, where we were able to apply this technique to the static BCS theory, the function $\phi(q)$ used here does not have a simple expression and its singularities in the complex plane are not known analytically. To move forward, we therefore place ourselves at vanishingly small temperature, in practice in the regime 
\begin{equation}
k_{\rm B} T\ll m c^2, \Delta, E_{\rm F}
\end{equation}
 The singularities closest to the real axis, which therefore lead to the slowest decreasing exponentials of $|\rr-\rr'|$, are then the poles ${q}_n{\simeq} 2\mathrm{i}\pi n k_{\rm B}T{/\hbar c} \ (n\in\mathbb{N}^*)$ of the thermal occupation number $\bar{n}_\mathbf{q} {\simeq [\exp(\beta\hbar c q)-1]^{-1}}$ of the phonon mode $\qq$ in expression (\ref{eq:exp3}), and the residue theorem finally gives in 3D:\footnote{The summation can be performed explicitly, $\sum_{n\geq 1} n^2 z^n=g_{-2}(z)=z(z+1)/(1-z)^3$ where $g_\alpha$ is a Bose function or polylogarithm.} 
\begin{equation}
\label{eq:asympnz3d}
\boxed{\delta g^{\rm ltd}_{\sigma\sigma'}(\rr,\rr') \stackrel{T\to 0^+}{\underset{|\rr-\rr'|\to +\infty}{\simeq}}\delta g_{\rm coll}(\rr,\rr') \stackrel{T\to 0^+}{\underset{|\rr-\rr'|\to +\infty}{\simeq}} -\frac{2\pi}{|\rr-\rr'|} \left(\frac{k_{\rm B}T}{\hbar c}\right)^3 \Phi'(0) \sum_{n=1}^{+\infty} n^2 \exp\left(-\frac{2\pi n k_{\rm B}T}{\hbar c} |\rr-\rr'|\right)}
\end{equation}
 As at zero temperature, the result is negative. Physically, to enter the regime of large distances, it is necessary that $|\rr-\rr'|\gg 1/k_{\rm F}$, but also that $|\rr-\rr'|\gg \hbar/m c$ in the BEC limit (this is the healing length $\xi$ of the dimer condensate) and that $|\rr-\rr'|\gg \hbar k_{\rm F}/m \Delta$ in the BCS limit (this is the size of a bound pair $\ell_{\rm pair}$). On the other hand, there is no need to impose any constraint on the ratio between $|\rr-\rr'|$ and the thermal wavelength of phonons $\lambda_\phi=\hbar c/k_{\rm B} T$. In particular, the two extreme situations are well described and continuously connected by (\ref{eq:asympnz3d}), which is what makes it so unique: (i) the case where $|\rr-\rr'|$ diverges faster than $\lambda_\phi$ and where the sum is dominated by the term $n=1$, which gives, up to a factor, the expected exponential decay with $\kappa_0=2\pi k_{\rm B}T/\hbar c$ (thermal asymptotic regime), and (ii) the case where $|\rr-\rr'|$ diverges more slowly than $\lambda_\phi$ and where the sum over $n$ can be replaced by an integral (quantum asymptotic regime). In the second case, the pair distribution functions do not see that the system is at non-zero temperature, and we find exactly the algebraic equivalent (\ref{eq:asympzero}) in $d=3$.\footnote{We give $\int_0^{+\infty}\mathrm{d} x\, x^2\exp(-x)=2$.}

Let us conclude with the more technical case $d=2$. The difficulty is that $J_0(q|\rr-\rr'|)$ is an even function of $q$, which makes the integrand of (\ref{eq:ipp2d}) odd and prevents the integration from being extended to the entire $\mathbb{R}$. We get out of this predicament by taking up the idea in section 8.7.3.2 of reference \cite{livre} and extending it to all orders in $1/|\rr-\rr'|$. To do this, we replace the Bessel function with its asymptotic expansion 
\begin{equation}
J_0(q|\rr-\rr'|) \underset{|\rr-\rr'|\to +\infty}{\simeq} \left(\frac{2}{\pi q|\rr-\rr'|}\right)^{1/2} \left[
\cos\left(q|\rr-\rr'|-\frac{\pi}{4}\right) u(q|\rr-\rr'|) - \sin\left(q|\rr-\rr'|-\frac{\pi}{4}\right) \frac{v(q|\rr-\rr'|)}{q|\rr-\rr'|}\right]
\end{equation}
where the functions $u(x)$ and $v(x)$ are given in paragraph \S8 .451(1) of reference \cite{GR} in the form of series presenting only even powers of $1/x$. We then show, as in reference \cite{livre}, that for any odd function $f_-(q)$ on $\mathbb{R}$ and for any even function $f_+(q)$ on $\mathbb{R}$, we have thanks to the square root $q^{1/2}$:\footnote{The first relation is the equation (8.349) of \cite{livre}; the second is deduced from it by taking the derivative with respect to $|\rr-\rr'|$. The intuitive idea is that the square root of a negative real number contains a factor $\exp(\mathrm{i}\pi/2)$, i.e., precisely the relative phase between the Fourier components of $\cos(|\rr-\rr'|-\pi/4)$ or $\sin(q|\rr-\rr'|-\pi/4)$ (up to a sign for the sine).} 
\begin{eqnarray}
\int_0^{+\infty} \frac{\mathrm{d} q}{q^{1/2}} f_-(q) \cos(q|\rr-\rr'|-\pi/4) &=& \eee^{-\mathrm{i} \pi/4} \int_{-\infty + \mathrm{i} 0^+}^{+\infty+\mathrm{i} 0^+} \frac{\mathrm{d} z}{2 z^{1/2}} f_-(z) \eee^{\mathrm{i} z|\rr-\rr'|} \\
\int_0^{+\infty} \frac{\mathrm{d} q}{q^{1/2}} f_+(q) \sin(q|\rr-\rr'|-\pi/4) &=& -\eee^{\mathrm{i} \pi/4} \int_{-\infty + \mathrm{i} 0^+}^{+\infty+\mathrm{i} 0^+} \frac{\mathrm{d} z}{2z^{1/2}} f_+(z) \eee^{\mathrm{i} z|\rr-\rr'|}
\end{eqnarray}
 which remains to be applied to $f_-(q)=\phi(q) u(q|\rr-\rr'|)$ and $f_+(q)=\phi(q) v(q|\rr-\rr'|)/q$. The sequence is similar to 3D. It is important to note that $u(\mathrm{i} x)+v(\mathrm{i} x)/x=(2 x/\pi)^{1/2} \exp(x) K_0(x)$, by virtue of the relation \S 8.451(6) of reference \cite{GR}. We ultimately find in 2D that 
\begin{equation}
\label{eq:asympnz2d}
\boxed{\delta g^{\rm ltd}_{\sigma\sigma'}(\rr,\rr') \stackrel{T\to 0^+}{\underset{|\rr-\rr'|\to +\infty}{\simeq}}\delta g_{\rm coll}(\rr,\rr') \stackrel{T\to 0^+}{\underset{|\rr-\rr'|\to +\infty}{\simeq}} -4\pi \left(\frac{k_{\rm B}T}{\hbar c}\right)^3 \Phi'(0) \sum_{n=1}^{+\infty} n^2 K_0\left(\frac{2\pi n k_{\rm B}T}{\hbar c} |\rr-\rr'|\right)}
\end{equation}
 As in 3D, this approximation, which to our knowledge is original, has negative values (we have $\Phi'(0)>0$ and $K_0(x)>0\,\forall x>0$) and continuously connects the asymptotic thermal regime to the asymptotic quantum regime. In particular, (i) by limiting the sum to the term $n=1$ and replacing the function $K_0$ with an asymptotic equivalent $K_0(x)\sim (\pi/2x)^{1/2}\exp(-x)$, we find the expected exponential decay (up to a power-law factor) 
\begin{equation}
\delta g_{\sigma\sigma'}^{\rm ltd}(\rr,\rr')\stackrel{|\rr-\rr'|\to +\infty}{\underset{{\rm then}\ T\to 0^+}{\sim}} -\frac{2\pi}{|\rr-\rr'|^{1/2}} \left(\frac{k_{\rm B}T}{\hbar c}\right)^{5/2} \Phi'(0)\, \exp(-\kappa_0|\rr-\rr'|) \quad \mbox{with}\quad \kappa_0=2\pi k_{\rm B}T/\hbar c\ ;
\end{equation}
 (ii) by replacing in (\ref{eq:asympnz2d}) the sum over $n$ with an integral, we find exactly the algebraic equivalent at zero temperature (\ref{eq:asympzero}) in $d=2$.\footnote{We give $\int_0^{+\infty} \mathrm{d} x \, x^2 K_0(x)=\pi/2$.}

\subsubsection{What quantum hydrodynamics says about it}\label{sec3_1_3}

Let us now take a step back and ask ourselves to what extent the beautiful approximations (\ref{eq:asympnz3d},\ref{eq:asympnz2d}) are specific to the linearized time-dependent BCS theory. It appears that, in the double limit of low temperatures $T\to 0^+$ and large distances $|\rr-\rr'|\to +\infty$ considered, we can make use of an effective low-energy theory, Landau and Khalatnikov's quantum hydrodynamics \cite{LK}. We expect this theory to be accurate at the dominant temperature order for any system — even strongly correlated ones such as liquid helium — (i) subject to short-range interactions and (ii) entirely superfluid in its ground state. Our unpolarized, contact-interacting spin-$1/2$ Fermi gas satisfies both of these conditions; see Appendix B of reference \cite{brouifer} for more details and reference \cite{qouv}. Thus, (i) in its ground state, the fluid has a linear acoustic excitation branch $\omega_\mathbf{q}\sim cq$, with a speed of sound $c$ linked to its exact equation of state {\it} by the hydrodynamic relation (\ref{eq:hydro}); (ii) at sufficiently low temperatures, the equilibrium state of phonons, which are the quanta of these acoustic excitations, can be assimilated to that of an ideal gas of bosons with zero chemical potential $\mu_\phi=0$, i.e., with many-body density operator $\hat{\sigma}_\phi\propto \exp(-\beta \sum_\mathbf{q} \hbar\omega_\mathbf{q} \hat{b}_\qq^\dagger \hat{b}_\qq)$; (iii) in a coarse-grained description, the operator-valued density fluctuations of the fluid are dominated—again at sufficiently low temperatures—by phonons and have modal expansion 
\begin{equation}
\delta\hat{\rho}(\rr) = \frac{1}{L^{d/2}} \sum_{\mathbf{q}\neq \mathbf{0}} \rho_\mathbf{q} \left(\hat{b}_\qq+\hat{b}_{-\mathbf{q}}^\dagger\right)\exp(\mathrm{i}\mathbf{q}\cdot\rr) \quad\mbox{with}\quad \rho_\qq=\left(\frac{\hbar\rho q}{2mc}\right)^{1/2}
\end{equation}
 Here, $\rho$ is the {\it exact} average density of the fluid in its ground state, $m$ is the mass of a particle of the fluid, and the creation $\hat{b}^\dagger_\qq$ and annihilation $\hat{b}_\qq$ operators of a phonon with wave vector $\qq$ obey the usual bosonic commutation relations, for example $[\hat{b}_\qq,\hat{b}^\dagger_{\qq'}]=\delta_{\mathbf{q}\qq'}$. Hence the correlation function of the total density at the thermodynamic limit in this theory, 
\begin{equation}
\label{eq:reshydro}
\langle\delta\hat{\rho}(\rr)\delta\hat{\rho}(\rr')\rangle_{\rm hydro} = \int_{q<\Lambda}\frac{\dd^dq}{(2\pi)^d} 2 \rho_\qq^2 \left(\bar{n}_\qq+\frac{1}{2}\right)\exp[\mathrm{i}\mathbf{q}\cdot(\rr-\rr')]
\end{equation}
 where the ultraviolet cutoff $\Lambda$ is $\gg k_{\rm B}T/\hbar c$ but $\ll m c/\hbar$. However, since the two internal states $\sigma'=\uparrow,\downarrow$ play symmetrical roles for a very distant spin $\sigma$, and the thermal dissociation of pairs $\uparrow\downarrow$ is exponentially negligible in our Fermi gas at low temperature, we must have at a great distance (at least much greater than the size of a bound pair, $|\rr-\rr'|\gg \ell_{\rm pair}$): 
\begin{equation}
\delta g_{\sigma\sigma'}(\rr,\rr') \simeq \frac{1}{4} \langle\delta\hat{\rho}(\rr)\delta\hat{\rho}(\rr')\rangle_{\rm hydro}
\end{equation}
 Comparison with (\ref{eq:exp3}) imposes 
\begin{equation}
\Phi(\qq)\underset{q\to 0^+}{\sim} \frac{\hbar \rho q}{4 m c}
\end{equation}
 Expression (\ref{eq:dphi0}) of $\Phi'(0)$ and the asymptotic approximations (\ref{eq:asympzero},\ref{eq:asympnz3d},\ref{eq:asympnz2d}) derived from it are therefore accurate and very general—the latter result from mathematical properties trivially satisfied by the integrand of (\ref{eq:reshydro}) — provided that the density and speed of sound BCS are replaced by their true values in the gas.

In conclusion, in our unpolarized Fermi gas at sufficiently low temperature, the pair distribution functions $g_{\sigma\sigma'}(\rr,\rr')$ are at certain points, in particular at sufficiently large distances $|\rr-\rr'|$, strictly below their asymptotic values. This property, in direct contradiction with the static BCS theory for $g_{\uparrow\downarrow}(\rr,\rr')$, see equation (\ref{eq:dgsta}), results from quantum or thermal fluctuations in the positions of the centers of mass of the bound pairs induced by phonons, which are conspicuously absent from this theory. It applies equally well to the linearized time-dependent BCS theory and to physical reality.

\subsection{Numerical results}\label{sec3_2}

We perform numerical calculations on the pair distribution functions $g^{\rm ltd}_{\sigma\sigma'}(\rr,\rr')$ of section \ref{sec2} using their decomposition (\ref{eq:sepa}) and the representation (\ref{eq:exp1}) of the collective part $\delta g_{\rm coll}(\rr,\rr')$—in reality, its Fourier transform $\delta\tilde{g}_{\rm coll}(\qq)$—by curvilinear integral in the complex plane. For simplicity, we set ourselves at order zero in $\exp(-\beta E_{\rm gap})$, i.e., we omit the fermionic thermal factors $f_{\mathbf{k}},f_{\pm}$ everywhere, in particular in the functions $\Sigma_{ij}(\qq,z)$ and therefore in the susceptibility $\tilde{\chi}_{\rm coll}(\qq,z)$.\footnote{\label{note:num} In the numerics, we have introduced a sufficiently large isotropic cutoff $k<k_{\rm max}$ on the internal wave vector $\kk$ in the functions $\Sigma_{ij}(\qq,z)$, with $k_{\rm max}=273(m\Delta/\hbar^2)^{1/2}$ in the left column (3D case) and $k_{\rm max}=157 q_{\rm dim}$ in the right column (2D case) of Figure~\ref{fig:compar}. The advantage is also that the branch cut on $z$ starting at $E_{\rm bord}(\qq)$ does not go beyond $\eps_{\rm max}\simeq \hbar^2(k^2_{\rm max}+q^4/4)-2\mu$, since the energy denominators in $\Sigma_{ij}(\qq,z)$ can no longer vanish if $z=\eps>0$ is too large; the integration path $C_{\rm h}$ can then join the real axis via a vertical segment between $z=\eps_{\rm max}+\mathrm{i}\eta$ (with $\eta=\pi k_{\rm B}T$ as stated below equation (\ref{eq:exp1})) and the terminal point $z=\eps_{\rm max}$. Furthermore, there is no point in exploring $C_{\rm h}$ below $\re z=-2E_{\rm gap}$ because of the exponential suppression by the thermal denominator in (\ref{eq:exp1}). In the integration over $\eps$, we took a step $\delta\eps=0.2\min(\Delta,\hbar c q, k_{\rm B}T,\eta)$ (see note \ref{note:matsu}). In the integration over the modulus of $\kk$, we took a step $\delta k=0.2 \min(\max((m\mu)^{1/2}/\hbar,(m\Delta)^{1/2}/\hbar),m\Delta/\hbar^2/(m\mu/\hbar^2)^{1/2})$ for $\mu>0$ and—case not explored—$\delta k=0.2\max((-m\mu)^{1/2}/\hbar,(m\Delta)^{1/2}/\hbar)$ for $\mu\leq 0$. In the integral on the angle $\theta$ between $\kk$ and $\qq$, reduced to the interval $[0,\pi/2]$ by symmetry, we took $30$ points. We verified that the results change little if we divide one of the integration steps $\delta\eps,\delta k$ or $\delta\theta$ by two. In the numerical calculation of the inverse Fourier transform giving $\delta g_{\rm coll}(\rr,\rr')$ from $\delta\tilde{g}_{\rm coll}(\qq)$, we put a cutoff $q_{\rm max}=k_{\rm max}/2$ on the wave number $q$, and we take a step $\delta q=0.1(m\Delta/\hbar^2)^{1/2}$ in 3D and $\delta q=0.13 q_{\rm dim}$ in 2D. The cutoff $q_{\rm max}$ chosen would be very insufficient for $g^{\rm ltd}_{\uparrow\downarrow}(\rr,\rr')$ if we performed the calculation of the inverse Fourier transform of $\delta\tilde{g}^{\rm ltd}_{\uparrow\downarrow}(\qq)$ directly without using the decomposition (\ref{eq:sepa}) ; in fact, the contribution $\delta\tilde{g}^{\rm sta}_{\uparrow\downarrow}(\qq)$ of static BCS theory decreases slowly with $q$, see equation (\ref{eq:stahbasymp}), while the collective contribution $\delta\tilde{g}_{\rm coll}(\qq)$ decreases as $1/q^4$, see equation (\ref{eq:collasymp}). Part $\delta g^{\rm sta}_{\sigma\sigma'}(\rr,\rr')$ is much easier to obtain through the anomalous and normal averages as in (\ref{eq:wickhb},\ref{eq:wickhh}); in 2D, these averages are known analytically, see equations (\ref{eq:moya2d},\ref{eq:moyn2d}); in 3D, they are given by the integrals over $\kk$ (\ref{eq:moya},\ref{eq:moyn}), whose numerical evaluation would also pose a problem of numerical truncation on $k$ \cite{Romero3D} if the integrand were not known analytically: we can evaluate the omitted part numerically using an asymptotic expansion of the integrand, or simply use a clever contour integral representation, see our note \ref{note:asymp} and reference \cite{Romero3D}. Finally, in 3D, the numerical precision on $\delta\tilde{g}_{\rm coll}(\qq)$ at large $q$ is not sufficient to accurately describe the departure of $g_{\uparrow\uparrow}(\rr,\rr')$ at short distances $|\rr-\rr'|$; we prefer to perform the calculation of the part $q>15(m\Delta/\hbar^2)^{1/2}$ of the inverse Fourier transform after substitution $\delta\tilde{g}_{\rm coll}(\qq)\simeq C/q^4+D/q^5+E/q^6$; coefficients $C$ and $D$ are given in equation (\ref{eq:coefcoll3d}), coefficient $E$ is obtained by fitting.} 

To test the numerical calculation of $\delta\tilde{g}_{\rm coll}(\qq)$, we compare it with the boundary behaviors (\ref{eq:appli2}) and (\ref{eq:collasymp}) obtained analytically; in the examples considered here, which are all at very low temperatures,
 the low-$q$ limit is negative, as is the high-$q$ limit, and we find that $\delta\tilde{g}_{\rm coll}(\qq)$ connects the two monotonically one to the other, thus remaining negative everywhere: 
\begin{equation}
\label{eq:estneg}
\delta\tilde{g}_{\rm coll}(\qq) \stackrel{\rm num}{<}0\quad\forall \mathbf{q}
\end{equation}

 There are two reasons for our numerical study: (i) to see whether $\delta g_{\uparrow\downarrow}(\rr,\rr')/\rho_{\uparrow}\rho_{\downarrow}$ can take significant negative values (relative to unity) in the linearized time-dependent BCS theory, a question that the asymptotic study in section \ref{sec3_1} cannot answer; (ii) to compare this theory with other approaches or with measurements in a gas of cold fermionic atoms.

\subsubsection{Three-dimensional case; comparison with Giorgini's quantum Monte Carlo}\label{sec3_2_1}
\begin{figure}[t]
\begin{center}
\includegraphics[width=6cm,clip=]{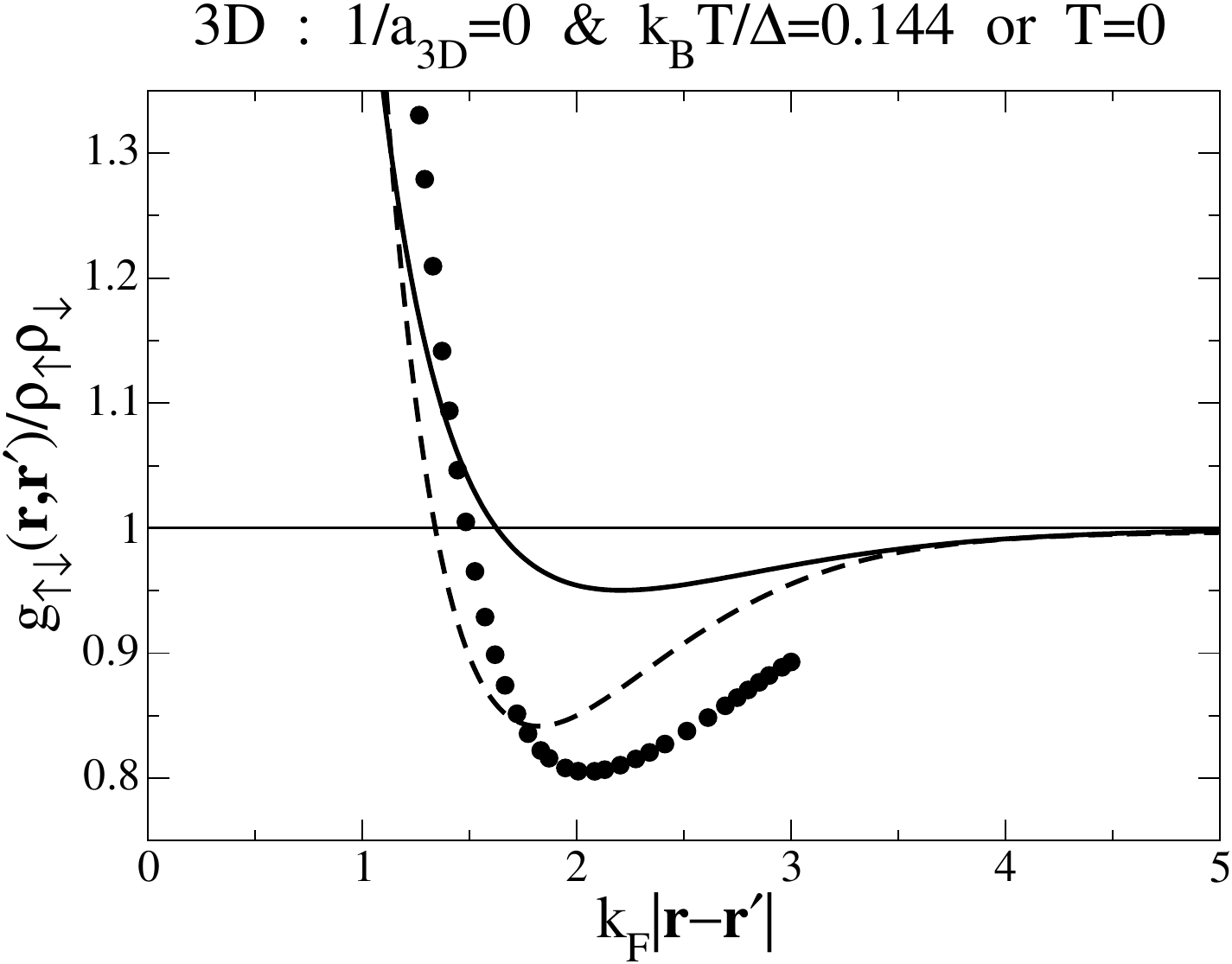}\hspace{1cm}\includegraphics[width=6cm,clip=]{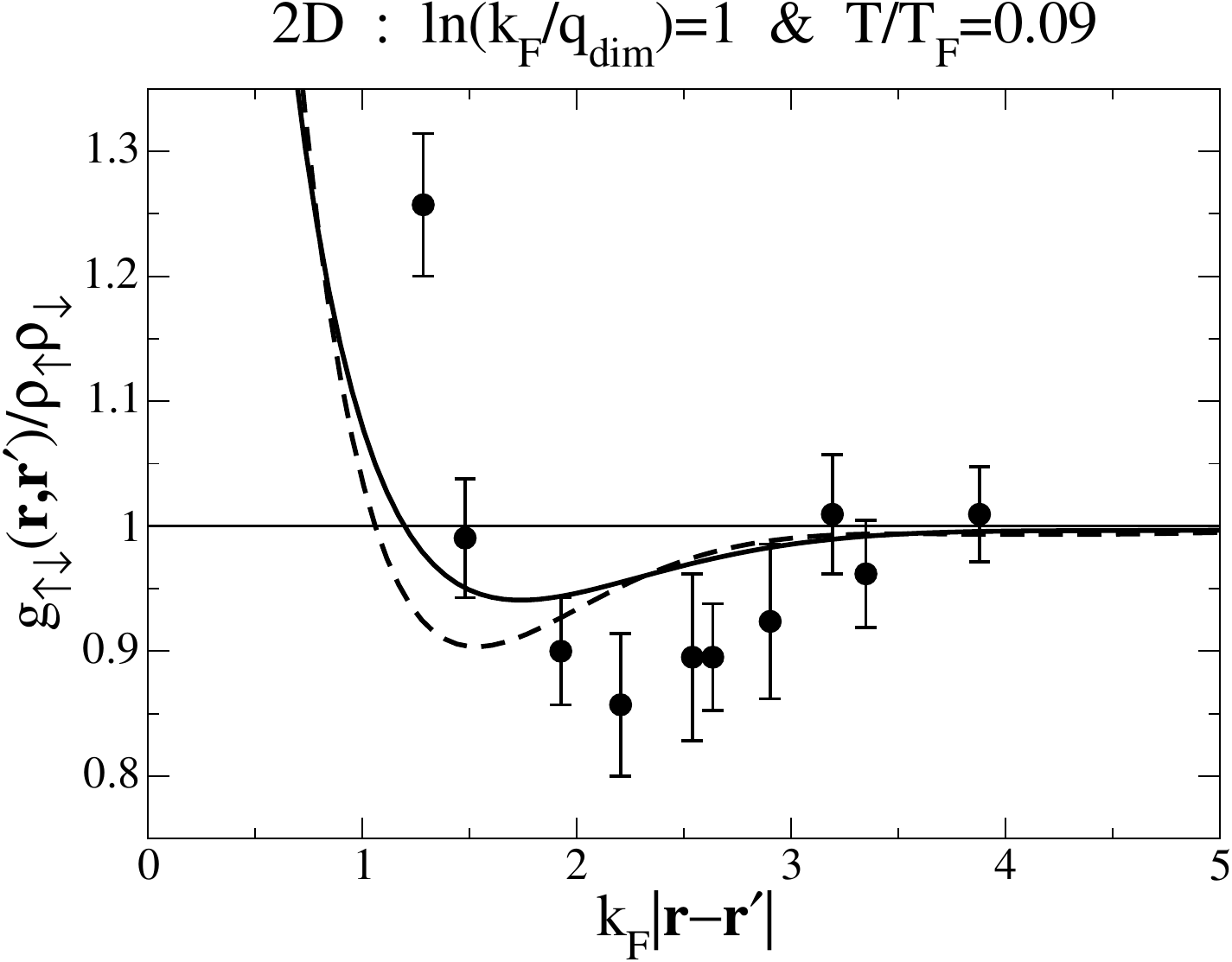}\\
\hspace{1cm}\\
\includegraphics[width=6cm,clip=]{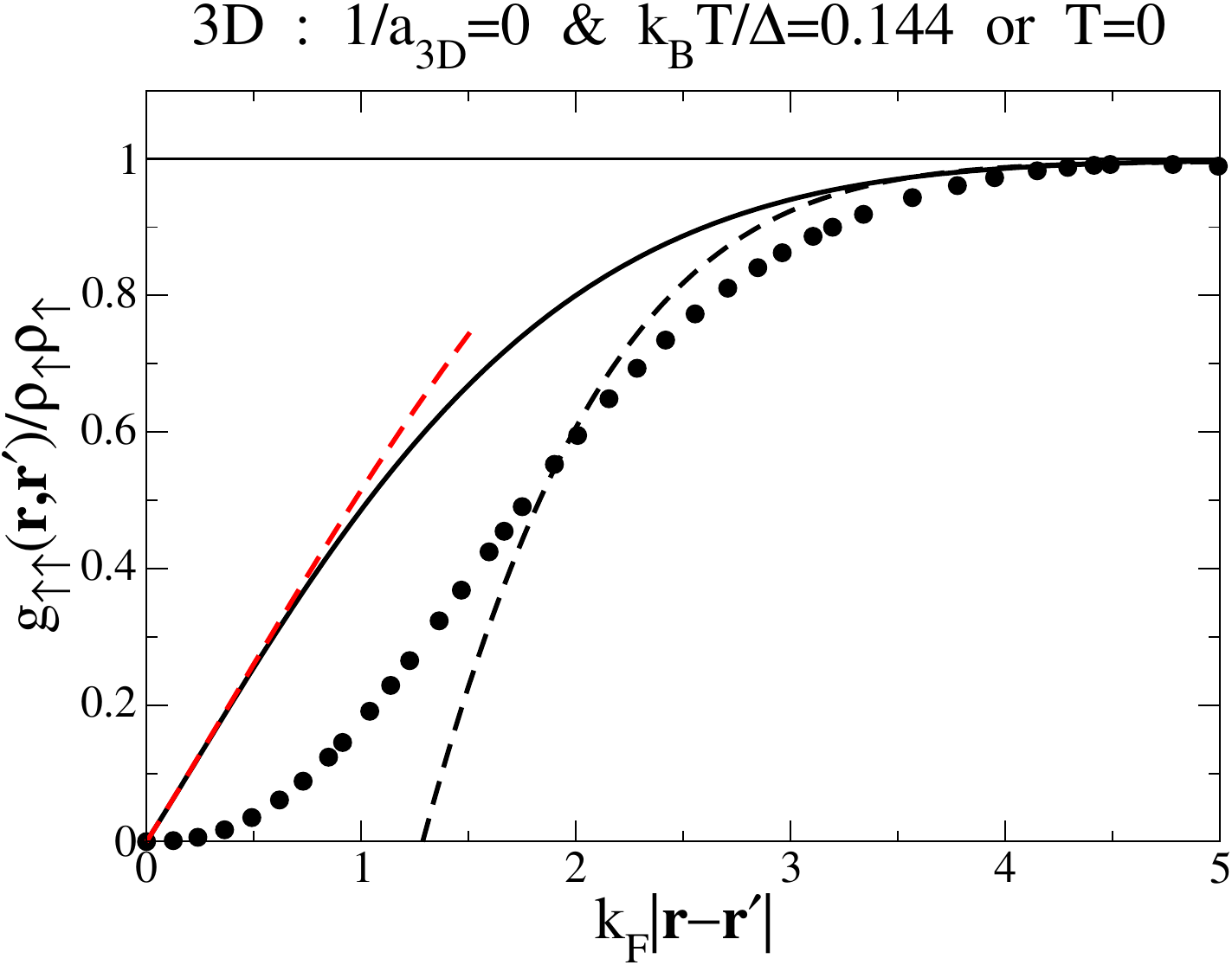}\hspace{1cm}\includegraphics[width=6cm,clip=]{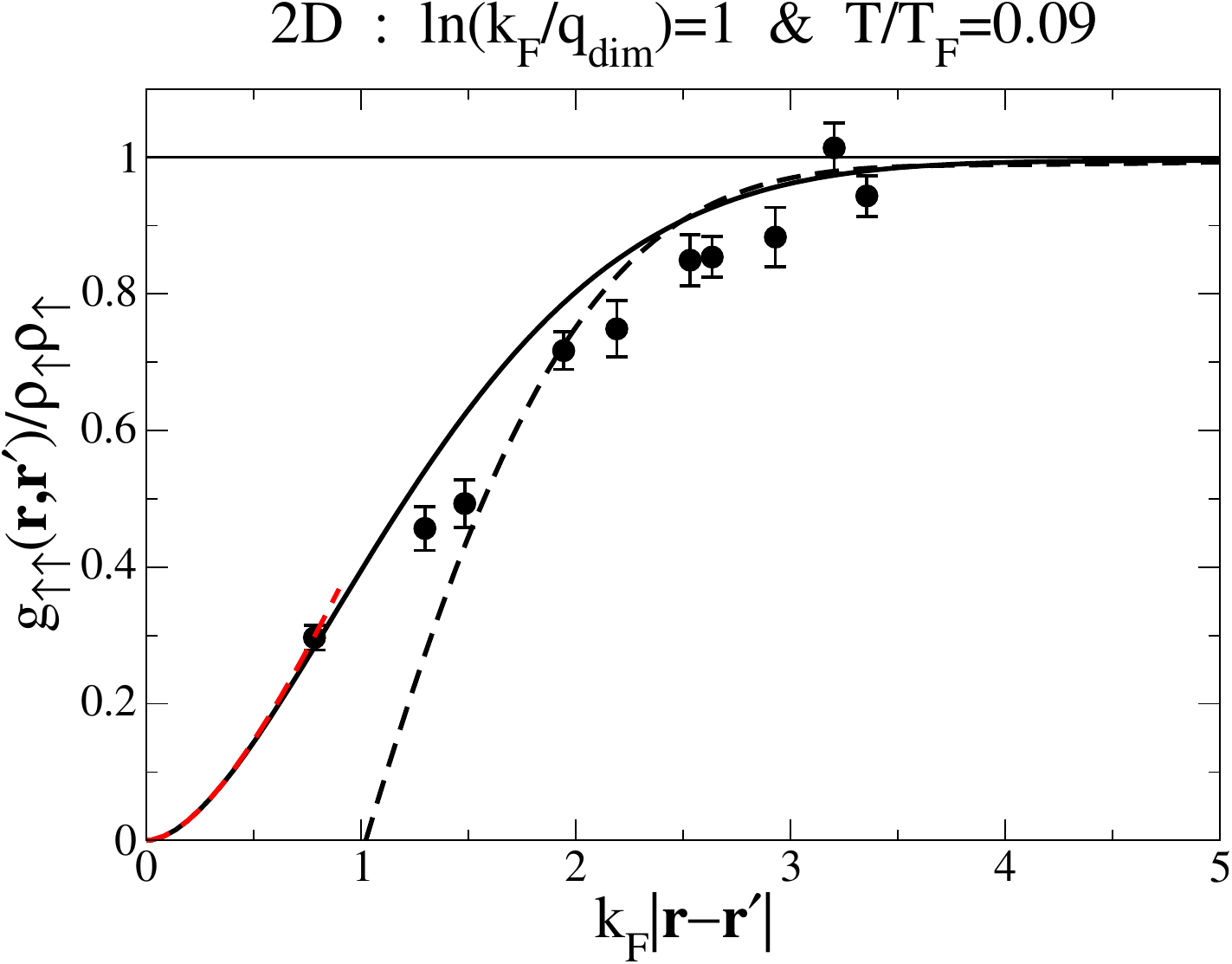}
\end{center}
\caption{In a spatially homogeneous, unpolarized Fermi gas of spin-$1/2$ fermions, pair distribution functions $g_{\uparrow\downarrow}$ (top row) and $g_{\uparrow\uparrow}$ (bottom row) in dimension $d=3$ at the unitary limit (left column) and in dimension $d=2$ for a parameter interaction $\ln(k_{\rm F}/q_{\rm dim})=1$ (right column). Dashed line: calculation of the deviations $\delta g_{\sigma\sigma'}$ of the $g_{\sigma\sigma'}$ from their asymptotic values by the linearized time-dependent BCS theory, then naive completion (\ref{eq:pas_bon}) of $\delta g_{\sigma\sigma'}$ in $g_{\sigma\sigma'}$ using the density $\rho^{\rm sta}$ of static BCS theory. Solid black line: same calculation of $\delta g_{\sigma\sigma'}$ in linearized time-dependent BCS theory, but correct completion (\ref{eq:bon}) using the modified density $\rho^{\rm ltd}$ in this theory. Red line at short distances: analytical predictions (\ref{eq:ghh3ddep},\ref{eq:ghh2ddep}) for the departure of $g_{\uparrow\uparrow}$, still in this theory. The total density $\rho=\rho_\uparrow+\rho_\downarrow=2\rho_\sigma$ and the resulting Fermi wave number $k_{\rm F}$ (\ref{eq:kf}) are considered fixed in the figure; the chemical potential $\mu$ in our grand canonical theory does not therefore have the same value for the dashed curves (for which $\rho=\rho^{\rm sta}(\mu,T)$) and the solid curves (for which $\rho=\rho^{\rm ltd}(\mu,T)$). Black discs: results of a 3D fixed-node diffusion quantum Monte Carlo calculation (zero temperature) \cite{GiorginiPRL,GiorginiRMP} smoothed by quintic polynomial fitting for the left column; experimental measurements using a microscope in a quasi-2D gas of cold atoms \cite{TarikHAL} for the right column (the error bars
 do not take into account the poorly known bias induced by the third dimension). The linearized time-dependent BCS theory is implemented at order zero in $\exp(-\beta E_{\rm gap})$, in 3D at a temperature $T=0.144\Delta/k_{\rm B}\simeq 0.251 mc^2/k_{\rm B}$ (low enough to allow comparison with zero temperature) and in 2D at a temperature $T=0.09T_{\rm F}$ (as in the experiment).}
\label{fig:compar}
\end{figure}

 In dimension $d=3$, we place ourselves at the unitary limit $a_{\rm 3D}^{-1}=0$; which is the maximum interaction regime accessible in the gas phase, and therefore a priori the most difficult to describe, but also the most studied, both theoretically and experimentally, in cold atomic gases, and the most interesting because it is scale invariant. Let us take a sufficiently low temperature in the numerics to be able to place ourselves at order zero in $\exp(-\beta E_{\rm gap})$, 
\begin{equation}
\frac{k_{\rm B}T}{\Delta}=0.144
\end{equation}
 where $\Delta$ is the BCS order parameter. Since the chemical potential is positive, we then have $E_{\rm gap}=\Delta$ and $\exp(-\beta E_{\rm gap})\simeq 10^{-3}\ll 1$. From equations (\ref{eq:delta3d},\ref{eq:rho3d}) of the static BCS theory and the definition (\ref{eq:kf}) of the Fermi wave number, we obtain ${\mu}/{\Delta} = 0.860\ 436\ldots$ and ${\mu}/{E_{\rm F}^{\rm sta}} =0.590\, 605\ldots$. The predicted ratio $\mu/E_{\rm gap}$ is close to the value measured at the lowest accessible temperatures, but the ratio $\mu/E_{\rm F}$ is quite far from it, since we find in experiments $E_{\rm gap}/E_{\rm F}\simeq 0.44$ \cite{Ketterle} and $\mu/E_{\rm F}\simeq 0.376(4)$ \cite{mit}, which leads to $\mu/\Delta\simeq 0.86$ (assuming that we also have $E_{\rm gap}=\Delta$ in physical reality). Fortunately, the modified equation of state (\ref{eq:bon}) of the time-dependent BCS theory does significantly better: it does not change the ratio $\mu/\Delta$, but gives 
\begin{equation}
\frac{\mu}{E_{\rm F}^{\rm ltd}} \simeq 0.401
\end{equation}
which is in perfect agreement with the Gaussian fluctuation approximation, see references \cite{Drummond,Randeria3D} and our note \ref{note:afg}. Our choice of temperature corresponds to $k_{\rm B}T/mc^2\simeq 0.251\ll 1$, so that the effect of thermal phonons should be small enough for the equation of state and the pair distribution functions $g_{\sigma\sigma'}(\rr,\rr')$ to be close to their values in the ground state (scale invariance imposes $m c^2=2\mu/3$ in the hydrodynamic expression (\ref{eq:hydro}) of the speed of sound).\footnote{\label{note:drhophi} Phonons change the grand potential of the system by an amount $\delta\Omega_\phi=k_{\rm B}T\sum_\mathbf{q} \ln[1-\exp(-\beta \hbar \omega_\qq)]$, therefore the density of the grand canonical Fermi gas by an amount $\delta\rho_\phi=-\partial_\mu \delta\Omega_\phi/L^d\sim-\int_{\mathbb{R}^d}[\dd^dq/(2\pi)^d] \hbar(\partial_\mu c)q/[\exp(\beta\hbar c q)-1]=-[\Omega_d/(2\pi)^d]\hbar(\partial_\mu c)(k_{\rm B}T/\hbar c)^{d+1}\Gamma(d+1)\zeta(d+1)$ where the dispersion relation $\omega_\qq$ is taken to be linear (\ref{eq:depart}) at sufficiently low temperature $T\to 0^+$, $\Omega_d$ is the total angle of space ($2\pi$ if $d=2$, $4\pi$ if $d=3$), $\Gamma$ is Euler's Gamma function, and $\zeta$ is Riemann's zeta function. In 3D at the unitary limit, we deduce that $\delta\rho_\phi/\rho{\underset{T\to 0^+}{\sim}} -(\pi^4/30)(k_{\rm B}T/mc^2)^4(\xi_{\rm B}/3)^{3/2}$ with $\xi_{\rm B}=\mu/E_{\rm F}$ the Bertsch parameter. For our parameters, this gives in the linearized time-dependent BCS theory $\delta\rho_\phi/\rho\simeq -6\cdot 10^{-4}\ll 1$. To see if the modified equation of state (\ref{eq:bon}) gives the correct value of $\delta\rho_\phi$ at low temperature, let us assume that the integral over $\qq$ in (\ref{eq:bon}) is a small correction to $\rho^{\rm sta}$; in this case, we can write $\rho^{\rm ltd}\simeq \rho^{\rm sta}-(2/\rho^{\rm sta})\int_{\mathbb{R}^d}[\dd^dq/(2\pi)^d]\delta\tilde{g}_{\rm coll}(\qq)$. We then find with (\ref{eq:exp3}) that $\delta\rho_{\phi}^{\rm ltd} = -\int_{\mathbb{R}^d}[\dd^dq/(2\pi)^d] (2\Phi'(0)/\rho^{\rm sta})q/[\exp(\beta\hbar c q)-1]$; given the expression (\ref{eq:dphi0}) of $\Phi'(0)$, there is agreement if $\hbar \partial_\mu c=\hbar/2 m c$, and therefore if $\partial_\mu(mc^2)=1$. This condition is not generally perfectly satisfied, except in 2D, see equation (\ref{eq:c2d}). In the case where $\rho^{\rm ltd}$ differs significantly from $\rho^{\rm sta}$, the previous expression of $\delta\rho_\phi^{\rm ltd}$ must be multiplied by $\rho^{\rm sta}/\rho^{\rm ltd}$ and the condition becomes $(\rho^{\rm ltd}/\rho^{\rm sta})\partial_\mu (mc^2)=1$, which gives in 3D at the unitary limit $(2/3)(\xi_{\rm B}^{\rm sta}/\xi_{\rm B}^{\rm ltd})^{3/2}=1$ or $1.19\,\#\, 1$.} It also corresponds to $T/T^{\rm ltd}_{\rm F}\simeq 0.067$, and therefore to a temperature significantly lower than that $T\simeq 0.1 T_{\rm F}$ of reference \cite{mit} but still higher than that $T\simeq 0.05 T_{\rm F}$ of reference \cite{recordKetterle}, the record in a gas of cold fermionic atoms.\footnote{\label{note:drhogam} For $T/T_{\rm F}\simeq 0.1$ and $\xi_{\rm B}\simeq 0.376$, as in \cite{mit}, we find a weak effect of phonons $\delta\rho_{\phi}/\rho\simeq -0.0036$, see our note \ref{note:drhophi}, but a pair dissociation factor $\exp(-\beta E_{\rm gap})\simeq 0.013$ that is not completely negligible a priori. To be sure, let us determine the change in density $\delta\rho_\gamma$ of the grand canonical Fermi gas due to the fermionic quasi-particles with Hamiltonian (\ref{eq:hsta}). Proceeding as in note \ref{note:drhophi} and approximating $\eps_\kk$ in the vicinity of its minimum by $\eps_\mathbf{k}\simeq E_{\rm gap}+\hbar^2(k-k_0)^2/2 m_*$, which is legitimate at low temperatures $T\to 0^+$, we find for $k_0>0$ that $\delta\rho_\gamma=(-2)\int_{\mathbb{R}^d}[\dd^dk/(2\pi)^d] \partial_\mu\eps_\kk/[1+\exp(\beta\eps_\kk)]\sim -2 (\partial_\mu E_{\rm gap}) \exp(-\beta E_{\rm gap})k_0^{{d-1}} (2\pi m_* k_{\rm B}T/\hbar^2)^{1/2}{\Omega_d/(2\pi)^d}$ where the overall factor $2$ is the number of internal states. In 3D at the unitary limit, where $\partial_\mu E_{\rm gap}=E_{\rm gap}/\mu$ due to scale invariance, the measured values $E_{\rm gap}=0.44 E_{\rm F},\ k_0\simeq 0.92 k_{\rm F}$ \cite{Ketterle} and the theoretical prediction $m_*/m\simeq 0.56$ \cite{Nishida} give $\delta\rho_\gamma/\rho\simeq -1.4\cdot 10^{-3}$ at the temperature of our numerical calculation $T=0.067 T_{\rm F}$ but $\delta\rho_\gamma/\rho\simeq -0.015$ at the experimental temperature $T=0.1 T_{\rm F}$. In the experiment, this leads to an overestimation of the Bertsch parameter $\xi_{\rm B}=(2m\mu/\hbar^2)/(3\pi^2\rho)^{2/3}$ by $\delta\xi_{\rm B}/\xi_{\rm B}=(-2/3)\delta\rho_\gamma/\rho\simeq 0.01$ in relative value, still within the error bars of reference \cite{mit}.}

In the left column of Figure~\ref{fig:compar}, we show the reduced pair distribution functions $g_{\sigma\sigma'}^{\rm ltd}(\rr,\rr')/\rho_\sigma\rho_{\sigma'}=1+\delta g^{\rm ltd}_{\sigma\sigma'}(\rr,\rr')/(\rho/2)^2$ as functions of the dimensionless distance $k_{\rm F}|\rr-\rr'|$, with $k_{\rm F}=(3\pi^2\rho)^{1/3}$ as in (\ref{eq:kf}). The figure is based on a fixed density $\rho$, as stated in the legend. However, its left column does not depend on the precise value of $\rho$ chosen because it is at the unitary limit, scale invariant; moreover, as we have set the ratio $k_{\rm B}T/\Delta$ or, which amounts to the same thing here, the ratio $k_{\rm B}T/\mu$, rather than $T/T_{\rm F}$, we consider it more convenient to discuss at a fixed chemical potential $\mu$ (the situation will be different for the right-hand column). After numerically determining the deviations $\delta g^{\rm ltd}_{\sigma\sigma'}(\rr,\rr')$ in the linearized grand canonical time-dependent BCS theory (the hard piece is of course $\delta g_{\rm coll}(\rr,\rr')$), we still need a calculation of the density $\rho$ (and therefore the Fermi wave number $k_{\rm F}$) to form the dimensionless quantities on the horizontal and vertical axes. The dashed curve corresponds to the tempting but incorrect choice of the static BCS equation of state $\rho=\rho^{\rm sta}(\mu,T)$ (\ref{eq:etat}), while the solid line corresponds to the correct choice of an equation of state corrected for collective excitations of the superfluid, $\rho=\rho^{\rm ltd}(\mu,T)$ (\ref{eq:bon}), as explained in section \ref{sec2_5} (in both cases, we are at order zero in $\exp(-\beta E_{\rm gap})$, which formally amounts to taking $\rho^{\rm sta}$ and the susceptibility $\tilde{\chi}_{\rm coll}(\qq,z)$ at zero temperature). Let us briefly comment on these results.

In case $\sigma=\sigma'=\uparrow$ (bottom row), the dashed curve is clearly catastrophically false, as it leads to negative values of $g_{\uparrow\uparrow}(\rr,\rr')$ in the vicinity of zero distance; however, we can clearly see from its definition (\ref{eq:defg}) that $g_{\uparrow\uparrow}(\rr,\rr')$ is positive—it is of the form $\langle \hat{A}^\dagger \hat{A}\rangle$ where $\hat{A}$ is an operator. The solid curve is much more convincing: it tends toward zero at zero distance, as it should for identical fermions $\uparrow\uparrow$ without direct interaction, and it reproduces the general shape of the Monte Carlo results (black discs). However, it differs dramatically from them in its linear departure in $|\rr-\rr'|$, rather than quadratic in Monte Carlo \cite{GiorginiPRL}. As we will see in section \ref{sec3_3}, this linear start is not unique to the linearized time-dependent BCS theory; it already exists in the static BCS theory (albeit with a lower slope), a fact that is not mentioned in references \cite{GiorginiPRL,GiorginiRMP}.

In case $\sigma=\uparrow,\sigma'=\downarrow$, the dashed curve, which is in reasonable agreement with the Monte Carlo results, gives us false hope, while the solid curve, with its much less pronounced dip in $g_{\uparrow\downarrow}(\rr,\rr')/\rho_\uparrow\rho_\downarrow$, gives us some disappointment. To understand why a relatively modest change in the equation of state ($\mu/E_{\rm F}$ changes from $0.59$ in static BCS to $0.40$ in linearized time-dependent BCS) has such a significant effect on the dip, let us write its depth (counted negatively) in Figure~\ref{fig:compar}: 
\begin{equation}
\label{eq:pour_comp}
\min_{|\rr-\rr'|} \frac{g_{\uparrow\downarrow}(\rr,\rr')}{\rho_\uparrow\rho_\downarrow}-1 = \frac{(3\pi^2)^2}{2} \left(\frac{\mu}{E_{\rm F}}\right)^3 \min_{|\rr-\rr'|} \frac{\delta g_{\uparrow\downarrow}(\rr,\rr')}{(m\mu/\hbar^2)^3}
\end{equation}
where the density has been expressed in terms of the Fermi wave number. The value of the minimum in the right-hand side of the equation, which depends solely on the thermodynamic variables $\mu$ and $T$, more precisely at the unitary limit on their ratio $k_{\rm B}T/\mu$ fixed here, is independent of the equation of state; on the other hand, its prefactor, proportional to the cube of the ratio $\mu/E_{\rm F}$, mechanically takes a factor $(0.40/0.59)^3\simeq 0.31$ when we move from the overly simplistic choice $\rho=\rho^{\rm sta}$ to the correct choice $\rho=\rho^{\rm ltd}$. The relative depth thus changes from $-0.16$ to $-0.05$, which is the true prediction of our theory, compared to $-0.2$ for Monte Carlo. The same analysis shows why the position of the minimum $|\rr-\rr'|_0$ changes very little (it is slightly shifted to the right): in the notation 
\begin{equation}
k_{\rm F} |\rr-{\rr'}|_0 = \left(\frac{E_{\rm F}}{\mu}\right)^{1/2} \left[(2m\mu/\hbar^2)^{1/2}|\rr-\rr'|_0\right]
\end{equation}
the quantity in brackets is independent of the equation of state, and the prefactor is multiplied only by $(0.59/0.40)^{1/2}\simeq 1.21$ between the dashed curve and the solid curve. The reduced position $k_{\rm F} |\rr-\rr'|_0$ of the minimum thus changes from $1.82$ to $2.21$, which is the true prediction of our theory, compared to $2.05$ for Monte Carlo.

Another way of reasoning is to take the opposite approach, i.e., to grand-canonize the results of references \cite{GiorginiPRL,GiorginiRMP} by expressing them in the natural grand-canonical wave number unit $(m\mu/\hbar^2)^{1/2}$. Since we have $\mu/E_{\rm F}\simeq 0.42$ in these references, we find by inverting relation (\ref{eq:pour_comp}) that 
\begin{equation}
\label{eq:jeugc}
\min_{|\rr-\rr'|} \frac{\delta g_{\uparrow\downarrow}(\rr,\rr')}{(m\mu/\hbar^2)^3} \simeq \left\{\begin{array}{ll} 
-6.0\cdot 10^{-3} & \mbox{in the Monte Carlo calculation of \cite{GiorginiPRL,GiorginiRMP}} \\
& \\
-1.8\cdot 10^{-3} & \mbox{in our linearized time-dependent BCS theory}
\end{array}
\right.
\end{equation}
 The disagreement is obvious, without any need to debate the equation of state, and the acceptable agreement between the dotted curve and the Monte Carlo points in the upper left panel of Figure~\ref{fig:compar} is only an illusion.

\subsubsection{Two-dimensional case; comparison with ENS experiments}\label{sec3_2_2}

In dimension $d=2$, we have recent experimental measurements of the pair distribution functions $g_{\sigma\sigma'}(\rr,\rr')$ in a spatially homogeneous gas of interacting cold fermionic atoms \cite{TarikHAL,ZwierMicro}, both for equal spins and opposite spins, thanks to a microscopy technique initially developed for atoms in an optical lattice \cite{Greiner,ZwierReseau,Gross} and then successfully extended to a continuous space \cite{TarikParf} (see reference \cite{KetterleMicro} for bosons). Here too, let us choose a fairly strong, if not maximum, interaction: in 2D, there is no strict equivalent of the unitary limit, but we can have a scattering length $a_{\rm 2D}$ or, if we prefer, a dimer size in free space $q_{\rm dim}^{-1}$ of the order of the average distance between particles; among the interaction strengths available in \cite{TarikHAL}, let us therefore take\footnote{The notation $\eta$ is the same as that of reference \cite{TarikHAL}; however, the quantity noted $a_{\rm 2D}$ in this reference is not {\sl not} our two-dimensional scattering length (\ref{eq:conva}) but rather the dimer size $q_{\rm dim}^{-1}$.} 
\begin{equation}
\label{eq:valeta}
\eta\equiv\ln(k_{\rm F}/q_{\rm dim})=1 \quad\mbox{that is}\quad\left\{\begin{array}{ccc}
\displaystyle \rho q_{\rm dim}^{-2}=\frac{1}{2\pi} \exp(2\eta)&\simeq& 1.1760 \\
&&\\
\displaystyle \frac{\eps_{\rm dim}}{E_{\rm F}}=2\exp(-2\eta)&\simeq& 0.2707
\end{array}
\right.
\end{equation}
 where the wave number $q_{\rm dim}$ is related to $a_{\rm 2D}$ and to the binding energy of the dimer $\eps_{\rm dim}$ as in (\ref{eq:edim2d}), and where $k_{\rm F}$ is the Fermi wave number (\ref{eq:kf}). We take the same temperature as in the experiment, 
\begin{equation}
\label{eq:valtemp}
\frac{T}{T_{\rm F}}=0.09 \quad \mbox{that is}\quad \frac{k_{\rm B}T}{\eps_{\rm dim}}=\frac{1}{2}\exp(2\eta)\frac{T}{T_{\rm F}}\simeq 0.3325
\end{equation}
It has the good taste of leading to a negligible thermal dissociation rate of bound pairs $\uparrow\downarrow$, 
\begin{equation}
\exp(-\beta E_{\rm gap})\simeq 10^{-3}
\end{equation}
 if we remember expression (\ref{eq:rel2d}) of the order parameter and anticipate value (\ref{eq:mucdtl}) of the chemical potential, which allows us to restrict the theory to zero order in $\exp(-\beta E_{\rm gap})$. Unlike the three-dimensional unitary limit, the density $\rho$ and the Fermi wave number $k_{\rm F}$ are now fixed by the choice (\ref{eq:valeta}); after dimensionless conversion of the wave numbers by $q_{\rm dim}$, the lengths by $q_{\rm dim}^{-1}$, and the energies by $\eps_{\rm dim}$, we must therefore numerically solve the equation 
\begin{equation}
\rho q_{\rm dim}^{-2} = \check{\rho}(\check{\mu}=\mu/\eps_{\rm dim},\check{T}=k_{\rm B}T/\eps_{\rm dim})
\end{equation}
 where the diacritical mark
 (a Czech accent) marks the dimensionless quantity, $\check{\rho}(\check{\mu},\check{T})$ is the reduced equation of state function, and $\check{\mu}$ is the only unknown (the other quantities are imposed by (\ref{eq:valeta},\ref{eq:valtemp})). The equation of state $\rho^{\rm sta}(\mu,T)$ of the static BCS theory is known analytically (\ref{eq:rel2d}), which immediately gives 
\begin{equation}
\label{eq:mucsta}
\check{\mu}^{\rm sta}=\frac{1}{2}[\exp(2\eta)-1]\simeq 3.1945
\end{equation}
 For $\rho^{\rm ltd}(\mu,T)$ (\ref{eq:bon}) of the linearized time-dependent BCS theory, we must numerically perform calculations for $\delta\tilde{g}_{\rm coll}(\qq)$ and its integral over $\qq$ for several values of the chemical potential; linear extrapolations and quadratic interpolations ultimately give 
\begin{equation}
\label{eq:mucdtl}
\check{\mu}^{\rm ltd}\simeq 2.0545
\end{equation}
which is very different and leads in expression (\ref{eq:c2d}), incidentally, to $mc^2/\eps_{\rm dim}\simeq 2.5545$ and therefore to $k_{\rm B}T/mc^2\simeq 0.130\ll 1$ (the effect of thermal phonons is very weak, $q_{\rm dim}^{-2}\delta\rho_\phi\sim [-\zeta(3)/2\pi](k_{\rm B}T/mc^2)^3mc^2/\eps_{\rm dim}\simeq -10^{-3}$ according to note \ref{note:drhophi}, and the experimental points are practically at zero temperature). Making the naive and incorrect choice (\ref{eq:mucsta}) in the grand canonical linearized time-dependent BCS calculation of the pair distribution functions leads to the dashed curves on the right half of Figure~\ref{fig:compar}, while the informed choice (\ref{eq:mucdtl}) leads to the solid curves. For function $g_{\uparrow\uparrow}(\rr,\rr')$, the dashed curve is disqualified as in 3D (it takes negative values), but the solid curve—our true prediction—is in fairly good agreement with the experiment. For function $g_{\uparrow\downarrow}(\rr,\rr')$, the trough of the dashed line has a depth (counted negatively) $\min_{|\rr-\rr'|} g_{\uparrow\downarrow}(\rr,\rr')/\rho_{\uparrow}\rho_{\downarrow}-1\simeq -0.0964$ closer to the experimental value $\simeq-0.13$ than that of the solid line $\simeq-0.0592\!$, but a position $k_{\rm F}|\rr-\rr'|_0\simeq 1.53$ further from the experimental value $\simeq 2.2$ than that of the solid line $\simeq 1.738$.\footnote{As the experimental data are quite noisy on $g_{\uparrow\downarrow}(\rr,\rr')$, we replace them with a cubic fit on the interval $1.4\leq k_{\rm F}|\rr-\rr'|\leq 3.5$ in order to determine the position and value of the minimum.}${}^{,}$\footnote{We do not reproduce the grand canonization game (\ref{eq:jeugc}) here because we do not know the chemical potential of the Fermi gas well: it has not been measured in reference \cite{TarikHAL}, and the values that could be drawn from theoretical or experimental literature for an interaction strength $\eta=1$, that is $\ln(k_{\rm F}a_{\rm 2D})\simeq 1.116$, are highly dispersed: $\mu/E_{\rm F}\simeq 0.43$ according to Figure~3a of \cite{VanLoon} (Gaussian fluctuation approximation), $\mu/E_{\rm F}\simeq 0.35$ by interpolation of the results in Table I in \cite{Giorgini2D} (fixed-node quantum Monte Carlo), $\mu/E_{\rm F}\simeq 0.29$ according to the interpolation polynomials of \cite{Zhang} (auxiliary-field quantum Monte Carlo on a lattice), $\mu/E_{\rm F}=0.14(2)$ in the data in Figure~1 of \cite{Enss} (cold atom experiment); for comparison, our equation (\ref{eq:mucdtl}) corresponds to $\mu/E_{\rm F}\simeq 0.556$.}

\subsection{Behavior of $g^{\rm ltd}_{\sigma\sigma'}(\rr,\rr')$ at short distances}\label{sec3_3}

We recall that the deviations from the asymptotic values $\delta g^{\rm ltd}_{\sigma\sigma'}(\rr,\rr')$ in the linearized time-dependent BCS theory differ from those $\delta g^{\rm sta}_{\sigma\sigma'}(\rr,\rr')$ in the static BCS theory by a common correction $\delta g_{\rm coll}(\rr,\rr')$ independent of the internal states $\sigma, \sigma'$, see equation (\ref{eq:sepa}). We first verify in section \ref{sec3_3_1} that $\delta g_{\rm coll}(\rr,\rr')$ has a finite limit when $|\rr-\rr'|\to 0^+$, which had been assumed in the derivation of the modified equation of state (\ref{eq:bon}) and in the elegant expression (\ref{eq:belle3}) of the common value of the differences $g^{\rm ltd}_{\sigma\sigma'}(\rr,\rr')-g^{\rm sta}_{\sigma\sigma'}(\rr,\rr')$. It follows that, at short distances, the linearized time-dependent BCS theory does not change the divergent behavior of $g_{\uparrow\downarrow}(\rr,\rr')$ predicted by static BCS at the same chemical potential, not even by an additive constant, see section \ref{sec_3_3_2}. On the other hand,
 as shown in section \ref{sec_3_3_3}, it significantly modifies the departure coefficient of $g_{\uparrow\uparrow}(\rr,\rr')$ in $|\rr-\rr'|$ in 3D and in $|\rr-\rr'|^2\ln|\rr-\rr'|$ in 2D predicted by static BCS, and reveals new but sub-dominant contributions proportional to $|\rr-\rr'|^2\ln|\rr-\rr'|$ in 3D and to $|\rr-\rr'|^2\ln(-\ln|\rr-\rr'|)$ in 2D. For simplicity, in this section \ref{sec3_3} we take the zero order in $\exp(-\beta E_{\rm gap})$.

\subsubsection{Study of $\delta\tilde{g}_{\rm coll}(\qq)$ for $q\to +\infty$}\label{sec3_3_1}

To determine the short-range behavior of a function defined by its Fourier transform, we need to know the behavior of the Fourier transform at large wave numbers $q$. 

With this preliminary remark made, let us return, in our linearized time-dependent grand canonical BCS theory, to the decomposition (\ref{eq:sepa}) of the pair distribution functions $g_{\sigma\sigma'}^{\rm ltd}(\rr,\rr')$, more precisely the deviations $\delta g_{\sigma\sigma'}^{\rm ltd}(\rr,\rr')$ from the asymptotic values $\rho^{\rm ltd}_\sigma \rho^{\rm ltd}_{\sigma'}$, into the sum of the deviations $\delta g_{\sigma\sigma'}^{\rm sta}(\rr,\rr')$ in the static BCS theory of the same chemical potential $\mu$ [but not the same density $\rho=2\rho_\sigma$, see equation (\ref{eq:bon})] and a single collective contribution $\delta g_{\rm coll}(\rr,\rr')$. In what follows, we will work at order zero in $\exp(-\beta E_{\rm gap})$ and neglect all Fermi-Dirac occupation numbers $f_\kk$ in the static BCS theory and in the functions $\Sigma_{ij}$ (\ref{eq:s11}--\ref{eq:s23}) of the linearized time-dependent BCS theory.

The deviations $\delta g_{\sigma\sigma'}^{\rm sta}(\rr,\rr')$ can be expressed, thanks to Wick's theorem, in terms of the anomalous and normal averages as in (\ref{eq:wickhb},\ref{eq:wickhh}), averages whose Fourier transforms are known explicitly, see equations (\ref{eq:moya},\ref{eq:moyn}), with a behavior at large wave numbers $k$ trivial to obtain. Their study at short distances will be simple to do, especially in 2D where we know how to write them in terms of Bessel functions (\ref{eq:moya2d},\ref{eq:moyn2d}), which even avoids the detour via Fourier space \cite{Romero2D}.

On the other hand, the Fourier transform $\delta\tilde{g}_{\rm coll}(\qq)$ of the collective part is not at all simple, since its expression (\ref{eq:exp2}) useful here (in practice the right-hand side of the second equality) is a contour integral over the complex energy $z$ of a rather daunting collective susceptibility $\tilde{\chi}_{\rm coll}(\qq,z)$
 (\ref{eq:defcoll}). Let us immediately give the result of the asymptotic expansions in Appendix \ref{ann:grandq}: 
\begin{equation}
\label{eq:collasymp}
\boxed{
\delta\tilde{g}_{\rm coll}(\qq)\underset{q\to +\infty}{=}\left\{\begin{array}{ll}
\displaystyle \frac{C_{\rm coll}}{q^4}+\frac{D_{\rm coll}}{q^5}+O\left(\frac{1}{q^6}\right) & \quad\mbox{in}\ \ d=3\\
&\\
\displaystyle\frac{C_{\rm coll}}{q^4}\frac{\ln(A_{\rm coll} q)}{\ln(B_{\rm coll} q)} + O\left(\frac{\ln^{\alpha}q}{q^6}\right) & \quad\mbox{in}\ \ d=2
\end{array}\right.
}
\end{equation}
 with the following coefficients: 
\begin{align}
\label{eq:coefcoll3d}
d&=3\ : & C_{\rm coll}&=-\frac{4\pi}{3\sqrt{3}} \frac{m^2\Delta^2}{\hbar^4}\rho^{\rm sta} &  &;\quad & D_{\rm coll}&=-2\left(\frac{4\pi}{9}-\frac{2\sqrt{3}}{3}\right) \frac{m^2\Delta^2}{\hbar^4} a_{\rm 3D}^{-1} \rho^{\rm sta} + \left(\frac{m\Delta}{\hbar^2}\right)^4\mathcal{D} \\
\label{eq:coefcoll2d}
d&=2\ : & C_{\rm coll}&=-\frac{2 m^2 \Delta^2}{\hbar^4}\rho^{\rm sta} &  &;\quad &  A_{\rm coll}&=\frac{2}{\sqrt{3}q_{\rm dim}}\neq B_{\rm coll}=\frac{\sqrt{3}}{2 q_{\rm dim}}
\end{align}
 Here, $\rho^{\rm sta}$ and $\Delta$ are the density (\ref{eq:etat}) and the order parameter (\ref{eq:delta}) in the static BCS theory at chemical potential $\mu$, $a_{\rm 3D}$ is the three-dimensional scattering length (\ref{eq:conva}) and $q_{\rm dim}$ is the two-dimensional dimer wave number (\ref{eq:edim2d}). We give the constant $\mathcal{D}$ of equation (\ref{eq:coefcoll3d}) in numerical form, 
\begin{equation}
\label{eq:drondenum}
\mathcal{D}=-0.549\, 686\ldots
\end{equation}
and in appendix \ref{ann:grandq} in integral form (\ref{eq:dronde}) ; the integrals involved could be calculated analytically, but at a cost that seemed excessive to us. In any case, the point to remember from the results (\ref{eq:collasymp}) is that they establish the existence and expression of the limit 
\begin{equation}
\lim_{|\rr-\rr'|\to 0^+} \delta g_{\rm coll}(\rr,\rr')= \int_{\mathbb{R}^d}\frac{\dd^d q}{(2\pi)^d} \delta \tilde{g}_{\rm coll}(\qq) < \infty
\end{equation}
 in section \ref{sec2_5} and thus give meaning to the modified equation of state (\ref{eq:bon}).

The laborious calculations in Appendix \ref{ann:grandq} are not without interest, as the results (\ref{eq:collasymp}) are quite surprising at first glance. Let us try to find them qualitatively using a simple argument of power counting $q^n$. In the limit $q\to +\infty$, the eigenenergy $\hbar\omega_\qq$ of the phonon mode (if it exists at large $q$, see note \ref{note:dom}) and the edge $E_{\rm bord}(\qq)$ (\ref{eq:ebord}) of the broken pair continuum diverge as $q^2$; we can therefore take a complex energy $z$ of order $q^2$ in the integration contour $C_{\rm d}$ of equation (\ref{eq:exp2}) [this also allows us to neglect the exponentially small thermal contribution $\sim \exp(-\beta z)$, which explains why the coefficients (\ref{eq:coefcoll3d},\ref{eq:coefcoll2d}) do not depend on temperature]. We therefore expect the typical wave number $k$ in integrals (\ref{eq:s11}--\ref{eq:s23}) giving the $\Sigma_{ij}(\qq,z)$ to be of order $q$, and the energies $\eps_\pm$ and $\xi_\pm$ of order $q^2$. Hence the power laws when $q\to +\infty$ at fixed $Z=z/E_\qq$ [with $E_\qq=\hbar^2 q^2/2m$ as in (\ref{eq:rac})]:
 
\begin{equation}
\label{eq:loipuis}
\Sigma_{11}(\qq,z),\Sigma_{12}(\qq,z),\Sigma_{22}(\qq,z) \approx q^{d-2} \quad ; \quad \Sigma_{13}(\qq,z),\Sigma_{23}(\qq,z) \approx q^{d-4} \quad ; \quad \tilde{\chi}_{\rm coll}(\qq,z)\approx q^{d-6}
\end{equation}
 and ultimately 
\begin{equation}
\label{eq:exprdgt}
\delta\tilde{g}_{\rm coll}(\qq)\underset{q\to +\infty}{\simeq} \int_{C_{\rm d}} \frac{\mathrm{d} z}{(-2\mathrm{i}\pi)} \tilde{\chi}_{\rm coll}(\qq,z) \stackrel{?}{\approx} q^{d-4}
\end{equation}
 since $\mathrm{d} z=E_{{\mathbf{q}}}\mathrm{d} Z$. This is at odds with the results (\ref{eq:collasymp}), fortunately. Otherwise, the integral of $\delta\tilde{g}_{\rm coll}(\qq)$ would have a logarithmic ultraviolet divergence in 2D, quadratic in 3D, and the modified equation of state (\ref{eq:bon}) would make no sense!

Let's explain this failure of power counting in dimension $d=2$. The complete calculation in Appendix \ref{ann:grandq} confirms, to the dominant order, that $\tilde{\chi}_{\rm coll}(\qq,z)$ is indeed $\approx q^{-4}$ (to within factors $\ln q$)\footnote{\label{note:ln} The logarithm $\ln q$ results from the fact that $\int_{k<\eta q}[\dd^2k/(2\pi)^2](1/\eps_\kk)=(m/\pi\hbar^2)\ln(\eta q/q_{\rm dim})+O(1/q^2)$ when $q\to +\infty$ at fixed $\eta\ll 1$ [we used (\ref{eq:rel2d})]; this integral is in fact a subpart of $\Sigma_{ij}$, it occurs in their restriction to a neighborhood of $\kk_-=\mathbf{0}$ with radius $\ll |\kk_+-\kk_-|=q$ (on this neighborhood, we can take $k_+\simeq q$).}, but each contribution—considered as a function of $z$—falls into one of the following two cases: (i) it has no singularity (pole or branch cut) on the positive real half-axis, or (ii) it has singularities, but only on the positive real half-axis (in particular, not on the negative real half-axis). In the first case, the loop $C_{\rm d}$ surrounds a domain of analyticity of the integrand, and the contour integral $\int_{C_{\rm d}} \mathrm{d} z$ is zero according to Cauchy's theorem. In the second case, one can close the loop $C_{\rm d}$ with a large circle at infinity, as shown in the diagram below, which has the same notation as Figure~\ref{fig:contour}: 
\begin{equation}
\includegraphics[width=3cm,clip=]{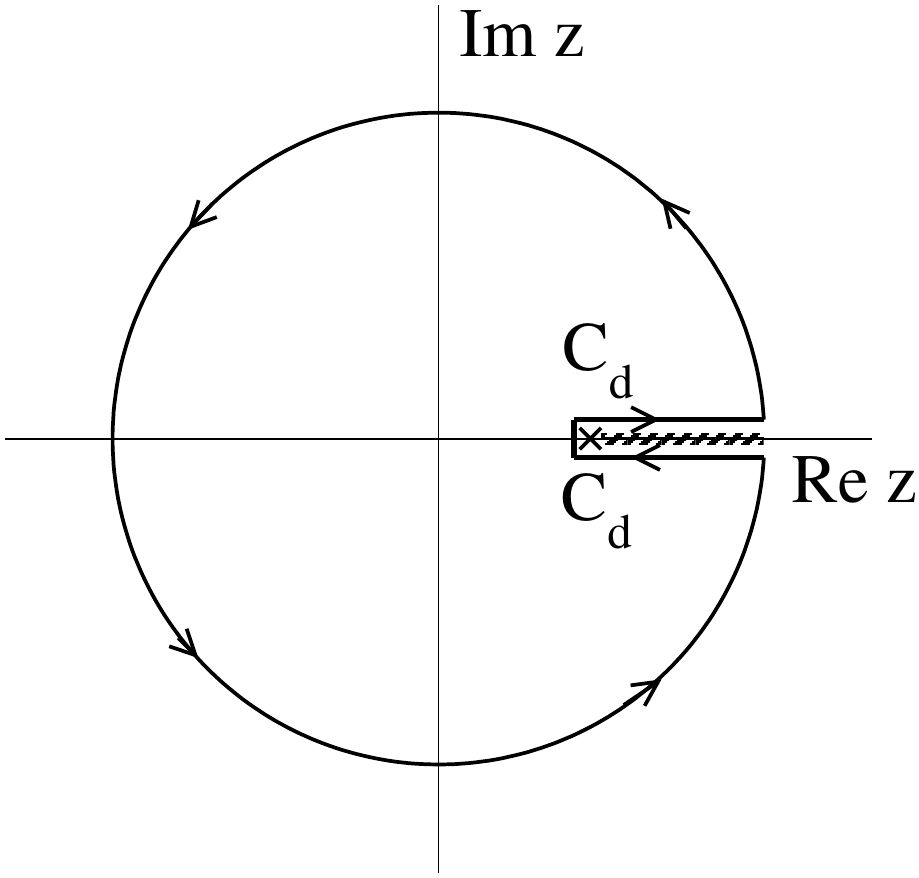}
\label{eq:schema}
\end{equation}
 The resulting contour no longer surrounds any singularities, and the integral is also zero there.\footnote{In this reasoning, we carefully check that the integrand is asymptotically $O(1/|z|^\alpha)$ with $\alpha>1$, so that the curvilinear integral on the great circle at infinity is zero.} We must therefore move to the next order in $q^{-1}$ in the susceptibility; since the expansion only involves even powers in 2D, we find in appendix \ref{ann:grandq} that 
\begin{equation}
\label{eq:devchi2d}
\tilde{\chi}_{\rm coll}(\qq,z=Z E_\qq) \stackrel{Z\ \mbox{\scriptsize fixed}}{\underset{q\to +\infty}{=}} \tilde{\chi}_{\rm coll}^{(-4)}(Z,\ln q) q^{-4} + \tilde{\chi}_{\rm coll}^{(-6)}(Z,\ln q) q^{-6} + O(q^{-8}) \quad \quad \quad (d=2)
\end{equation}
 where the coefficients depend weakly on $q$ through its logarithm $\ln q$. The new coefficient $\tilde{\chi}_{\rm coll}^{(-6)}(Z)$ has singularities on both $\mathbb{R}^+$ and $\mathbb{R}^-$, in practice a branch cut plus a pole on $\mathbb{R}^-$ and a pole on $\mathbb{R}^+$ or the opposite depending on the term considered, and its integral on the loop $C_{\rm d}$ is no longer zero. This leads to equation (\ref{eq:collasymp}) for $d=2$.

The same considerations extend to dimension $d=3$, except that (i) there is no longer $\ln q$ but the asymptotic expansion now has both even and odd powers of $q^{-1}$,\footnote{\label{note:pimp} The reasoning in note \ref{note:ln} gives $\int_{k<\eta q}[\dd^3k/(2\pi)^3](1/\eps_\kk)=(m/2\pi\hbar^2)[-a_{\rm 3D}^{-1}+(2/\pi)\eta q+O(1/\eta q)]$ this time. We quickly arrive at this result by writing $\int_{k<\eta q}[\dd^3k/(2\pi)^3](1/\eps_\kk)=\int_{k<\eta q}[\dd^3k/(2\pi)^3](1/E_\kk)+\int_{\mathbb{R}^3}[\dd^3k/(2\pi)^3](1/\eps_\kk-1/E_\kk)-\int_{k>\eta q}[\dd^3k/(2\pi)^3](1/\eps_\kk-1/E_\kk)$ and using equation (\ref{eq:delta}) on the order parameter. Outside the unitary limit, this is sufficient to explain the coexistence of even and odd powers of $q^{-1}$. } (ii) we start further away,
 the dominant term in $\tilde{\chi}_{\rm coll}(\qq,z)$ being of order $q^{-3}$ as power counting correctly indicates, but the first non-zero integral term on $C_{\rm d}$ being of order $q^{-6}$. As we are going in (\ref{eq:collasymp}) to the subdominant order, we must also know the term $q^{-7}$ in $\tilde{\chi}_{\rm coll}(\qq,z)$, 
\begin{equation}
\label{eq:devchi3d}
\tilde{\chi}_{\rm coll}(\qq,z=Z E_\qq) \stackrel{Z\ \mbox{\scriptsize fixed}}{\underset{q\to +\infty}{=}} \sum_{n=3}^{7} \tilde{\chi}_{\rm coll}^{(-n)}(Z) q^{-n} +O(q^{-8}) \quad \quad \quad (d=3)
\end{equation}
so in principle determine the $\Sigma_{ij}$ to the relative order $q^{-4}$; fortunately, a few clever considerations allow us to simplify the calculation, see Appendix \ref{ann:grandq}. In any case, we find that $\int_{C_{\rm d}}\mathrm{d} Z \tilde{\chi}_{\rm coll}^{(-n)}(Z) = 0\  \forall n\in\{3;4;5\}$ but $\neq 0\ \forall n\in\{6;7\}$, hence the asymptotic law (\ref{eq:collasymp}) in 3D (we still have $\mathrm{d} z=E_\mathbf{q} \mathrm{d} Z$, hence an additional factor of $q^2$ in $\delta\tilde{g}_{\rm coll}(\qq)$).

\subsubsection{Study of $g_{\uparrow\downarrow}^{\rm ltd}(\rr,\rr')$ for $|\rr-\rr'|\to 0^+$}\label{sec_3_3_2}

In our linearized time-dependent BCS theory, let us determine, up to an infinitesimal, the short-distance behavior of the distribution function of pairs of fermions with opposite spins. Since relation (\ref{eq:belle3}) taken for $\sigma=\uparrow$ and $\sigma'=\downarrow$ imposes 
\begin{equation}
g_{\uparrow\downarrow}^{\rm ltd}(\rr,\rr')-g_{\uparrow\downarrow}^{\rm sta}(\rr,\rr')=o(1)\quad \mbox{when}\quad |\rr-\rr'|\to 0^+
\end{equation}
we see that it suffices to know that of the static BCS pair distribution function $g_{\uparrow\downarrow}^{\rm sta}(\rr,\rr')$ with the same precision. This amounts, after applying Wick's theorem (\ref{eq:wickhb}), to calculating the anomalous mean $\langle\hat{\psi}_\downarrow(\rr')\hat{\psi}_\uparrow(\rr)\rangle_{\rm sta}$ to within $o(|\rr-\rr'|)$ in 3D and to within $o(|\rr-\rr'|^\alpha)\ (\alpha>0)$ in 2D, because the dominant contribution to the average diverges as $|\rr-\rr'|^{-1}$ in the first case and as $\ln|\rr-\rr'|$ in the second, and the average appears squared in (\ref{eq:wickhb}). We ultimately find that 
\begin{equation}
\label{eq:ghbcourte}
\boxed{
g^{\rm ltd}_{\uparrow\downarrow}(\rr,\rr') \underset{|\rr-\rr'|\to 0^+}{=} \left\{\begin{array}{ll}
\displaystyle \frac{1}{4}\left[\rho^{\rm sta}(\mu)\right]^2+\left(\frac{m\Delta}{4\pi\hbar^2}\right)^2\left[\left(\frac{1}{|\rr-\rr'|}-\frac{1}{a_{\rm 3D}}\right)^2-\frac{2 m \mu}{\hbar^2}\right]+o(1) & \quad\mbox{in}\ \ d=3\\
&\\
\displaystyle \frac{1}{4}\left[\rho^{\rm sta}(\mu)\right]^2+\left(\frac{m\Delta}{2\pi\hbar^2}\right)^2\ln^2\left(\frac{|\rr-\rr'|}{a_{\rm 2D}}\right)+o(1) & \quad\mbox{in}\ \ d=2
\end{array}
\right.
}
\end{equation}
 see justifications below.

These expressions (\ref{eq:ghbcourte}) recognize the square of the two-body wave function $\psi_0(\rr,\rr')$ of equation (\ref{eq:conva}): as might be expected, the pair distribution function $g_{\uparrow\downarrow}(\rr,\rr')$ reproduces at short distances the density structure of the stationary state of two fermions $\uparrow$ and $\downarrow$ with zero energy. Up to a factor, the prefactor of $|\psi_0(\rr,\rr')|^2$ is called the contact coefficient $\mathcal{C}$, or simply "contact" in Anglo-American literature,\footnote{The convention of \cite{WernerGenFer} is that $g_{\uparrow\downarrow}(\rr,\rr')\sim \mathcal{C}/(4\pi|\rr-\rr'|)^2$ in 3D and $g_{\uparrow\downarrow}(\rr,\rr')\sim [\mathcal{C}/(2\pi)^2]\ln^2(|\rr-\rr'|/a_{\rm 2D})$ in 2D, in which case we have the exact relation $\hbar^2L^d\mathcal{C}/m=4\pi\,\partial_{(-1/a_{\rm 3D})}\Omega(\mu,T) \ (d=3) \ \mbox{or}\ 2\pi\,\partial_{\ln a_{\rm 2D}}\Omega(\mu,T)\ (d=2)$ with $\Omega(\mu,T)$ the grand potential.} and has been the subject of extensive literature placing it at the heart of several exact relationships between different observables, see the foundational references \cite{Olsha} in 1D and \cite{Tan2} in 3D. We know that the value $\mathcal{C}^{\rm sta}\propto (m\Delta/\hbar^2)^2$ predicted by BCS theory in (\ref{eq:ghbcourte}) is only a lower bound on the exact value $\mathcal{C}$, see section V.K of reference \cite{relgenaug}, and that it becomes catastrophically bad in the BCS limit. Thus, in 3D where, in this limit, $k_{\rm F}\to 0^+$ at fixed scattering length $a_{\rm 3D}\!\!\in]-\infty,0[$, $\mathcal{C}^{\rm sta}(T=0)$ tends like $\Delta$ exponentially rapidly towards zero with $1/k_{\rm F}$, while the exact value vanishes algebraically, $\mathcal{C}(T=0)\sim (2\pi\rho a_{\rm 3D})^2$ \cite{Tarruell}. To remedy this shortcoming, reference \cite{Palestini} takes into account the quantum and thermal fluctuations of the pairing field—absent from static BCS—by means of well-chosen ladder diagrams in the many-body $T$ matrix;\footnote{We thank Pierbiagio Pieri for bringing this reference to our attention. } the resulting expression of $g_{\uparrow\downarrow}(\rr,\rr')$, an integral over wave vectors $\kk,\kk',\qq$ and a sum over associated Matsubara angular frequencies $\omega,\omega',\Omega$, see its equation (20), has in its integrand a collective part with the same denominator $\Sigma_{11}(\qq,\mathrm{i}\hbar\Omega)\Sigma_{22}(\qq,\mathrm{i}\hbar\Omega)-\Sigma_{12}^2(\qq,\mathrm{i}\hbar\Omega)$ as in our linearized time-dependent BCS theory; an important difference, however, is that its Fourier phase factor $\exp[\ii(\kk-\kk')\cdot(\rr-\rr')]$ involves the internal wave vectors $\kk,\kk'$ rather than the wave vector $\qq$ of the center of mass of the bound pairs $\uparrow\downarrow$ set in motion by sound waves, as $\exp[\mathrm{i}\mathbf{q}\cdot(\rr-\rr')]$ in the fluctuation-dissipation theorem (\ref{eq:expdg}). This results in a better description of $g_{\uparrow\downarrow}(\rr,\rr')$ at short distances but a poorer description of $\delta g_{\uparrow\downarrow}(\rr,\rr')$ at long distances; in particular, the prediction (20) of \cite{Palestini} violates the sum rule (\ref{eq:relsomhb}) in our section \ref{sec1}, whereas ours $g_{\uparrow\downarrow}^{\rm ltd}(\rr,\rr')$ complies perfectly with it, see Figure~14 of \cite{Palestini} and our appendix \ref{sec:rel_som}.

In 3D, to obtain (\ref{eq:ghbcourte}), it was sufficient to perform a few manipulations on the Fourier representation (\ref{eq:moya}) of the anomalous mean, written here at order zero in $\exp(-\beta E_{\rm gap})$ as we said, 
\begin{multline}
\langle\hat{\psi}_\downarrow(\rr')\hat{\psi}_\uparrow(\rr)\rangle_{\rm sta} = -\int_{\mathbb{R}^3} \frac{\dd^3k}{(2\pi)^3} \frac{\Delta}{2 E_\mathbf{k}} \eee^{\mathrm{i}\mathbf{k}\cdot(\rr-\rr')} - \int_{\mathbb{R}^3} \frac{\dd^3k}{(2\pi)^3} \left(\frac{\Delta}{2\eps_\mathbf{k}}-\frac{\Delta}{2E_\mathbf{k}}\right)  \\
- \int_{\mathbb{R}^3} \frac{\dd^3k}{(2\pi)^3} \left(\frac{\Delta}{2\eps_\mathbf{k}}-\frac{\Delta}{2E_\mathbf{k}}\right)\left(\eee^{\mathrm{i}\mathbf{k}\cdot(\rr-\rr')}-1\right)\underset{|\rr-\rr'|\to 0^+}{=} -\frac{m\Delta}{4\pi\hbar^2}\left(\frac{1}{|\rr-\rr'|}-\frac{1}{a_{\rm 3D}}\right)+ \frac{m^2\mu\Delta}{4\pi\hbar^{{4}}}|\rr-\rr'|[1+o(1)]
\end{multline}
 In the second equality, we remembered equation (\ref{eq:delta}) on the order parameter $\Delta$ and the inverse Fourier transform of $1/k^2$, $\int_{\mathbb{R}^3}[\dd^3k/(2\pi)^3](1/k^2)\exp[\mathrm{i}\mathbf{k}\cdot(\rr-\rr')]=1/4\pi|\rr-\rr'|$, then we were able to apply theorem (\ref{eq:theo3d})
 from Appendix \ref{ann:courtedis} to the third integral in the right-hand side of the first equality because $(2\eps_\kk)^{-1}-(2E_\kk)^{-1}{=}C/k^4+O(1/k^6)$ when $k\to +\infty$, with $C=2m^2\mu/\hbar^4$. A simple squaring gives (\ref{eq:ghbcourte}) in dimension $d=3$, which is indeed (qualitatively, with inexact coefficients) the general form expected for a spin-$1/2$ Fermi gas with contact interaction, see relation (3a) of table V in reference \cite{WernerGenFer}, also written to within $o(1)$.

In 2D, it was sufficient to transfer the known expansions of the Bessel functions $K_0(x)=-\ln(x\eee^\gamma/2)+O(x^2\ln x)$ ($\gamma=0.577\ldots$ is the Euler-Mascheroni constant) and $J_0(x)=1+O(x^2)$ into the analytical expression (\ref{eq:moya2d}) of the anomalous mean, then to relate $q_{\rm dim}$ to the two-dimensional scattering length as in (\ref{eq:edim2d}) to obtain (\ref{eq:ghbcourte}) in dimension $d=2$, which unfortunately causes disagreement (even qualitatively) with reference \cite{WernerGenFer}, see relation (3b) of its table V: this states in general that for contact interactions, we have 
\begin{equation}
g_{\uparrow\downarrow}(\rr,\rr')\stackrel{d=2}{\underset{|\rr-\rr'|\to 0^+}{=}}\frac{\mathcal{C}}{(2\pi)^2} \ln^2\left(\frac{|\rr-\rr'|}{a_{\rm 2D}}\right)+O(r^2\ln^2r)
\end{equation}
 where $\mathcal{C}$ is the exact contact coefficient, here for two-dimensional gas; there should therefore be no non-zero constant term in (\ref{eq:ghbcourte}) for $d=2$.

For the sake of completeness, let us ask ourselves to what extent we could have obtained the laws (\ref{eq:ghbcourte}) without using the anomalous mean but by working directly on the Fourier representation of $\delta g_{\uparrow\downarrow}^{\rm sta}(\rr,\rr')$ in the sense of (\ref{eq:tfdgcoll}). The large-$q$ expansion of the integral over the internal wave vector $\kk$ giving $\delta\tilde{g}_{\uparrow\downarrow}^{\rm sta}(\qq)$ in (\ref{eq:fourhb}) is not completely trivial: it is not enough, after the obvious scaling $\kk=q\KK$, to expand the integrand in powers of $q^{-1}$ at fixed $\KK$, because this reveals a singularity at $\KK=\mp\hat{\mathbf{q}}/2$ ($\hat{\mathbf{q}}=\qq/q$ is the direction of $\qq$) that is severe enough in 2D to be non-integrable (we have $\eps_\pm\sim E_\mathbf{q} (\mathbf{K}\pm\hat{\mathbf{q}}/2)^2$). The neighborhood of the singularities $k_\pm<\eta q,\ \eta\ll 1$ must in fact be treated separately, by expanding the integrand in powers of $q$ at fixed $\kk_\pm$, even in dimension $d=3$. We refer the reader to section 4.3.5.4 of reference \cite{livre}, and give the result directly here:\footnote{\label{note:metdev}The integral over the complementary neighborhood, $k_\pm>\eta q$ or rather $K_\pm\equiv|\mathbf{K}\pm\hat{\mathbf{q}}/2|>\eta$ since we have scaled $\kk$, is often difficult to perform a calculation for, as noted in reference \cite{livre}. In 3D, we encounter $I_{\rm 3D}=\int_0^{+\infty} \mathrm{d} K_-\int_0^{\theta_{\rm m}(K_-)}\sin\theta\mathrm{d}\theta/(1+2K_-\cos\theta+K_-^2)$ or $\theta_{\rm m}(K)=\arccos(\max(-1,-1/2K))$; the symbolic computation gives an explicit result $I_{\rm 3D}=[2\pi^2-6\ln^22+\ln 3\ln(64/27)-3 g_2(1/4)-6g_2(1/3)]/6$ in terms of the Bose function or dilogarithm $g_2(z)$; known relations on $g_2(z)$ have allowed us to show that we have more simply $I_{\rm 3D}=\pi^2/4$. In 2D, we encounter the difficult part $I_{\rm 2D}=\int_{1/2}^{+\infty}(\mathrm{d} K_-/K_-)\int_{-\theta_{\rm m}(K_-)}^{+\theta_{\rm m}(K_-)}\mathrm{d}\theta/(1+2K_-\cos\theta+K_-^2)$; by performing a calculation in the limit $\eta\to 0^+$ of the auxiliary integral $J=\int_{\mathbf{K}\cdot\hat{\mathbf{q}}>0}\dd^2K/\{[(\KK+\hat{\mathbf{q}}/2)^2+\eta^2][(\KK-\hat{\mathbf{q}}/2)^2+\eta^2]\}$ in two different ways, in variable $\KK$—integrating over the half-space $\mathbf{K}\cdot\hat{\mathbf{q}}>0$, which then more simply amounts to integrating over the entire space and dividing the result by $2$—then in variable $\KK_-$, we find that $J=-2\pi \ln\eta + o(1)$ then that $J=-\pi\ln 3+I_{\rm 2D}-2\pi\ln\eta+o(1)$, hence $I_{\rm 2D}=\pi\ln 3$.} 
\begin{equation}
\label{eq:stahbasymp}
\delta\tilde{g}^{\rm sta}_{\uparrow\downarrow}(\qq)\underset{q\to +\infty}{=}\left\{\begin{array}{ll}
\displaystyle\frac{m^2\Delta^2}{8\hbar^4 q}-\frac{m^2\Delta^2}{2\pi\hbar^4 a_{\rm 3D} q^2}+O\left(\frac{1}{q^4}\right) & \quad\mbox{in}\ \ d=3\\
&\\
\displaystyle\frac{m^2\Delta^2}{\pi\hbar^4 q^2}\ln\left(\frac{q}{q_{\rm dim}}\right) + O\left(\frac{{\ln^\alpha q}}{q^4}\right) & \quad\mbox{in}\ \ d=2\end{array}\right.
\end{equation}
 An inverse Fourier transform gives (\ref{eq:ghbcourte}) but only to within unknown additive constants, those coming from the neglected $O(1/q^4)$.\footnote{In 3D, we regularize the inverse Fourier transform (IFT) of $1/q$ by the usual factor $\exp(-\eta q)$ where $\eta\to 0^+$, to obtain $\mathrm{IFT}(1/q)=1/2\pi^2r^2$ by taking $\rr'=\mathbf{0}$ for short. In 2D, we introduce as an intermediate the wave function of the dimer $\phi_{\rm dim}(\rr)$, whose Fourier transform (FT) is known, $\tilde{\phi}_{\rm dim}(\kk)=2\pi^{1/2}q_{\rm dim}/(k^2+q_{\rm dim}^2)$. This allows us to perform an analytical calculation of the FT of the probability density of the dimer, $\tilde{\rho}_{\rm dim}(\qq)=\mathrm{FT}[\phi_{\rm dim}^2(\rr)]=\ln[1+Z+Y(Y+4+Z)/2]/Z$ with $Z=\sqrt{Y(Y+4)}$ and $Y=q^2/q_{\rm dim}^2$; we then verify that $\tilde{\rho}_{\rm dim}(\qq)=2\ln Y/Y+O(\ln Y/Y^2)=4\ln(q/q_{\rm dim})/(q^2/q_{\rm dim}^2)+O({\ln q}/q^4)$ when $q\to +\infty$. However, $\phi_{\rm dim}(\rr)=(q_{\rm dim}/\pi^{1/2})K_0(q_{\rm dim}r)=(q_{\rm dim}/\pi^{1/2})[-\ln(r/a_{\rm 2D})]+O(r^2\ln r)$ when $r\to 0^+$, which allows us to conclude.} It would be difficult to go beyond this.

\subsubsection{Study of $g_{\uparrow\uparrow}^{\rm ltd}(\rr,\rr')$ for $|\rr-\rr'|\to 0^+$}\label{sec_3_3_3}

Let us now determine, to second order, the short-range behavior of the distribution function of pairs of fermions of same spin $\uparrow$ in our linearized time-dependent BCS theory. To this end, let us specialize the relation (\ref{eq:belle3}) to the case $\sigma=\sigma'=\uparrow$, 
\begin{equation}
\label{eq:chaine2}
g_{\uparrow\uparrow}^{\rm ltd}(\rr,\rr')-g_{\uparrow\uparrow}^{\rm sta}(\rr,\rr')=\int_{\mathbb{R}^d} \frac{\dd^d q}{(2\pi)^d} \delta\tilde{g}_{\rm coll}(\qq)\left[\eee^{\mathrm{i}\mathbf{q}\cdot(\rr-\rr')}-1\right]
\end{equation}
 and write the normal average (\ref{eq:moyn}) appearing in the Wick expression (\ref{eq:wickhh}) of $g_{\uparrow\uparrow}^{\rm sta}(\rr,\rr')$ in a similar Fourier form (here at order zero in $\exp(-\beta E_{\rm gap})$, after multiplying the top and bottom by the conjugate quantity $\eps_\kk+\xi_\kk$ to make a constant numerator appear under the integral sign): 
\begin{multline}
\label{eq:moyn2}
g_{\uparrow\uparrow}^{\rm sta}(\rr,\rr')=-\left\{\rho^{\rm sta}+\int_{\mathbb{R}^d} \frac{\dd^dk}{(2\pi)^d} \frac{\Delta^2}{2\eps_\kk(\eps_\kk+\xi_\kk)}\left[\eee^{\mathrm{i}\mathbf{k}\cdot(\rr-\rr')}-1\right]\right\}\int_{\mathbb{R}^d} \frac{\dd^dk}{(2\pi)^d} \frac{\Delta^2}{2\eps_\kk(\eps_\kk+\xi_\kk)}\left[\eee^{\mathrm{i}\mathbf{k}\cdot(\rr-\rr')}-1\right]
\end{multline}
 To expand these expressions to order $|\rr-\rr'|^2$, we can use the same theorems (\ref{eq:theo2d}) and (\ref{eq:theo3d}) in Appendix \ref{ann:courtedis}—initially intended for (\ref{eq:chaine2})—since the integrand of (\ref{eq:moyn2}) also decreases as the inverse of the wave number to the power of four; as equation (\ref{eq:stahhasymp}) will show, the analogy with (\ref{eq:collasymp}) is complete if we set in any dimension 
\begin{equation}
\label{eq:csta}
C_{\rm sta}\stackrel{\forall d}{=}-\frac{m^2\Delta^2}{\hbar^4}\rho^{\rm sta}
\end{equation}
 So we will be brief.

In dimension $d=3$, we obtain the expansion 
\begin{equation}
\label{eq:ghh3ddep}
\boxed{g^{\rm ltd}_{\uparrow\uparrow}(\rr,\rr') \stackrel{d=3}{\underset{|\rr-\rr'|\to 0^+}{=}} -\frac{1}{8\pi} \left(C_{\rm sta}+C_{\rm coll}\right)|\rr-\rr'|+\frac{1}{6} |\rr-\rr'|^2\left[\mathcal{B}_{\rm sta}+\mathcal{B}_{\rm coll}+\frac{D_{\rm coll}}{2\pi^2} \ln(k_\Delta|\rr-\rr'|) +o(1)\right]}
\end{equation}
 with the wave number $k_\Delta=(m\Delta/\hbar^2)^{1/2}$ and the following quantities, coming from static BCS theory (index “sta”) or from the collective part of linearized time-dependent BCS theory (index "coll"): 
\begin{eqnarray}
\label{eq:brondesta}
\mathcal{B}_{\rm sta} &\stackrel{d=3}{=}& \frac{3 k_0 C_{\rm sta}}{8\pi} -6\left(\frac{C_{\rm sta}}{8\pi\rho^{\rm sta}}\right)^2 + \int_{\mathbb{R}^3} \frac{\dd^3k}{(2\pi)^3} k^2 \left[\frac{1}{2}\left(1-\frac{\xi_\mathbf{k}}{\eps_\mathbf{k}}\right)\rho^{\rm sta}+\frac{C_{\rm sta}}{(k^2+k_0^2)^2}\right] \\
\nonumber
\mathcal{B}_{\rm coll} &\stackrel{d=3}{=}& \frac{3 q_0 C_{\rm coll}}{8\pi} +\frac{D_{\rm coll}}{2\pi^2} \left(-\frac{13}{12}+\gamma+\ln\frac{q_0}{k_\Delta}\right) - \int_{\mathbb{R}^3} \frac{\dd^3q}{(2\pi)^3} q^2 \left[\delta\tilde{g}_{\rm coll}(\qq)-\frac{C_{\rm coll}}{(q^2+q_0^2)^2}-\frac{q D_{\rm coll}}{(q^2+q_0^2)^3}\right]\\
\label{eq:brondecol}
&&
\end{eqnarray}
 Recall that $\gamma=0.577\, 215\ldots$ is the Euler-Mascheroni constant and that the coefficients $C_{\rm coll}, D_{\rm coll}$ are given in equation (\ref{eq:coefcoll3d}). By construction, the quantities (\ref{eq:brondesta},\ref{eq:brondecol}) do not depend on the arbitrary wave numbers $k_0$ or $q_0>0$ (this can be verified by checking that their derivatives with respect to $k_0$ or $q_0$ are identically zero). At the unitary limit, we find numerically that $\mathcal{B}_{\rm sta}\simeq 5.07\cdot 10^{-3}k_{\Delta}^8$ and $\mathcal{B}_{\rm coll}\simeq -1.29\cdot 10^{-2} k_{\Delta}^8$, which is useful for the lower left panel of Figure~\ref{fig:compar} — the law (\ref{eq:ghh3ddep}) is shown in red. It should be emphasized that the linear departure of $g_{\uparrow\uparrow}(\rr,\rr')$ is not specific to the linearized time-dependent BCS theory; it preexists in the static BCS theory, albeit with a significantly smaller slope since $C_{\rm coll}/C_{\rm sta}=4\pi/3\sqrt{3}\simeq 2.42$ independently of the interaction strength.
 The linear departure is therefore significantly reinforced but not created by collective excitations. On the other hand, the contribution proportional to $\ln|\rr-\rr'|$ in the coefficient of the quadratic term is new, coming from the term $1/q^5$ (\ref{eq:collasymp}) in $\delta\tilde{g}_{\rm coll}(\qq)$ at large $q$, absent from $\delta\tilde{g}^{{\rm sta}}_{\uparrow\uparrow}(\qq)$ as we will see in equation (\ref{eq:stahhasymp}).

In dimension $d=2$, we obtain the expansion\\ 
\begin{equation}
\label{eq:ghh2ddep}
\fbox{
$
\begin{array}{r}
\displaystyle
g^{\rm ltd}_{\uparrow\uparrow}(\rr,\rr') \stackrel{d=2}{\underset{|\rr-\rr'|\to 0^+}{=}}\frac{1}{4}|\rr-\rr'|^2\left\{
{\mathcal{B}_{\rm sta}+\mathcal{B}_{\rm coll}}+\frac{C_{\rm sta}+C_{\rm coll}}{2\pi}\left[\ln(q_{\rm dim}|\rr-\rr'|/2)+\gamma-1\right]\right.\hspace{1.7cm} \\
\displaystyle\left. -\frac{C_{\rm coll}}{2\pi}\ln\left(\frac{A_{\rm coll}}{B_{\rm coll}}\right) \ln(-\ln(q_{\rm dim}|\rr-\rr'|)) +o(1)\right\}
\end{array}
$}
\end{equation}
 As before, the index of the coefficients indicates their origin, “sta” for the static BCS theory and “coll” for the collective part of linearized time-dependent BCS. We had to introduce the quantities 
\begin{eqnarray}
\label{eq:brondesta2d}
\mathcal{B}_{\rm sta} &\stackrel{d=2}{=}& \frac{C_{\rm sta}}{2\pi}\ln\left(\frac{k_0}{q_{\rm dim}}\right)+\int_{\mathbb{R}^2}\frac{\dd^2k}{(2\pi)^2}\left[\frac{k^2}{2}\left(1-\frac{\xi_\mathbf{k}}{\eps_\mathbf{k}}\right)\rho^{\rm sta}+\frac{C_{\rm sta}}{k^2+k_0^2}\right]=\frac{C_{\rm sta}}{2\pi}\left(\frac{1}{2}-\frac{\pi\rho^{\rm sta}}{2 q_{\rm dim}^2}\right)\\
\nonumber
\mathcal{B}_{\rm coll} &\stackrel{d=2}{=}& \frac{C_{\rm coll}}{2\pi}\left[\ln\left(\frac{q_0}{q_{\rm dim}}\right)+\ln\left(\frac{A_{\rm coll}}{B_{\rm coll}}\right)\ln(\ln(B_{\rm coll}q_0))\right] \\
&&\hspace{3cm} -\int_{\mathbb{R}^2}\frac{\dd^2q}{(2\pi)^2} \left[q^2\delta\tilde{g}_{\rm coll}(\qq)-\frac{C_{\rm coll}}{q^2+q_0^2}\frac{\ln[A_{\rm coll}^2(q^2+q_0^2)]}{\ln[B_{\rm coll}^2(q^2+q_0^2)]}\right]
\label{eq:brondecoll2d}
\end{eqnarray}
 involving the coefficients $A_{\rm coll}, B_{\rm coll}$ and $C_{\rm coll}$ of equation (\ref{eq:coefcoll2d}). Independent of wave number $k_0>0$ or wave number $q_0$ provided that we have $q_0>1/B_{{\rm coll}}$, these quantities (\ref{eq:brondesta2d},\ref{eq:brondecoll2d}) are the two-dimensional equivalents of (\ref{eq:brondesta},\ref{eq:brondecol}), hence the identical notation. We have given the very simple explicit expression of $\mathcal{B}_{\rm sta}$ (in 3D, it would show elliptic integrals).\footnote{\label{note:compli} Direct application of the results in Appendix \ref{ann:courtedis} to function ${\delta}\tilde{g}^{\rm sta}_{\uparrow\uparrow}(\qq)$ of equation (\ref{eq:fourhh}) written to order zero in $\exp(-\beta E_{\rm gap})$ would lead, given (\ref{eq:stahhasymp}), to $\mathcal{B}_{\rm sta}\stackrel{}{=}(C_{\rm sta}/2\pi) \ln(q_0/q_{\rm dim}) - \int_{\mathbb{R}^2}[\dd^2q/(2\pi)^2] [q^2{\delta}\tilde{g}_{\uparrow\uparrow}^{\rm sta}(\qq)-C_{\rm sta}/(q^2+{q}_0^2)]$, an expression independent of $q_0>0$ and in principle equivalent, but much more difficult to perform a calculation on (we have nevertheless succeeded in doing so); the same remark could be made in 3D.} For an interaction strength $\ln(k_{\rm F}^{\rm ltd}/q_{\rm dim})=1$, we find numerically that $\mathcal{B}_{\rm coll}\simeq -0.291 q_{\rm dim}^6$, which allowed us to represent (\ref{eq:ghh2ddep}) in red on the lower right panel of Figure~\ref{fig:compar}. There was already a logarithmic contribution to the coefficient of the quadratic term in static BCS; however, it is tripled in linearized time-dependent BCS, since $C_{\rm coll}=2 C_{\rm sta}$. On the contrary, the double logarithmic contribution is a novelty, due to the logarithmic factors in the tail in $1/q^4$ (\ref{eq:collasymp}) of $\delta\tilde{g}_{\rm coll}(\qq)$, absent from $\delta\tilde{g}^{{\rm sta}}_{\uparrow\uparrow}(\qq)$ as we shall see in equation (\ref{eq:stahhasymp}).

For the sake of completeness and symmetry between $g_{\uparrow\uparrow}^{\rm ltd}(\rr,\rr')$ and $g_{\uparrow\uparrow}^{\rm sta}(\rr,\rr')$, let us see to what extent we can expand $g_{\uparrow\uparrow}^{\rm sta}(\rr,\rr')$ at short distance without going through the normal mean (\ref{eq:moyn}) but by directly applying the theorems in Appendix \ref{ann:courtedis} to its integral representation, which automatically takes into account its zero limit at zero distance: 
\begin{equation}
\label{eq:repint}
g_{\uparrow\uparrow}^{\rm sta}(\rr,\rr')=\int_{\mathbb{R}^d} \frac{\dd^dq}{(2\pi)^d} \delta\tilde{g}^{\rm sta}_{\uparrow\uparrow}(\qq)\left[\eee^{\mathrm{i}\mathbf{q}\cdot(\rr-\rr')}-1\right]
\end{equation}
 where the Fourier transform $\delta\tilde{g}^{\rm sta}_{\uparrow\uparrow}(\qq)$ is given by equation (\ref{eq:fourhh}). By the same method that led to (\ref{eq:stahbasymp}), we find independently of the dimension that 
\begin{equation}
\label{eq:stahhasymp}
\delta\tilde{g}^{\rm sta}_{\uparrow\uparrow}(\qq)\stackrel{\forall d}{\underset{q\to +\infty}{=}}
\frac{C_{\rm sta}}{q^4}+O\left(\frac{1}{q^6}\right)
\end{equation}
 The introduction of the coefficient $C_{\rm sta}$ was anticipated by equation (\ref{eq:csta}). For the rest, the static BCS theory is poorer at large wave numbers than the linearized time-dependent BCS theory in the sense that, in the notations used in the asymptotic laws (\ref{eq:collasymp}) on $\delta\tilde{g}_{\rm coll}(\qq)$,
 law (\ref{eq:stahhasymp}) corresponds in dimension $d=3$ to a coefficient $D_{\rm sta}=0$ (no subdominant contribution in $1/q^5$) and in dimension $d=2$ to equal coefficients $A_{\rm sta}=B_{\rm sta}$ (no logarithmic factors in the dominant term $1/q^4$).\footnote{\label{note:plussimple} The situation is simpler than that of $\delta\tilde{g}^{\rm sta}_{\uparrow\downarrow}(\qq)$ (see note \ref{note:metdev} and the text that refers to it) because function $1-\xi_\kk/\eps_\kk=\Delta^2/\eps_\kk(\eps_\kk+\xi_\kk)$ is integrable over $\mathbb{R}^d$ even for $d=3$; the integral over $\kk$ defining $\delta\tilde{g}^{\rm sta}_{\uparrow\uparrow}(\qq)$ in (\ref{eq:fourhh}) is then dominated at large $q$ by neighborhoods of $\kk_\pm=\mathbf{0}$ with radius $\eta q$, $\eta\ll 1$ fixed: we can replace $\kk_\mp$ with ${\mp}\qq$ in the rest of the integrand and we obtain a law $q^{-4}$. By counting powers, we verify that the integration domain $k_\pm\approx q$ gives a subdominant contribution $O(q^{d-8})$ (we make the variable change $\kk=q\KK$ then we pass to the limit $q\to +\infty$ at $\KK$ fixed). This argument is sufficient to show that $D_{\rm sta}=0$ in 2D, but not in 3D ($3-8=-5$). For $d=3$, we must do the complete calculation, the cancellation of $D_{\rm sta}$ resulting from a subtle compensation between the proximity zone $k_+\ \mathrm{or}\ k_-<\eta q$ and the complementary asymptotic zone $k_+\ \mbox{and}\ k_->\eta q$. Mathematically, this stems from the a priori non-obvious fact that a certain integral expression $I(\eta)$ has a zero limit when the cutoff parameter $\eta<1/2$ separating the two zones tends towards $0^+$, namely $I(\eta)=-2\eta^{-1}+\int_\eta^{+\infty} (\mathrm{d} K_-/K_-^2) \int_0^{\theta_{\rm m}(K_-)} (\sin\theta\mathrm{d}\theta)/(K_-^2+2K_-\cos\theta+1)^2=-\eta/(1-\eta^2)-3\argth \eta$ where the angle $\theta_{\rm m}(K)$ is defined in note \ref{note:metdev} (the term $-2\eta^{-1}$ comes from the proximity zone, the rest from the asymptotic zone).} The application of the theorems in any case directly gives the contributions including a factor $C_{\rm sta}$ in (\ref{eq:ghh3ddep},\ref{eq:ghh2ddep}) but leads to expressions of $\mathcal{B}_{\rm sta}$ that are of little practical use because they are difficult to integrate, see note \ref{note:compli}.

\section{General conclusion}\label{sec4}
 We have studied at low temperature the distribution functions $g_{\sigma\sigma'}(\rr,\rr')$ of a two-dimensional or three-dimensional spin-$1/2$ Fermi gas with contact interaction, unpolarized and spatially homogeneous, in what we believe to be the minimum mean field theory acceptable for their calculation, the linearized time-dependent BCS theory, whose density-density response functions $\chi_{\sigma\sigma'}(\rr,t;\rr',t)$ derived from classical field dynamics lead to entirely quantum predictions on the $g_{\sigma\sigma'}(\rr,\rr')$ thanks to the fluctuation-dissipation theorem. On the other hand, the static BCS theory, which completely ignores collective excitations, in particular the Bogoliubov-Anderson acoustic branch—and therefore phonons—, is disqualified. The study is conducted in the grand canonical ensemble of chemical potential $\mu$ and temperature $T$, and the density of the gas at equilibrium is deduced from the equation of state $\rho(\mu,T)$.

The most glaring shortcoming of static BCS appears in the distribution function $g_{\uparrow\downarrow}(\rr,\rr')$ of pairs of fermions with opposite spins and its deviation $\delta g_{\uparrow\downarrow}(\rr,\rr')$ from its asymptotic value $[\rho(\mu,T)/2]^2$: this theory actually describes only the spatial structure of a bound pair $\uparrow\downarrow$ of the pair condensate, since $\delta g_{\uparrow\downarrow}^{\rm sta}(\rr,\rr')$ is proportional to $|\phi_0(\rr,\rr')|^2$ with $\phi_0(\rr,\rr')$ the wave function of a bound pair, see section \ref{sec1}; thus $\delta g_{\uparrow\downarrow}(\rr,\rr')$ would be a function that is positive everywhere, tending monotonically towards zero, with an exponential decay constant $\kappa_0$ with a non-zero limit at zero temperature, see note \ref{note:asymp}. The linearized time-dependent BCS theory shows that, under the effect of quantum or thermal fluctuations in the phonon modes, the reality is quite different: at low temperature, 
\begin{itemize}\item[(i)] The function $g_{\uparrow\downarrow}^{\rm ltd}(\rr,\rr')$ has an absolute minimum strictly below its asymptotic value, i.e., a trough, with a depth fairly close to that of the quantum Monte Carlo in reference \cite{GiorginiPRL} for the 3D unitary gas, and that in the experiment in reference \cite{TarikHAL} in 2D for an interaction strength $\ln(k_{\rm F}/q_{\rm dim})=1$, if we use the static BCS equation of state $\rho^{\rm sta}(\mu,T)$ to move from $\delta g_{\uparrow\downarrow}(\rr,\rr')$ to $g_{\uparrow\downarrow}(\rr,\rr')$; however, the agreement deteriorates catastrophically in 3D if we use
 — as should be done in principle — the equation of state modified under the effect of collective excitation modes $\rho^{\rm ltd}(\mu,T)$, see section \ref{sec3_2}; this results from a lack of self-consistency in the linearized time-dependent BCS theory.
\item[(ii)] The function $\delta g_{\uparrow\downarrow}^{\rm ltd}(\rr,\rr')$ tends towards zero at infinity by negative values, with an exponential decay constant $\kappa_0\sim 2\pi k_{\rm B}T/\hbar c$ where $c$ is the speed of sound; at zero temperature, $\delta g^{\rm ltd}_{\uparrow\downarrow}(\rr,\rr')$ tends towards zero as the inverse of the distance $|\rr-\rr'|$ to the power of $d+1$ where $d\in\{2;3\}$ is the dimension of space, again by negative values. 
\end{itemize}
In the limit $T\to 0^+$ and $|\rr-\rr'|\to +\infty$, section \ref{sec3_1} gives more precisely an original approximation for $\delta g^{\rm ltd}_{\uparrow\downarrow}(\rr,\rr')$ in the form of a series that is negative everywhere, connecting the quantum asymptotic regime $|\rr-\rr'| \ll \hbar c/2\pi k_{\rm B} T$, where the system appears to be in its ground state, to the thermal asymptotic regime $|\rr-\rr'|\gg \hbar c/2\pi k_{\rm B} T$. Apart from a renormalization of the density and speed of sound, the approximation is consistent with Landau and Khalatnikov's quantum hydrodynamics, an effective low-energy theory that is thought to be accurate at the dominant temperature order. It also holds for $\delta g_{\uparrow\uparrow}(\rr,\rr')$, which is equivalent to $\delta g_{\uparrow\downarrow}(\rr,\rr')$ at distances much larger than the size of a bound pair. On the other hand, at short distances $|\rr-\rr'|\to 0^{{+}}$, the behavior — diverging, of course, toward $+\infty$ — of $g^{\rm ltd}_{\uparrow\downarrow}(\rr,\rr')$ remains the same as that of $g^{\rm sta}_{\uparrow\downarrow}(\rr,\rr')$ at the same chemical potential, the difference being only $o(1)$, see section \ref{sec3_3}, which constitutes a limitation of the theory.

With regard to the function $g_{\uparrow\uparrow}(\rr,\rr')$ taken at short distances, the difference between static BCS and linearized time-dependent BCS with the same chemical potential is only quantitative at the dominant order in $|\rr-\rr'|$: in 3D, the coefficients of the linear departure $|\rr-\rr'|$ differ by a factor of $1+4\pi/3\sqrt{3}\simeq 3.4$; in 2D, those of the logarithmic-quadratic departure $|\rr-\rr'|^2\ln|\rr-\rr'|$ differ by a factor of $3$. The weakest coefficients are those of static BCS, and the ratios given are independent of the strength of the interactions. Qualitative differences appear at the subdominant order, with the linearized time-dependent BCS theory containing terms proportional to $|\rr-\rr'|^2\ln|\rr-\rr'|$ in 3D and to $|\rr-\rr'|^2\ln(-\ln|\rr-\rr'|)$ in 2D, which are completely absent from static BCS, see section \ref{sec3_3}. Here too, the modified equation of state must be used in the calculation of $g^{\rm ltd}_{\uparrow\uparrow}(\rr,\rr')$, the fluctuation-dissipation theorem giving only the deviations $\delta g_{\sigma\sigma'}(\rr,\rr')$. The change in the equation of state from $\rho^{\rm sta}(\mu,T)$ to $\rho^{\rm ltd}(\mu,T)$ is welcome this time: instead of degrading the agreement with theory or experiment, it avoids an absurdity, that of negative values of $g^{\rm ltd}_{\uparrow\uparrow}(\rr,\rr')$ over a neighborhood of zero distance! However, it does not resolve a clear disagreement with quantum Monte Carlo on the 3D unitary gas, according to which the departure of $g_{\uparrow\uparrow}(\rr,\rr')$ is quadratic rather than linear, see section \ref{sec3_2}.

This article is only a first foray into the subject of pair distribution functions of Fermi gases. It lays the foundations for the minimal acceptable BCS theory but leaves two important questions unanswered: (i) a comparison with a more elaborate theory based on path integrals over an auxiliary bosonic field (after summing over fermionic degrees of freedom), taken in the so-called Gaussian fluctuation approximation \cite{SdMTc,RanderiaBEC}, (ii) an explicit calculation at all orders in the thermal dissociation factor of bound pairs $\exp(-\beta E_{\rm gap})$, in particular in the vicinity of the superfluid transition, which may leave an observable imprint in the pair distribution functions, for example in 3D through an increasingly slow exponential approach to their asymptotic values when $T\to T_{\rm c}$, as is the case for Bose-Einstein condensation of ideal gas.\footnote{Mathematically, we do not exclude that, in the case of fermions, this exponential relaxation is oscillatory, i.e., of the form $\exp(\mathrm{i} k_0|\rr-\rr'|)$ plus complex conjugate with a real part $\re k_0$ that is not infinitesimal but an imaginary part $\kappa_0=\im k_0>0$ tending towards zero at the transition.} The article seemed long enough to us that the resolution of these questions could be postponed until later.

\paragraph{Note} During the proofreading, we became aware of the very recent preprint \cite{KPsdM}, which performs a numerical calculation of the pair distribution functions in 2D at zero temperature in the Gaussian fluctuations approximation.

\appendix

\section{Low-$q$ expansion of $\delta\tilde{g}_{\rm coll}(\qq)$ to order zero in $\exp(-\beta E_{\rm gap})$}\label{ann:deccont}

At order zero in $\exp(-\beta E_{\rm gap})$ (Fermi-Dirac occupation numbers are set to zero), equation (\ref{eq:exp3}) separates the Fourier transform $\delta\tilde{g}_{\rm coll}(\qq)$ of the collective part of the pair distribution functions into the contribution $\Phi(\qq)(\bar{n}_\qq+1/2)$ of the phonon mode and the contribution $\delta\tilde{g}^{C^0}_{\rm coll}(\qq)$ of the broken pair continuum by decomposing the integration path over complex energy $z$ into a circle $c_{\rm d}$ surrounding the phonon pole $z=\hbar\omega_\qq$ and a loop $L_{\rm d}$ surrounding the branch cut $z\in[E_{\rm bord}(\qq),+\infty[$, see Figure~\ref{fig:contour}c. Here, we determine the behavior of functions $\Phi(\qq)$ and $\delta\tilde{g}^{C^0}_{\rm coll}(\qq)$ at low wave numbers $q$;
 remember that, in this limit, the existence of the phonon mode is guaranteed, see note \ref{note:dom}.
\subsection{What the broken pair continuum gives}
 Let's start with the contribution of the broken pair continuum and begin with expression (\ref{eq:defcoll}) of the collective susceptibility $\tilde{\chi}_{\rm coll}(\qq,z)$. Since the loop $L_{\rm d}$ remains at a non-infinitesimal distance from the singularities on the real axis, we can expand the functions $\Sigma_{ij}(\qq,z)$ in powers of $q$ at fixed $z$, which only involves even powers (the integrands are smooth and even functions of the wave vector $\qq$): 
\begin{equation}
\label{eq:dsij}
\Sigma_{ij}(\qq,z)\stackrel{z\,\mbox{\scriptsize fixed}}{\underset{q\to 0^+}{=}}\Sigma_{ij}^{(0)}(z)+q^2\Sigma_{ij}^{(2)}(z)+O(q^4)
\end{equation}
 We note that the zero orders only involve two independent functions: 
\begin{equation}
\label{eq:dsig01}
\sigma_0(z)=\int_{\mathbb{R}^d}\frac{\dd^dk}{(2\pi)^d} \frac{1}{\eps_\kk(z^2-4\eps_\kk^2)}\quad ;\quad
\sigma_1(z)=\int_{\mathbb{R}^d}\frac{\dd^dk}{(2\pi)^d} \frac{\xi_\mathbf{k}}{\eps_\kk(z^2-4\eps_\kk^2)}
\end{equation}
 We have 
\begin{eqnarray}
\Sigma_{11}^{(0)}(z)&=&\frac{1}{2}z^2\sigma_0(z) \quad ;\quad
\Sigma_{22}^{(0)}(z)=\frac{1}{2}(z^2-4\Delta^2)\sigma_0(z) \quad ;\quad
\Sigma_{12}^{(0)}(z)=z\,\sigma_1(z) \\
\Sigma_{13}^{(0)}(z)&=&z\Delta\,\sigma_0(z) \quad \ \ ;\quad \Sigma_{23}^{(0)}(z)=2\Delta\,\sigma_1(z) 
\end{eqnarray}
 This leads to a nice simplification in the expansion of $\tilde{\chi}_{\rm coll}(\qq,z)$:\footnote{A nice mathematical consequence of (\ref{eq:dev}) is that the limits $q\to 0^+$ and $z\to 0$ do not commute for the function $\tilde{\chi}_{\rm coll}(\qq,z)$ (here still at order zero in $\exp(-\beta E_{\rm gap})$). From equation (\ref{eq:dev}) we first derive $\lim_{z\to 0}\lim_{q\to 0^+} \tilde{\chi}_{\rm coll}(\qq,z) = -\Delta ^2 \sigma_0(0)$. From the definition (\ref{eq:defcoll}) we then derive, for $\qq$ not equal to zero, the expression $\tilde{\chi}_{\rm coll}(\qq,0)=-\Sigma_{23}^2(\qq,0)\Sigma_{11}(\qq,0)/2\Sigma_{11}(\qq,0)\Sigma_{22}(\qq,0)=-\Sigma_{23}^2(\qq,0)/2\Sigma_{22}(\qq,0)$ knowing that $\Sigma_{11}(\qq,0)>0$ [see note \ref{note:sig11zero}], hence $\lim_{q\to 0^+}\lim_{z\to 0} \tilde{\chi}_{\rm coll}(\qq,z) = \sigma_1^2(0)/\sigma_0(0)$. We deduce, equivalently but more elegantly, that $\lim_{z\to 0}\lim_{q\to 0^+} \tilde{\chi}_{\uparrow\downarrow}^{\rm ltd}(\qq,z)=0$ while $\lim_{q\to 0^+}\lim_{z\to 0} \tilde{\chi}_{\uparrow\downarrow}^{\rm ltd}(\qq,z)=-\rho^{\rm sta}/4 mc^2$ anticipating the result (\ref{eq:deriv_mu}).} 
\begin{equation}
\label{eq:dev}
\tilde{\chi}_{\rm coll}(\qq,z)\stackrel{z\,\mbox{\scriptsize fixed}}{\underset{q\to 0^+}{=}} -\Delta^2\sigma_0(z) -\frac{q^2}{z^2}\left[(z^2-2\Delta^2)\Sigma_{11}^{(2)}(z)-z^2\Sigma_{22}^{(2)}(z)\right] + O(q^4)
\end{equation}
 using the second equality of (\ref{eq:lien}) to eliminate $\Sigma_{13}^{(2)}(z)$ in order two. Since the zero order is simply $-(1/2)[\Sigma_{11}^{(0)}(z)-\Sigma_{22}^{(0)}(z)]$, we deduce in particular that $\tilde{\chi}^{\rm ltd}_{\uparrow\downarrow}(\qq,z)$ has a quadratic start $\approx q^2$, if we remember equation (\ref{eq:chihb}) and the first equality of (\ref{eq:lien}). It remains to integrate (\ref{eq:dev}) over $L_{\rm d}$ as in (\ref{eq:exp3}). To do this, we contract the loop on the real axis and use relations $\Sigma_{ij}(\qq,z=\eps+\mathrm{i} 0^+)-\Sigma_{ij}(\qq,z=\eps-\mathrm{i} 0^+)=2\mathrm{i}\im[\Sigma_{ij}(\qq,z=\eps+\mathrm{i} 0^+)],\ \eps\in\mathbb{R}$ and $\im[1/(x+\mathrm{i} 0^+)]=-\pi\delta(x),\ x\in\mathbb{R}$ (after performing a partial fraction decomposition of the integrand of $\Sigma_{ij}(\qq,z)$ with respect to the variable $z$) for the calculation of the integral over $\eps\in[E_{\rm bord}({\mathbf{q}}),+\infty[$.\footnote{\label{note:pra}In practice, it suffices to know that $\im\{(\eps_++\eps_-)/[(\eps+\mathrm{i} 0^+)^2-(\eps_++\eps_-)^2]\}=(\pi/2)[\delta(\eps+\eps_++\eps_-)-\delta(\eps-\eps_+-\eps_-)].$} In the integrals over the remaining wave vector $\kk$, we expand into powers of $\qq$ under the sum sign and then integrate term by term to obtain the curvilinear integrals of $\Sigma_{ij}^{(n)}(z)$ useful for equation (\ref{eq:dev}): 
\begin{multline}
\label{eq:intis11ms22}
\int_{L_{\rm d}}\frac{\mathrm{d} z}{(-2\mathrm{i}\pi)} \left[\Sigma_{11}(\qq,z)-\Sigma_{22}(\qq,z)\right]= 
{\int_{\mathbb{R}^d}\frac{\dd^dk}{(2\pi)^d}\frac{\Delta^2}{2\eps_+\eps_-}\underset{q\to 0^+}{=}\int_{\mathbb{R}^d}\frac{\dd^dk}{(2\pi)^d} \frac{\Delta^2}{2\eps_\kk^2}}\\
+\Delta^2 E_\mathbf{q} \int_{\mathbb{R}^d}\frac{\dd^dk}{(2\pi)^d} \frac{2u^2E_\kk(\xi_\kk^2-\Delta^2)-\eps_\kk^2\xi_\mathbf{k}}{4\eps_\kk^6}+O(q^4)
\end{multline}
 \begin{multline}
\label{eq:intis11surz2}
\int_{L_{\rm d}}\frac{\mathrm{d} z}{(-2\mathrm{i}\pi)} \frac{\Sigma_{11}(\qq,z)}{z^2} = \int_{\mathbb{R}^d}\frac{\dd^dk}{(2\pi)^d} \frac{\eps_+\eps_-+\xi_+\xi_-+\Delta^2}{4\eps_+\eps_-(\eps_++\eps_-)^2} \\
\underset{q\to 0^+}{=}\int_{\mathbb{R}^d}\frac{\dd^dk}{(2\pi)^d} \frac{1}{8\eps_\kk^2} - E_\mathbf{q} \int_{\mathbb{R}^d}\frac{\dd^dk}{(2\pi)^d} \frac{\eps_\kk^2\xi_\kk+4\Delta^2E_\mathbf{k} u^2}{16\eps_\kk^6} + O(q^4)
\end{multline}
 where $u$ is the cosine of the angle between $\qq$ and $\kk$, and we recall that $E_\kk=\hbar^2k^2/2m$. After collecting the various contributions in (\ref{eq:dev}), we obtain 
\begin{equation}
\label{eq:repq}
\delta\tilde{g}_{\rm coll}^{C^0}(\qq)\underset{q\to 0^+}{=} -\int_{\mathbb{R}^d}\frac{\dd^dk}{(2\pi)^d} \frac{\Delta^2}{4\eps_\kk^2} +E_\mathbf{q}\int_{\mathbb{R}^d}\frac{\dd^dk}{(2\pi)^d} \frac{\Delta^2\xi_\kk(\eps_\kk^2-4u^2E_\mathbf{k}\xi_\kk)}{8\eps_\kk^6}+O(q^4)
\end{equation}
which answers the question. We cannot help but notice that the result is much simpler for the distribution of pairs of fermions with opposite spins, where we have 
\begin{equation}
\label{eq:compcont}
\delta\tilde{g}_{\uparrow\downarrow}^{C^0}(\qq)\underset{q\to 0^+}{=}-E_\mathbf{q}\int_{\mathbb{R}^d}\frac{\dd^dk}{(2\pi)^d}\frac{\Delta^2 u^2E_\mathbf{k}}{4\eps_\kk^4}+O(q^4)
\end{equation}

\subsection{What the Bogoliubov-Anderson phonon gives}

Let us now move on to the expansion of the function $\Phi(\qq)$ in the contribution of the Bogoliubov-Anderson phonon to $\delta\tilde{g}_{\rm coll}(\qq)$, starting from its expression (\ref{eq:nqPhi}). We reuse the expansions (\ref{eq:dsij}) of $\Sigma_{ij}(\qq,z)$ in powers of $q$, adding an expansion in powers of $z=\hbar\omega_\qq=\hbar c q+O(q^3)$ ($c$ is the speed of sound), which can be done since $z\ll E_{\rm bord}(\qq)$ is very far from any singularity of $\Sigma_{ij}$. Limiting ourselves to the orders useful here, we obtain: 
\begin{eqnarray}
\Sigma_{11}(\qq,z)&\stackrel{z=O(q)}{\underset{q\to 0^+}{=}}& \frac{1}{2}z^2\sigma_0(0)+q^2\Sigma_{11}^{(2)}(0)+O(q^4) \quad;\quad \Sigma_{22}(\qq,z)\stackrel{z=O(q)}{\underset{q\to 0^+}{=}}-2\Delta^2\sigma_0(0)+O(q^2) \\
\Sigma_{12}(\qq,z)&\stackrel{z=O(q)}{\underset{q\to 0^+}{=}}& z\sigma_1(0)+O(q^3)\hspace{2cm}\quad;\quad \Sigma_{13}(\qq,z)\stackrel{z=O(q)}{\underset{q\to 0^+}{=}} z\Delta\sigma_0(0)+O(q^3) \\
\Sigma_{23}(\qq,z)&\stackrel{z=O(q)}{\underset{q\to 0^+}{=}}& 2\Delta\sigma_1(0)+O(q^2) 
\end{eqnarray}
Equation (\ref{eq:valpro}) on the eigenenergy of the phonon mode is then reduced to 
\begin{equation}
\label{eq:vprodev}
0=\left[-|\Delta\sigma_0(0)+\mathrm{i}\sigma_1(0)|^2 z^2-2\Delta^2\sigma_0(0)\Sigma_{11}^{(2)}(0)q^2\right]\Big|_{z=\hbar c q}+O(q^4)
\end{equation}
which allows us to eliminate the coefficient $\Sigma_{11}^{(2)}(0)$ in the following;\footnote{\label{note:sig11zero} Equation (\ref{eq:vprodev}) for $z$ has a real solution only if $\Sigma_{11}^{(2)}(0)$ is positive. To verify this, still at order zero in $\exp(-\beta E_{\rm gap})$, we show that $\Sigma_{11}(\qq,0)=\int_{\mathbb{R}^d}[\dd^dk/(2\pi)^d] (\xi_+-\xi_-)^2/4\eps_+\eps_-(\eps_++\eps_-)$ using the symmetrized version of the counterterm [see note \ref{note:symct}] and then putting the integrand in the same denominator. However, we simply have $\xi_+-\xi_-=\hbar^2 k q \cos\theta/m$ where $\theta$ is the angle between $\kk$ and $\qq$, so that $\Sigma_{11}^{(2)}(0)=\int_{\mathbb{R}^d}[\dd^dk/(2\pi)^d] (\hbar^2 k\cos\theta/m)^2/8\eps_\kk^3>0$. The same method gives for $z$ non-zero $\Sigma_{11}(\qq,z)=\int_{\mathbb{R}^d}[\dd^dk/(2\pi)^d] (\eps_++\eps_-)[(\xi_+-\xi_-)^2-z^2]/4\eps_+\eps_-[(\eps_++\eps_-)^2-z^2]$, hence the identity (19) of \cite{CKS}.} we deduce in particular that 
\begin{equation}
\Sigma_{11}(\qq,z=\hbar\omega_\qq)\underset{q\to 0^+}{=} \frac{(\hbar c)^2\sigma_1^2(0)}{2\Delta^2|\sigma_0(0)|}q^2+O(q^4)
\end{equation}
 and that its sign $s_{11}$ is equal to $1$. We were able to replace the quantity $\sigma_0(0)$ by $-|\sigma_0(0)|$ since it is negative, as can be clearly seen from its definition (\ref{eq:dsig01}) ; on the other hand, the sign of the quantity $\sigma_1(0)$, and therefore of $\Sigma_{12}(\qq,\hbar\omega_\qq)$, is indeterminate, but it is also that of $\Sigma_{23}(\qq,\hbar\omega_\qq)$, so it factors out and disappears in the square in the numerator of $\Phi(\qq)$ in (\ref{eq:nqPhi}). The denominator of $\Phi(\qq)$ in (\ref{eq:nqPhi}) is obtained by simple differentiation with respect to $z$ of the expression in brackets in (\ref{eq:vprodev}). We ultimately arrive at the equivalent of the sought-after $\Phi(\qq)$: 
\begin{equation}
\label{eq:avantsimp}
\Phi(\qq)\underset{q\to 0^+}{=} \frac{\hbar c q}{|\sigma_0(0)|} |\Delta\sigma_0(0)+\mathrm{i}\sigma_1(0)|^2+O(q^3)
\end{equation}
 with, remember 
\begin{equation}
\sigma_0(0)=-\int_{\mathbb{R}^d}\frac{\dd^dk}{(2\pi)^d} \frac{1}{4\eps_\kk^3} \quad ; \quad 
\sigma_1(0)=-\int_{\mathbb{R}^d}\frac{\dd^dk}{(2\pi)^d} \frac{\xi_\mathbf{k}}{4\eps_\kk^3}
\end{equation}
 To obtain the much more elegant form (\ref{eq:dphi0}) of $\Phi'(0)$ used in section \ref{sec3_1}, simply take the derivative with respect to the chemical potential of equation (\ref{eq:delta}) on the order parameter and of equation (\ref{eq:etat}) on the density at fixed scattering length $a_{d{\rm D}}$, with $f_\mathbf{k}\equiv 0$ here, then recall the hydrodynamic relation (\ref{eq:hydro}) satisfied by the static BCS theory: 
\begin{equation}
\label{eq:deriv_mu}
\Delta\frac{\mathrm{d}}{\mathrm{d}\mu}\Delta = \frac{\sigma_1(0)}{\sigma_0(0)}\quad ; \quad \frac{\rho^{\rm sta}}{mc^2}=\frac{\mathrm{d}}{\mathrm{d}\mu} \rho^{\rm sta}=-4\left[\Delta^2\sigma_0(0)+\Delta\left(\frac{\mathrm{d}}{\mathrm{d}\mu}\Delta\right)\sigma_1(0)\right]=\frac{4}{|\sigma_0(0)|}|\Delta\sigma_0(0)+\mathrm{i}\sigma_1(0)|^2
\end{equation}
 Equation (\ref{eq:avantsimp}) is then simplified as announced: 
\begin{equation}
\label{eq:apressimp}
\Phi(\qq)\underset{q\to 0^+}{=} \frac{\hbar\rho^{\rm sta}}{4mc} q + O(q^3)
\end{equation}

\subsection{Applications: sum rules}\label{sec:rel_som}

In the first application, we verify that, in our linearized time-dependent BCS theory, the Bogoliubov-Anderson phonon saturates at the dominant order in $q$ the well-known sum rule on the density-density susceptibility \cite{Stringari_livre} 
\begin{equation}
\mathcal{R}(\qq)\equiv\int_{\mathbb{R}} \frac{\mathrm{d}\eps}{(-\pi)} \eps \im\left[\tilde{\chi}_{\rho\rho}(\qq,\eps+\mathrm{i} 0^+)\right] = \rho \frac{\hbar^2 q^2}{m}
\end{equation}
 (the relation would catastrophically not be verified by the static BCS theory at low $q$ \cite{Minguzzi}). From note \ref{note:plutot} and expressions (\ref{eq:chihb},\ref{eq:chihh}) and then equation (\ref{eq:lien}), we deduce that $\tilde{\chi}^{{\rm ltd}}_{\rho\rho}(\qq,z)=2\Sigma_{33}(\qq,z)+4\tilde{\chi}_{\rm coll}(\qq,z)=2(\Sigma_{11}-\Sigma_{22}+\Sigma_{44})(\qq,z)+4\tilde{\chi}_{\rm coll}(\qq,z)$. The function $z\mapsto \tilde{\chi}^{{\rm ltd}}_{\rho\rho}(\qq,z)$ therefore has a pole at $z=\hbar\omega_\qq$ with residue $2\Phi(\qq)$ and a pole at $z=-\hbar\omega_\qq$ with opposite residue (it is an even function), if we return to the definition of $\Phi(\qq)$ in the text preceding equation (\ref{eq:nqPhi}). As $\im[1/(\eps\mp\hbar\omega_\qq+\mathrm{i} 0^+)]=-\pi\delta(\eps\mp\hbar\omega_\qq)$, the contribution of the phonon to the sum rule is written as 
\begin{equation}
\mathcal{R}^{\phi}(\qq) = 4\Phi(\qq) \hbar\omega_\mathbf{q}\underset{q\to 0^+}{=}\rho^{\rm sta}\frac{\hbar^2 q^2}{m}+O(q^4)
\end{equation}
 It therefore remains to be shown that the contribution of the continuum is $O(q^4)$. Using the expansion (\ref{eq:dev}) and expression (\ref{eq:s44}) of $\Sigma_{44}(\qq,z)$, we find that $\Sigma_{44}^{(0)}(z){\equiv}0$ and that 
\begin{equation}
\frac{1}{4}\tilde{\chi}^{{\rm ltd}}_{\rho\rho}(\qq,z)\stackrel{z\,\mbox{\scriptsize fixed}}{\underset{q\to 0^{{+}}}{=}} \frac{2 q^2\Delta^2}{z^2}\Sigma_{11}^{(2)}(z)+\frac{q^2}{2}\left[-\Sigma^{(2)}_{11}(z)+\Sigma^{(2)}_{22}(z)+\Sigma^{(2)}_{44}(z)\right] + O(q^4)
\end{equation}
 Integrating over $\eps$ as in equations (\ref{eq:intis11ms22},\ref{eq:intis11surz2}), see note \ref{note:pra}, and noting that the part at $\eps<0$ doubles that at $\eps>0$, we arrive at 
\begin{multline}
\int_{\mathbb{R}\setminus [-E_{\rm bord}(\qq),E_{\rm bord}(\qq)]}\frac{\mathrm{d}\eps}{(-\pi)}\,{\eps}\,\im[\tilde{\chi}^{{\rm ltd}}_{\rho\rho}(\qq,\eps+\mathrm{i} 0^+)]\underset{q\to 0^+}{=} \int_{\mathbb{R}^d}\frac{\dd^dk}{(2\pi)^d}\left\{\frac{4\Delta^2(\eps_+\eps_-+\xi_+\xi_-+\Delta^2)}{(\eps_++\eps_-)\eps_+\eps_-}\right.\\
\left.+\frac{(\eps_++\eps_-)}{\eps_+\eps_-}\left[-(\eps_+\eps_-+\xi_+\xi_-+\Delta^2)+(\eps_+\eps_-+\xi_+\xi_+-\Delta^2)+(\eps_+\eps_--\xi_+\xi_+-\Delta^2)\right]\right\}+O(q^4)
\end{multline}
 This is zero to order $q^2$, even before integration over the angle and over the modulus of $\kk$, as shown by a Taylor expansion of the integrand. We therefore have 
\begin{equation}
\mathcal{R}^{C^0}(\qq)\underset{q\to 0^+}{=} O(q^4)
\end{equation}

 In a second application, we subject the various BCS theories to the test of the following exact sum rules on the deviations from the asymptotic values,\footnote{With the fermionic anticommutation relations on the field operators, we reduce it to showing that $\int_{\mathbb{R}^d} \dd^d r\, C_{\sigma\sigma'}(\rr,\rr')=k_{\rm B} T \partial_{\mu_{\sigma'}}\rho_\sigma$ where $C_{\sigma\sigma'}(\rr,\rr')=\langle\hat{\rho}_\sigma(\rr)\hat{\rho}_{\sigma'}(\rr')\rangle-\langle\hat{\rho}_\sigma(\rr)\rangle\langle\hat{\rho}_{\sigma'}(\rr')\rangle$ and $\hat{\rho}_\sigma(\rr)\equiv \hat{\psi}^\dagger_\sigma(\rr)\hat{\psi}_\sigma(\rr)$. To this end, one might be tempted to place oneself in the usual quantization box $[0,L]^d$ with periodic boundary conditions \cite{Palestini}. By integrating the correlation function over $\rr$ and $\rr'$ in the volume $\mathcal{V}$ of the box, one would then obtain $I_{\sigma\sigma'}(\mathcal{V})\equiv\mathcal{V}^{-1}\int_{\mathcal{V}}\dd^dr'\int_{\mathcal{V}}\dd^d r\, C_{\sigma\sigma'}(\rr,\rr')=\mathcal{V}^{-1}\mathrm{Cov}\,(\hat{N}_\sigma,\hat{N}_{\sigma'})$ where $\mathrm{Cov}$ is the covariance and $\hat{N}_\sigma$ the total number operator of spin-$\sigma$ fermions. By spatial homogeneity, we would also have $I_{\sigma\sigma'}(\mathcal{V})=\int_{\mathcal{V}} \dd^dr\, C_{\sigma\sigma'}(\rr,\rr')$ for any fixed $\rr'$. At the thermodynamic limit in the grand canonical ensemble, we would arrive at the expected result since $\mathrm{Cov}\,(\hat{N}_\sigma,\hat{N}_{\sigma'})=-k_{\rm B}T\partial_{\mu_\sigma}\partial_{\mu_{\sigma'}}\Omega$ and $\langle\hat{N}_\sigma\rangle=-\partial_{\mu_\sigma}\Omega$ with $\Omega=-k_{\rm B}T\ln \Xi$ the grand potential and $\Xi=\mathrm{Tr}\, \exp[-\beta(\hat{H}-\mu_\uparrow\hat{N}_\uparrow-\mu_\downarrow\hat{N}_\downarrow)]$ the grand canonical partition function. But in the canonical ensemble, we would have $I_{\sigma\sigma'}(\mathcal{V})\equiv 0$ because $N_\sigma$ are fixed. The proof is therefore clearly not valid. The reason for this is given in note 64 of section 8.7.2.2 of reference \cite{livre}: in order to obtain the universal result (\ref{eq:relsomdg}), the integral over $\rr$ must be performed {\sl after} taking the thermodynamic limit, i.e., for an infinite system, which we will now indicate with the symbol $\infty$. In fact, $C^{\infty}_{\sigma\sigma'}(\rr,\rr')\equiv\mathrm{thermo.}\,\mathrm{lim.}\, C_{\sigma\sigma'}(\rr,\rr')$ at $\rr,\rr'$ fixed does not depend on the statistical ensemble chosen. But we can salvage the previous simple reasoning by applying it to a finite subvolume $B$ of an infinite gas, for example a ball with radius $R$. We then have 
\begin{equation}
\label{eq:grandI}
I_{\sigma\sigma'}(B)\equiv \frac{1}{\mathcal{V}_B} \int_{B}\dd^d r'\int_{B}\dd^d r\,  C^\infty_{\sigma\sigma'}(\rr,\rr') = \frac{1}{\mathcal{V}_B}\, \mathrm{Cov}_\infty(\hat{N}_\sigma^B,\hat{N}_{\sigma'}^B)
\end{equation}
where $\hat{N}_\sigma^B$ is this time the operator number of spin-$\sigma$ fermions in the subvolume $B$ of measure $\mathcal{V}_B$. Two remarks in the limit $R\to +\infty$ separate us from the result. (i) At a fixed position $\rr'$, the integral $\int\dd^d r C^\infty_{\sigma\sigma'}(\rr,\rr')$ converges absolutely over a finite distance $\ell$ from $\rr'$, see the asymptotic behaviors of $\delta g_{\sigma\sigma'}(\rr,\rr')$ in section \ref{sec3_1}; in the right-hand side of the first equality in (\ref{eq:grandI}), we can therefore replace $\int_{B}\dd^d r$ with $\int_{\mathbb{R}^d}\dd^d r$, in which case the integral over $\rr$ no longer depends on $\rr'$, unless $\rr'$ is at a distance $\lesssim \ell$ from the edge; we deduce that $I_{\sigma\sigma'}(B)=\int_{\mathbb{R}^d}\dd^d r\, C^\infty_{\sigma\sigma'}(\rr,\rr') + o(1)$, and the edge effects related to the cut defining $B$ become negligible (at worst, i.e., at $T=0$, $o(1)$ for a ball of radius $R$ is $\sim 3\Phi'(0)\ln(R/\xi)/2\pi R$ in 3D and $\sim \Phi'(0)\ln(R/\xi)/\pi R$ in 2D where $\xi=\hbar/mc$, see section 8.7.2.2 of \cite{livre} based on expressions (46,47) of \cite{Varenna}). (ii) Furthermore, subvolume $B$ exhibits fluctuations in its particle number that are {\sl grand canonical} to within $o(R^{d/2})$ since, by construction, the subsystem in $B$ can exchange energy and particles with the rest of the gas, which is infinite and in equilibrium, thus perfectly playing the role of a grand canonical reservoir. This explains why the correct covariance to use in the simple reasoning is that of the grand canonical ensemble \cite{Bell}. The right-hand side of the second equality in (\ref{eq:grandI}) is thus reduced to $k_{\rm B}T \partial_{\mu_{\sigma'}}\rho_\sigma$ to within $o(1)$, and we find (\ref{eq:relsomdg}) when $R\to +\infty$. At $T=0$, this allows us to rewrite the estimates of $o(1)$ in the ball of radius $R$ in a more meaningful way, 
\begin{equation}
\frac{1}{\mathcal{V}_{\rm B}}\, \mathrm{Cov}_\infty(\hat{N}_\sigma^B,\hat{N}_{\sigma'}^B)|_{T=0} \stackrel{d=3}{\underset{R\to +\infty}{\sim}} \frac{3\rho\ln(mcR/\hbar)}{8\pi mc R/\hbar} \quad ; \quad \frac{1}{\mathcal{V}_{\rm B}}\, \mathrm{Cov}_\infty(\hat{N}_\sigma^B,\hat{N}_{\sigma'}^B)|_{T=0} \stackrel{d=2}{\underset{R\to +\infty}{\sim}} \frac{\rho\ln(mcR/\hbar)}{4\pi mc R/\hbar}
\end{equation}
 after replacing $\Phi'(0)$ with its exact value $\hbar\rho/4 mc$, see section \ref{sec3_1_3}. } 
\begin{equation}
\label{eq:relsomdg}
\int_{\mathbb{R}^d} \dd^dr\, \delta g_{\sigma\sigma'}(\rr,\rr')= -\rho_\sigma \delta_{\sigma\sigma'} + k_{\rm B} T \partial_{\mu_{\sigma'}}\rho_\sigma(\mu_\uparrow,\mu_\downarrow) \stackrel{\mu_\uparrow=\mu_\downarrow=\mu}{=} -\frac{1}{2}\rho\delta_{\sigma\sigma'} + \frac{1}{4} k_{\rm B} T \partial_\mu \rho
\end{equation}
 where $\rr'$ takes an arbitrary fixed value, for example $\rr'=\mathbf{0}$, and we have taken into account the fact, in the second equality, that $\rho(\mu)=2\rho_\sigma(\mu,\mu)$ in the unpolarized case considered in this article. The relation (\ref{eq:relsomhb}) in section \ref{sec1} is only a special case, written for $\sigma\neq\sigma'$ and at zero temperature. It is convenient here to reformulate (\ref{eq:relsomdg}) as a constraint on the limit of Fourier transforms with zero wave vector, knowing that 
\begin{equation}
\lim_{q\to 0^+} \delta\tilde{g}_{\sigma\sigma'}(\qq) = \int_{\mathbb{R}^d} \dd^dr\, \delta g_{\sigma\sigma'}(\rr,\rr') 
\end{equation}
 As before, we place ourselves at order zero in $\exp(-\beta E_{\rm gap})$, i.e., formally at zero temperature for static BCS theory. Then, given expressions (\ref{eq:fourhb},\ref{eq:fourhh}) of the Fourier transforms, this theory clearly violates the relation (\ref{eq:relsomdg}) in all possible spin configurations: 
\begin{eqnarray}
\label{eq:limstahb}
\lim_{q\to 0^+} \delta\tilde{g}^{\rm sta}_{\uparrow\downarrow}(\qq)&=&+\int_{\mathbb{R}^d} \frac{\dd^dk}{(2\pi)^d} \frac{\Delta^2}{4\eps_\kk^2} > 0 \\
\label{eq:limstahh}
\lim_{q\to 0^+} \delta\tilde{g}^{\rm sta}_{\uparrow\uparrow}(\qq)&=&-\int_{\mathbb{R}^d} \frac{\dd^dk}{(2\pi)^d} \frac{1}{4}\left(1-\frac{\xi_\mathbf{k}}{\eps_\mathbf{k}}\right)^2 > -\frac{1}{2}\rho^{\rm sta}=-\int_{\mathbb{R}^d} \frac{\dd^dk}{(2\pi)^d} \frac{1}{2} \left(1-\frac{\xi_\mathbf{k}}{\eps_\mathbf{k}}\right)
\end{eqnarray}
 the inequality in (\ref{eq:limstahh}) resulting from the identity 
\begin{equation}
\label{eq:evident}
\Delta^2 + (\eps_\kk-\xi_\kk)^2 = 2\eps_\mathbf{k} (\eps_\kk-\xi_\kk)
\end{equation}
 obvious given (\ref{eq:ampmodener}). This shortcoming of static BCS was already highlighted in $\delta g_{\rho\rho}(\rr,\rr')$ in reference \cite{Bell}. What about time-dependent linearized BCS? To see this, we perform the calculation of the limit of $\delta\tilde{g}_{\rm coll}(\qq)$ at zero wave vector, by simply combining results (\ref{eq:compcont}) and (\ref{eq:apressimp}) in expression (\ref{eq:exp3}) : 
\begin{equation}
\label{eq:appli2}
\lim_{q\to 0^+} \delta\tilde{g}_{\rm coll}(\qq)=\frac{k_{\rm B}T}{4mc^2}\rho^{\rm sta}-\int_{\mathbb{R}^d}\frac{\dd^dk}{(2\pi)^d} \frac{\Delta^2}{4\eps_\kk^2}
\end{equation}
 with 
\begin{equation}
\int_{\mathbb{R}^d}\frac{\dd^dk}{(2\pi)^d} \frac{\Delta^2}{4\eps_\kk^2} = \left\{\begin{array}{ll}
\displaystyle\frac{m\Delta}{8\pi\hbar^2}\left(\frac{\pi}{2}+\arctan\frac{\mu}{\Delta}\right) & \quad \mbox{if}\ \ d=2 \\
&\\
\displaystyle\frac{m^{3/2}\Delta}{8\pi\hbar^3} \left(\mu+\sqrt{\mu^2+\Delta^2}\right)^{1/2} & \quad \mbox{if}\ \ d=3
\end{array}
\right.
\end{equation}
 We note in passing that the limit is negative at sufficiently low temperatures. By adding its integral expression (\ref{eq:appli2}) to the limits (\ref{eq:limstahb},\ref{eq:limstahh}) of static theory, we obtain 
\begin{equation}
\lim_{q\to 0^+} \delta\tilde{g}^{\rm ltd}_{\sigma\sigma'}(\qq) = -\frac{1}{2}\rho^{\rm sta}\delta_{\sigma\sigma'} + \frac{1}{4} k_{\rm B} T \partial_\mu \rho^{\rm sta}
\end{equation}
remembering (\ref{eq:evident}) and the hydrodynamic expression (\ref{eq:hydro}) for the speed of sound, leading here to $\rho^{\rm sta}/mc^2=\partial_\mu\rho^{\rm sta}$. There is therefore perfect agreement with the sum rule (\ref{eq:relsomdg}) for the static BCS equation of state.

\section{Large-$q$ expansion of $\delta\tilde{g}_{\rm coll}(\qq)$ to zero order in $\exp(-\beta E_{\rm gap})$}\label{ann:grandq}

As explained in section \ref{sec3_3_1}, we are at order zero in $\exp(-\beta E_{\rm gap})$ (all Fermi-Dirac occupation numbers $f_\kk,f_\pm$ are set to zero) and we must expand the collective part $\tilde{\chi}_{\rm coll}(\qq,z)$ (\ref{eq:defcoll}) of the density-density susceptibilities in the limit $q\to +\infty$ in powers of $q^{-1}$ at a fixed reduced complex energy $Z=z/E_\qq$ (here $E_\qq=\hbar^2q^2/2m$ as in equation (\ref{eq:rac}) and we take $Z\in\mathbb{C}\setminus\mathbb{R}$ to avoid singularities—poles or branch cuts—on the real axis), going to a high enough order that the contribution corresponding to $\delta\tilde{g}_{\rm coll}(\qq)$—obtained by curvilinear integration on the path $C_{\rm d}$ as in (\ref{eq:exprdgt})—is non-zero.

To do this, we use the multiscale expansion method from section 4.3.5.4 of reference \cite{livre}, already outlined in our section \ref{sec3_3} [see notes \ref{note:ln}, \ref{note:pimp}, \ref{note:metdev}, and \ref{note:plussimple}], (i) by limiting the integration over $\kk$ to the half-space $\mathbf{k}\cdot\qq>0$ at the cost of a global factor of two, (ii) by expanding the integrand in powers of $q^{-1}$ at fixed $\mathbf{k}_-\equiv\kk-\qq/2$ in the ball $k_-<\eta q$, $\eta\ll 1$, (iii) by expanding the integral over the complement of the ball in powers of $q^{-1}$ at fixed $\KK_-=\kk_-/q$, (iv) by reexpanding the whole in powers of $q^{-1}$, in particular by writing $\int_0^{\eta q}\mathrm{d} q\, f(q)=\int_0^{+\infty}\mathrm{d} q\, f(q)-\int_{\eta q}^{+\infty}\mathrm{d} q\, f(q)$ and replacing $f(q)$ with its asymptotic expansion in the third integral, and finally (iv) by taking the limit $\eta\to 0^+$ in the coefficients of $q^{-n}$.\footnote{\label{note:symct} To implement this method, we replace the counterterm $1/2\eps_\kk$ with $1/4\eps_++1/4\eps_-$ in the definition (\ref{eq:s11},\ref{eq:s22}) of $\Sigma_{11}(\qq,z)$ and $\Sigma_{22}(\qq,z)$, so that the only singular points in the limit $q\to +\infty$ at fixed $\kk/q$ are $\kk=\pm\qq/2$ (and not $\kk=\mathbf{0}$). We are allowed to do this by the identity $\int_{\mathbb{R}^d}[\dd^dk/(2\pi)^d](2\eps_\kk^{-1}-\eps_+^{-1}-\eps_-^{-1})=0$, which we prove by introducing $\Phi(\rr)=\int_{\mathbb{R}^d}[\dd^dk/(2\pi)^d](2\eps_\kk^{-1}-\eps_+^{-1}-\eps_-^{-1})\exp(\mathrm{i}\mathbf{k}\cdot\rr)=2[1-\cos(\mathbf{q}\cdot\rr/2)]\phi(\rr)$ with $\phi(\rr)=\int_{\mathbb{R}^d}[\dd^dk/(2\pi)^d]\exp(\mathrm{i}\mathbf{k}\cdot\rr)/\eps_\kk$, then noting that $\Phi(\rr)\to 0$ when $r\to 0^{{+}}$ since $\phi(\rr)=O(\ln r)$ in 2D and $\phi(\rr)=O(1/r)$ in 3D. }

\subsection{Two simplifying tricks and a summary table}

The calculations to come are greatly simplified by two tricks. 

First, we can advantageously transform the expression (\ref{eq:defcoll}) of $\tilde{\chi}_{\rm coll}(\qq,z)$ by introducing half-sums and half-differences as follows: 
\begin{eqnarray}
\alpha(\qq,z)&\equiv&\frac{1}{2}\left[\Sigma_{23}(\qq,z)+\Sigma_{13}(\qq,z)\right] \quad ; \quad
D_{12}(\qq,z)\equiv\frac{1}{2}\left[\Sigma_{11}(\qq,z)-\Sigma_{22}(\qq,z)\right] \quad \quad \\
S_{12}(\qq,z)&\equiv&\frac{1}{2}\left[\Sigma_{11}(\qq,z)+\Sigma_{22}(\qq,z)\right]+\Sigma_{12}(\qq,z)
\end{eqnarray}
 We then have, taking into account the even or odd parity of the functions $\Sigma_{ij}(\qq,z)$ in the variable $z$: 
\begin{equation}
\tilde{\chi}_{\rm coll}(\qq,z)=-\frac{\alpha^2(\qq,z)S_{12}(\qq,-z)+\alpha^2(\qq,-z)S_{12}(\qq,z)+2\alpha(\qq,z)\alpha(\qq,-z)D_{12}(\qq,z)}{S_{12}(\qq,z)S_{12}(\qq,-z)-D_{12}^2(\qq,z)}
\end{equation}
 However, it turns out that at the order in $q^{-1}$ where we are going, the square of $D_{12}$ in the denominator will remain negligible and we can use the simplified form 
\begin{equation}
\label{eq:utile}
\tilde{\chi}_{\rm coll}(\qq,z)\simeq -\frac{\alpha^2(\qq,z)}{S_{12}(\qq,z)}-\frac{\alpha^2(\qq,-z)}{S_{12}(\qq,-z)}-\frac{2\alpha(\qq,z)\alpha(\qq,-z)D_{12}(\qq,z)}{S_{12}(\qq,z)S_{12}(\qq,-z)}
\end{equation}
 Let us give the equivalent of the power laws (\ref{eq:loipuis}) for the new combinations introduced, obtained by the same power-counting argument: 
\begin{equation}
\label{eq:loipuis2}
S_{12}(\qq,z)\approx q^{d-2}\quad;\quad\alpha(\qq,z)\approx q^{d-4}\quad;\quad D_{12}(\qq,z)\approx q^{d-6}
\end{equation} Thus, $D^2_{12}$ in the denominator is a correction of relative order $q^{2d-12}/q^{2d-4}=q^{-8}$, whereas we will only go to relative order $q^{-2}$ in 2D, see equation (\ref{eq:devchi2d}), and to relative order $q^{-4}$ in 3D, see equation (\ref{eq:devchi3d}). We also find that the third term in (\ref{eq:utile}), of relative order $q^{-4}$ with respect to the first, can be neglected in 2D but not in 3D. To conclude this point, note that the first and second terms of (\ref{eq:utile}) have exactly the same curvilinear integral over the loop $C_{\rm d}$:\footnote{\label{note:relgen} Consider a general function $F(z)$ of the complex variable analytic outside the real intervals $[a,+\infty[$ and $]-\infty,-a]$ with $a>0$, and show that we have in all generality 
\begin{equation}
\label{eq:prop}
\int_{C_{\rm d}}\mathrm{d} z\, F(-z)=\int_{C_{\rm d}}\mathrm{d} z\, F(z)
\end{equation}
 Here, as in Figure~\ref{fig:contour}b, the oriented loop $C_{\rm d}$ surrounds the singularities of $F(z)$ on the positive real half-axis, and the oriented loop $C_{\rm g}$ used below, which is deduced by the action of $z\mapsto -z$, surrounds those on the negative real half-axis. We reason in two steps. (i) The change of variable $z=-z'$ gives $\int_{C_{\rm d}}\mathrm{d} z\, F(-z)=\int_{C_{\rm g}}(-\mathrm{d} z')F(z')$. (ii) Let us close the double loop $C_{\rm g}\cup C_{\rm d}$ with a semicircle at infinity $DC_{\rm h}$ in the upper half-plane and a semicircle at infinity $DC_{\rm b}$ in the lower half-plane; the contour obtained $C_{\rm d}\cup DC_{\rm h}\cup C_{\rm g}\cup DC_{\rm b}$ does not surround any singularity of the integrand, so according to Cauchy's theorem, the corresponding curvilinear integral is zero, $\int_{C_{\rm d}}\mathrm{d} z\, F(z) + \int_{DC_{\rm h}}\mathrm{d} z\, F(z)+\int_{C_{\rm g}}\mathrm{d} z\, F(z)+\int_{DC_{\rm b}}\mathrm{d} z\, F(z)=0$. If the contributions of the semicircles are zero (for example because $F(z)$ decreases to infinity like $1/|z|^\alpha,\,\alpha>1$), this imposes $\int_{C_{\rm d}}\mathrm{d} z\, F(z)+\int_{C_{\rm g}}\mathrm{d} z\, F(z)=0$. It remains to combine (i) and (ii) to obtain the announced property (\ref{eq:prop}) ($z'$ is a dummy variable in point (i)).} 
\begin{equation}
\label{eq:double}
\int_{C_{\rm d}} \mathrm{d} z \left[-\frac{\alpha^2(\qq,-z)}{S_{12}(\qq,-z)}\right]= \int_{C_{\rm d}} \mathrm{d} z \left[-\frac{\alpha^2(\qq,z)}{S_{12}(\qq,z)}\right]
\end{equation}
so that the contribution of the second term at $\delta\tilde{g}_{\rm coll}(\qq)$ exactly doubles that of the first.

Secondly, the series expansion in powers of $q^{-1}$ is greatly simplified by the following procedure: (i) we temporarily assume that the chemical potential $\mu$ is negative and set $\mu=-\nu E_\qq,\ \nu>0$ (since $\mu$ is fixed, $\nu$ tends towards zero like $1/q^2$); (ii) we construct calculation aids $f^{\eps\to\xi}(\qq,z)$ by replacing $\eps_\pm$ with $\xi_\pm$ in the integral expressions of each function $f(\qq,z)$ (here in practice $f=\alpha,S_{12},D_{12}$); (iii) it turns out that we know how to perform calculation $f^{\eps\to\xi}(\qq,z)$ analytically, which allows us to expand them into powers of $\nu$ and therefore $q^{-2}$; (iv) it remains to expand the remainders $\delta f(\qq,z)\equiv f(\qq,z)-f^{\eps\to\xi}(\qq,z)$ using the multiscale method mentioned above; (v) finally, we collect the results and formally extend them to the case $\mu>0$. Let us give the integral expressions of the useful functions and their auxiliaries:\footnote{The symmetrization of the counterterm in equation (\ref{eq:s12eax}) is useful for the multiscale expansion of $\delta S_{12}(\qq,z)$ in powers of $q^{-1}$. It is not useful for the analytical calculation of $S_{12}^{\eps\to\xi}(\qq,z)$ (it only complicates the angular integration without changing the final result).} 
\begin{eqnarray}
\label{eq:alphaexpli}
\alpha(\qq,z)&=&\int_{\mathbb{R}^d}\frac{\dd^dk}{(2\pi)^d} \frac{\Delta(\eps_++\eps_-)(z+\xi_++\xi_-)}{4\eps_+\eps_-[z^2-(\eps_++\eps_-)^2]}  \\
\label{eq:aeax}
\alpha^{\eps\to\xi}(\qq,z)&=&\int_{\mathbb{R}^d}\frac{\dd^dk}{(2\pi)^d} \frac{\Delta(\xi_++\xi_-)}{4\xi_+\xi_-[z-(\xi_++\xi_-)]} \\
\label{eq:s12expli}
S_{12}(\qq,z)&=&\int_{\mathbb{R}^d}\frac{\dd^dk}{(2\pi)^d}\left\{\frac{(\eps_++\eps_-)(\eps_+\eps_-+\xi_+\xi_-)+z(\eps_+\xi_-+\eps_-\xi_+)}{2\eps_+\eps_-[z^2-(\eps_++\eps_-)^2]}+\frac{1}{4\eps_+}+\frac{1}{4\eps_-}\right\}\\
\label{eq:s12eax}
S_{12}^{\eps\to\xi}(\qq,z)&=&\int_{\mathbb{R}^d}\frac{\dd^dk}{(2\pi)^d} \left[\frac{1}{z-(\xi_++\xi_-)}+\frac{1}{4\xi_+}+\frac{1}{4\xi_-}\right]\\
D_{12}(\qq,z)&=&\int_{\mathbb{R}^d}\frac{\dd^dk}{(2\pi)^d}\frac{(\eps_++\eps_-)\Delta^2}{2\eps_+\eps_-[z^2-(\eps_++\eps_-)^2]} \\
D_{12}^{\eps\to\xi}(\qq,z)&=&\int_{\mathbb{R}^d}\frac{\dd^dk}{(2\pi)^d}\frac{(\xi_++\xi_-)\Delta^2}{2\xi_+\xi_-[z^2-(\xi_++\xi_-)^2]}
\end{eqnarray}
 As we can see, the auxiliary functions $\alpha^{\eps\to\xi}(\qq,z)$ and $S_{12}^{\eps\to\xi}(\qq,z)$ have as their only singularity in the complex plane the branch cut $z\in \{\xi_++\xi_-\,|\,\mathbf{k}\in \mathbb{R}^d\}=[E_\qq/2-2\mu,+\infty[$, i.e., $Z\in[1/2+2\nu,+\infty[$; in particular, they have no singularities on the negative real axis, and their contributions to the first two terms of (\ref{eq:utile}) fall under the reasoning illustrated by diagram (\ref{eq:schema}) (their curvilinear integral over $C_{\rm d}$ is zero); this is of course not the case for the remainders $\delta \alpha(\qq,z)$ and $\delta S_{12}(\qq,z)$, if their expansion in powers of $q^{-1}$ is pushed far enough. We will see that it is the cross terms (or interference terms) between $\alpha^{\eps\to\xi}(\qq,z)$ and $\delta\alpha(\qq,z)$ that give the dominant non-zero contribution to the contour integral of the first two terms of (\ref{eq:utile}). The problem does not arise in the third term of (\ref{eq:utile}): this automatically presents singularities on both $\mathbb{R}^+$ and $\mathbb{R}^-$ because it contains a factor $\alpha(\qq,z)\alpha(\qq,-z)$, but also a factor $D_{12}(\qq,z)$, whose auxiliary function has the union of intervals $Z\in]-\infty,-1/2-2\nu]\cup[1/2+2\nu,+\infty[$ as branch cuts. Finally, we can see that the functions introduced are not independent, in the sense that 
\begin{equation}
\label{eq:lien2}
D_{12}(\qq,z)=\frac{\Delta}{z}[\alpha(\qq,z)-\alpha(\qq,-z)]\quad\mbox{and therefore}\quad
D^{\eps\to\xi}_{12}(\qq,z)=\frac{\Delta}{z}[\alpha^{\eps\to\xi}(\qq,z)-\alpha^{\eps\to\xi}(\qq,-z)]
\end{equation}
 which is simply a reformulation of the second relation in (\ref{eq:lien}).

To make things easier, we summarize in the table below the dominant orders and the orders a priori required for the different functions [need we remind you that
 we have $z=ZE_\qq$ and $q\to +\infty$ with $Z$ fixed, and we recall equations (\ref{eq:devchi2d},\ref{eq:devchi3d}) and the first equality giving $\delta\tilde{g}_{\rm coll}(\qq)$ in the form of a contour integral in (\ref{eq:exprdgt})] : 
\begin{equation}
\label{eq:table}
\begin{tabular}{|c||c|c||c|c|}
\hline
& dom. $d=2$ & req. $d=2$ & dom. $d=3$ & req. $d=3$  \\
\hline
$\alpha(\qq,z)$ & $q^{-2}$ & $q^{-4}$ & $q^{-1}$ & $q^{-5}$ \\
$S_{12}(\qq,z)$ & $q^{+0}$ & $q^{-2}$ & $q^{+1}$ & $q^{-3}$ \\
$D_{12}(\qq,z)$ & $q^{-4}$ & -- & $q^{-3}$ & $q^{-3}$ \\
\hline
$\tilde{\chi}_{\rm coll}(\qq,z)$ & $q^{-4}$ & $q^{-6}$ & $q^{-3}$ & $q^{-7}$ \\
$1^{\rm st}$ or $2^{\rm nd}$ term in (\ref{eq:utile}) & $q^{-4}$ & $q^{-6}$ & $q^{-3}$ & $q^{-7}$ \\
$3^{\rm rd}$ term in (\ref{eq:utile}) & $q^{-8}$ & -- & $q^{-7}$ & $q^{-7}$ \\
\hline
$\delta\tilde{g}_{\rm coll}(\qq)$ & $q^{-4}$ & $q^{-4}$ & $q^{-4}$ & $q^{-5}$\\
\hline
\end{tabular}
\end{equation}

\subsection{ Two-dimensional case }

In dimension $d=2$, we obtain the explicit expressions of the auxiliary functions after angular integration, then radial integration using the change of variable $x=k^2/q^2$:\footnote{We also set $x=\frac{1}{4}-\nu+\nu^{1/2}(t-t^{-1})/2$ with $t\in[2\nu^{1/2},+\infty[$ in the integral giving $\alpha^{\eps\to\xi}(\qq,z)$ in order to transform the difficult part $[(x+\nu+1/4)^2-x]^{1/2}=\nu^{1/2}(1+t^2)/2t$ into a rational fraction. Despite appearances, the square root in the denominator of (\ref{eq:aex2d}) does not introduce any branch cut because the numerator is—like the denominator—an odd function of this square root (a square root transforms into its opposite when it crosses its branch cut). The logarithms in (\ref{eq:aex2d}) do have $Z\in[\frac{1}{2}+2\nu,+\infty[$ as a branch cut.} 
\begin{eqnarray}
\label{eq:aex2d}
\alpha^{\eps\to\xi}(\qq,z)&=&\frac{m\Delta}{4\pi\hbar^2E_\mathbf{q}}\frac{\ln\left[4\nu+1-Z-\sqrt{4\nu+(Z-1)^2}\right]-\ln\left[4\nu+1-Z+\sqrt{4\nu+(Z-1)^2}\right]}{\sqrt{4\nu+(Z-1)^2}}\\
\label{eq:sex2d}
S_{12}^{\eps\to\xi}(\qq,z)&=&\frac{m}{4\pi\hbar^2}\left[\ln\left(2\nu+\frac{1}{2}-Z\right)-\ln(2\nu)\right]
\end{eqnarray}
 We then obtain the required orders of the expansions of the rest $\delta f(\qq,z)$:\footnote{To make certain integrals explicit, we used the equation of state (\ref{eq:etat}) and the equation on the order parameter (\ref{eq:delta}) from the static BCS theory at zero temperature. For example, $\int_{\mathbb{R}^2}[\dd^2k/(2\pi)^2](\eps_\kk^{-1}-\xi_\kk^{-1})=(m/2\pi\hbar^2)\ln(\nu q^2/q_{\rm dim}^2)$.} 
\begin{eqnarray}
\nonumber
\delta\alpha(\qq,z)&\stackrel{Z\,\mbox{\scriptsize fixed}}{\underset{q\to +\infty}{=}}&
\frac{m\Delta}{4\pi\hbar^2E_\mathbf{q}}\frac{2\ln(q/q_{\rm dim})+\ln\nu}{Z-1}
+\boxed{\frac{\rho^{\rm sta}\Delta}{(2E_\qq)^2}\frac{1}{Z+1}}+\frac{\Delta}{(2E_\qq)^2}\frac{1}{(Z-1)^3}\times\\
\label{eq:da2d}
&&\hspace{2cm}\times\int_{\mathbb{R}^2}\frac{\dd^2k}{(2\pi)^2} (\eps_\kk^{-1}-\xi^{-1}_\kk)[2E_\kk+(Z^2-1)\xi_\mathbf{k}]+O(q^{-6}\ln q)\\
\label{eq:ds2d}
\delta S_{12}(\qq,z)&\stackrel{Z\,\mbox{\scriptsize fixed}}{\underset{q\to +\infty}{=}}&\frac{m}{4\pi\hbar^2}\ln(\nu q^2/q_{\rm dim}^2)-\frac{1}{E_\mathbf{q}}\frac{\rho^{\rm sta}}{Z-1} + O(q^{-4})
\end{eqnarray}
 The most interesting contribution in all these results is the one framed in $\delta\alpha(\qq,z)$: it is the only one to present a singularity on the negative real half-axis (a pole at $Z=-1$). We note it $\alpha_{\rm sing}^{\rm dom}(\qq,z)$ in the following, 
\begin{equation}
\alpha_{\rm sing}^{\rm dom}(\qq,z)=\frac{\rho^{\rm sta}\Delta}{(2E_\qq)^2}\frac{1}{Z+1}=\frac{\rho^{\rm sta}\Delta}{4E_\qq(z+E_\qq)}\end{equation}
 because it it is the dominant order in $q^{-1}$ of the part $\alpha_{\rm sing}(\qq,z)$ of $\alpha(\qq,z)$ singular on $\mathbb{R}^-$.\footnote{This can be easily found by transposing note \ref{note:argsimsing} to the two-dimensional case.} Everything else will give a zero contribution after integration on the contour $C_{\rm d}$, see the discussion around diagram (\ref{eq:schema}). We can therefore, in the first term of (\ref{eq:utile}), make the substitution 
\begin{equation}
\label{eq:subs}
-\frac{\alpha^2(\qq,z)}{S_{12}(\qq,z)} \longrightarrow -\frac{2\alpha^{\rm dom}(\qq,z)\alpha_{\rm sing}^{\rm dom}(\qq,z)}{S_{12}^{\rm dom}(\qq,z)}
\end{equation}
 with the dominant behaviors of the functions $\alpha(\qq,z)$ and $S_{12}(\qq,z)$:\footnote{These are obtained by expanding (\ref{eq:aex2d},\ref{eq:sex2d}) to order $\nu^0$ and collecting with (\ref{eq:da2d}, \ref{eq:ds2d}); we verify that the logarithms $\ln\nu$ cancel each other out exactly.} 
\begin{eqnarray}
\alpha^{\rm dom}(\qq,z) &=& \frac{m\Delta}{4\pi\hbar^2E_\mathbf{q}}\frac{1}{Z-1}[2\ln(q/q_{\rm dim})+2\ln(1-Z)-\ln(1-2Z)] \\
S_{12}^{\rm dom}(\qq,z)&=& \frac{m}{4\pi\hbar^2} \left[2\ln\left(\frac{q}{2 q_{\rm dim}}\right)+\ln(1-2Z)\right]
\end{eqnarray}
 After the substitution (\ref{eq:subs}), the curvilinear integral (\ref{eq:exprdgt}) of the first term of (\ref{eq:utile}) can be easily calculated by closing the loop $C_{\rm d}$ with a large circle at infinity, as shown in the diagram (\ref{eq:schema}); the only singularity surrounded is then the pole of $\alpha_{\rm sing}^{\rm dom}(\qq,z)$ at $Z=-1$. The line integral over $C_{\rm d}$ of the second term (subject to the same substitution) can be deduced directly from the residue theorem without distorting the loop, the pole of $\alpha_{\rm sing}^{\rm dom}(\qq,-z)$ at $Z=+1$ being the only singularity surrounded by $C_{\rm d}$; it exactly doubles the contribution of the first term, as already indicated by equation (\ref{eq:double}). The third term of (\ref{eq:utile}), as we have said, is negligible. This leaves 
\begin{equation}
\delta\tilde{g}_{\rm coll}(\qq) \stackrel{d=2}{\underset{q\to +\infty}{=}} 2\times \frac{\rho^{\rm sta}\Delta}{2E_\mathbf{q}} \frac{\alpha^{\rm dom}(\qq,-E_\qq)}{S_{12}^{\rm dom}(\qq,-E_{\mathbf{q}})}+O(\ln^\alpha/q^6)
\end{equation}
 in agreement with equations (\ref{eq:collasymp},\ref{eq:coefcoll2d}), which was to be demonstrated.

\subsection{Three-dimensional case}

In dimension $d=3$, we also manage to integrate analytically (\ref{eq:aeax},\ref{eq:s12eax}):\footnote{\label{note:beaucal} Let us explain our rather tricky calculation of $\alpha^{\eps\to\xi}(\qq,z)$. After angular averaging in spherical coordinates with polar axis $\qq$, we extend the radial integral to $\mathbb{R}$ by formal parity of the integrand, then apply a contour integral method. More precisely, after the change of variable $k=qK$, we arrive at the integral $I_\alpha=\int_{-\infty}^{+\infty}\mathrm{d} K\, K [\ln P(K)-\ln P(-K)]/Q(K)$ with the trinomials $P(K)=(K-1/2)^2+\nu$ and $Q(K)=2K^2+1/2+2\nu-Z$. Calling $u=1/2+\mathrm{i}\nu^{1/2}$ and $z_0=(\ii/2)\sqrt{1+4\nu-2Z}$ the roots of $P$ and $Q$ of positive imaginary part (the others are $u^*$ and $-z_0$), using the identity $\ln[(K-u)(K-u^*)]=\ln(K-u)+\ln(K-u^*)\, \forall K\in\mathbb{R}$, then decomposing $K/Q(K)$ into partial fractions, we find that $I_\alpha=(1/4)\int_{-\infty}^{+\infty}\mathrm{d} K[(K+z_0)^{-1}+(K-z_0)^{-1}]\{[\ln(K-u)-\ln(K+u^*)]+[\ln(K-u^*)-\ln(K+u)]\}$. We are reduced to integrals of the form $J=\int_{-\infty}^{+\infty}\mathrm{d} K [\ln(K-a)-\ln(K-b)]/(K-c)$ with non-zero imaginary parts $\im a$ and $\im b$ of the same sign. We perform the calculation for $J$ using the residue theorem, closing the integration path with a semicircle at infinity in the lower ($s=-1$) or upper ($s=+1$) complex half-plane $P_s=P_{\not\owns a,b}$ not containing the branch cuts $a+\mathbb{R}^-$ and $b+\mathbb{R}^-$ of the logarithms. Then $J=0$ if $c\notin P_{\not\owns a,b}$ and $J=2\mathrm{i}\pi s[\ln(c-a)-\ln(c-b)]$ otherwise.} 
\begin{eqnarray}
\label{eq:aex3d}
\alpha^{\eps\to\xi}(\qq,z)&=&\frac{q^3\Delta}{(4\pi E_\qq)^2}\mathcal{F}\left(u=2\mathrm{i}\nu^{1/2}+\mathrm{i}\sqrt{1+4\nu-2Z}\right) \\
S_{12}^{\eps\to\xi}(\qq,z)&=& \frac{m q}{8\pi\hbar^2}\left[\left(1+4\nu-2Z\right)^{1/2}-2\nu^{1/2}\right]
\end{eqnarray}
 To arrive at a compact notation, we had to introduce\footnote{Contrary to appearances, the function $\mathcal{F}(u)$ only has the segment $u\in[-1;1]$ as a branch cut (the discontinuities of the first two logarithms at the crossing of the interval $]-\infty,-1[$, or those of the last two at the crossing of the interval $]1,+\infty[$, compensate each other exactly). In any case, the variable $u$ in (\ref{eq:aex3d}) never reaches the real axis when $Z$ describes $\mathbb{C}\setminus\mathbb{R}$ because the square root of a complex number $\notin\mathbb{R}^{-}$ is always positive in its real part; the branch cut of (\ref{eq:aex3d}) is therefore simply that of $\sqrt{1+4\nu-2Z}$, namely $[\frac{1}{2}+2\nu,+\infty[$ as it should be.} 
\begin{equation}
\mathcal{F}(u)=\frac{\mathrm{i}\pi}{2}[\ln(u-1)-\ln(u+1)+\ln(1-u)-\ln(-1-u)]\end{equation}
 The missing function $D_{12}^{\eps\to\xi}$ can be deduced from (\ref{eq:aex3d}) through equation (\ref{eq:lien2})
 It remains to expand the remainders $\delta\alpha(\qq,z)$ and $\delta S_{12}(\qq,z)$ into powers of $q^{-1}$, which is quite cumbersome. To see how far we need to go, let's start again from the useful expression (\ref{eq:utile}) of the collective part of the susceptibility and analyze it term by term and factor by factor, knowing that we must determine up to order $q^{-7}$ the parts that are singular on both $\mathbb{R}^-$ and $\mathbb{R}^+$: 
\begin{itemize}\item The third term of (\ref{eq:utile}) is immediately of order $q^{-7}$ [see summary (\ref{eq:table})] and bilaterally singular; it therefore requires only knowledge of the dominant orders $\alpha^{\rm dom}(\qq,z)\propto q^{-1}$ and $S^{\rm dom}_{12}(\qq,z)\propto q^{+1}$. We can do in (\ref{eq:utile}) the substitution 
\begin{equation}
\label{eq:subs3}
-\frac{2\alpha(\qq,z)\alpha(\qq,-z)D_{12}(\qq,z)}{S_{12}(\qq,z)S_{12}(\qq,-z)}\longrightarrow
-\frac{2\alpha^{\rm dom}(\qq,z)\alpha^{\rm dom}(\qq,-z)D^{\rm dom}_{12}(\qq,z)}{S^{\rm dom}_{12}(\qq,z)S^{\rm dom}_{12}(\qq,-z)}
\end{equation}
 \item The first singularity of $\alpha(\qq,z)$ on $\mathbb{R}^-$ appears at order $q^{-4}$, see note \ref{note:argsimsing}; we call it $\alpha_{\rm sing}^{\rm dom}(\qq,z)$. It does not exhaust the required order on $\alpha(\qq,z)$, which is $q^{-5}$ [see summary (\ref{eq:table})]; at this order, we will also find a singular contribution on $\mathbb{R}^-$, which we will logically note as $\alpha_{\rm sing}^{\rm sdom}(\qq,z)$ (the index “sdom” means “subdominant”). \item The first singularity of $S_{12}(\qq,z)$ on $\mathbb{R}^-$ appears at order $q^{-6}$, see note \ref{note:argsimsing}, which goes beyond the required order on this function; we can therefore ignore it. \item In the first term of (\ref{eq:utile}), at order $q^{-7}$, we can therefore replace $\alpha^2(\qq,z)$ in the numerator with the double product of the part of $\alpha(\qq,z)$ that is regular on $\mathbb{R}^-$ (written at subdominant order $q^{-2}$) and the part of $\alpha(\qq,z)$ that is singular on $\mathbb{R}^-$ (also written at subdominant order $q^{-5}$); in the denominator, we can limit $S_{12}(\qq,z)$ to the subdominant order $q^0$. Hence the substitution: 
\begin{equation}
\label{eq:substi}
-\frac{\alpha^2(\qq,z)}{S_{12}(\qq,z)} \longrightarrow -\frac{2[\alpha^{\rm dom}(\qq,z)+\alpha^{\rm sdom}(\qq,z)][\alpha_{\rm sing}^{\rm dom}(\qq,z)+\alpha_{\rm sing}^{\rm sdom}(\qq,z)]}{S_{12}^{\rm dom}(\qq,z)+S_{12}^{\rm sdom}(\qq,z)}
\end{equation}
 \item We can apply exactly the same reasoning and the same modification to the second term of (\ref{eq:utile}), simply changing $z$ to $-z$ in (\ref{eq:substi}). In any case, this second term gives the same contribution as the first to $\delta\tilde{g}_{\rm coll}(\qq)$ by virtue of equation (\ref{eq:double}); we will take this into account with a factor of $2$.
\item The parts with singularity on $\mathbb{R}^-$ can of course only come from the remainders $\delta f(\qq,z)$. 
\end{itemize}
For the sake of brevity, we do not give the rather long expansions of $\delta\alpha(\qq,z)$ and $\delta S_{12}(\qq,z)$, which
 are, after all, only intermediate calculations. Let us simply say that that of $\delta\alpha(\qq,z)$ begins at order $q^{-2}$ and that the next term is $q^{-4}$ (both terms are $O(1/|Z|)$ to infinity); that of $\delta S_{12}(\qq,z)$ begins at order $q^0$, with a subsequent term $q^{-2}$ (the first is independent of $Z$, the second is proportional to $1/(Z-1)$). It should also be noted that the dependency on $\nu$ disappears in the sum of $\alpha^{\eps\to\xi}(\qq,z)$ and $\delta\alpha(\qq,z)$ at order $q^{-2}$, and in that of $S_{12}^{\eps\to\xi}(\qq,z)$ and $\delta S_{12}(\qq,z)$ at order $q^0$, by virtue of the equality $\int_{\mathbb{R}^3}[\dd^3k/(2\pi)^3](\eps_\kk^{-1} -\xi_\kk^{-1})=(m/2\pi\hbar^2)(\nu^{1/2}q-a_{\rm 3D}^{-1})$ deduced from equation (\ref{eq:delta}). Finally, for the parts that are regular on $\mathbb{R}^-$, we have: 
\begin{align}
\alpha^{\rm dom}(\qq,z) &=\frac{m^2\Delta}{4\pi^2\hbar^4}\left[F(Z)-\frac{\pi^2}{2}\right]q^{-1}&;&\quad\quad \alpha^{\rm sdom}(\qq,z) = -\frac{m^2\Delta a_{\rm 3D}^{-1}}{2\pi\hbar^4}\frac{1}{Z-1} q^{-2}\\
S_{12}^{\rm dom}(\qq,z)&=\frac{m}{8\pi\hbar^2}(1-2Z)^{1/2}q &;&\quad\quad S_{12}^{\rm sdom}(\qq,z)=-\frac{m a_{\rm 3D}^{-1}}{4\pi\hbar^2}
\end{align}
 The sub-subdominant contributions are given in note \ref{note:infos} below. For brevity, we introduced function\footnote{\label{note:infos} In $F(Z)$, the argument of the logarithm is a negative real number $-x^2$, $x\geq 0$, and therefore lies on the branch cut of the logarithm, if $Z=1-2x/(1+x)^2$ (which describes $[1/2,1]$ when $x$ describes $\mathbb{R}^+$) or if $Z=1+2x/(x-1)^2$ (which describes $[1,+\infty[$ when $x$ describes $\mathbb{R}^+$); we obtain the predicted branch cut $Z\in[1/2,+\infty[$, which is also that of the square root under the logarithm. Furthermore, we verify that $F(Z)=-\pi Z-\pi Z^2+O(Z^3)$ when $Z\to 0$ and, above all, that $F(Z)-\pi^2/2{\sim -\pi(-2/Z)^{1/2}}=O(1/|Z|^{1/2})$ when $|Z|\to +\infty$. The second property allows us to verify that the circle at infinity in the diagram (\ref{eq:schema}) has zero contribution, and therefore that the curvilinear integral over $C_{\rm d}$ of the coefficients $\tilde{\chi}_{\rm coll}^{(-n)}(Z)$ of the expansion (\ref{eq:devchi3d}) is zero up to order $n=5$. To convince the reader, let us give the contribution of the first term of (\ref{eq:utile}) to these coefficients: $\tilde{\chi}_{\rm coll}^{(-3)}(Z)|_{\rm 1st}=-\Delta^2(m/2\pi\hbar^2)^3[\pi^2-2F(Z)]^2/\sqrt{1-2Z}$, $\tilde{\chi}_{\rm coll}^{(-4)}(Z)|_{\rm 1st}=-2\Delta^2a_{\rm 3D}^{-1}(m/2\pi\hbar^2)^3[\pi^2-2F(Z)]\{[\pi^2-2F(Z)]/(1-2Z)+4\pi/[(Z-1)\sqrt{1-2Z}]\}$, $\tilde{\chi}_{\rm coll}^{(-5)}(Z)|_{\rm 1st}=-4\Delta^2(m/2\pi\hbar^2)^3(1-2Z)^{-3/2}\{a_{\rm 3D}^{-2}[2F(Z)-\pi^2+2\pi\sqrt{1-2Z}/(1-Z)]^2+(m\mu/\hbar^2)[2F(Z)-\pi^2][2F(Z)-\pi^2-4\pi Z\sqrt{1-2Z}/(1-Z)^2]\}$. The calculation of $\tilde{\chi}_{\rm coll}^{(-5)}(Z)$ requires the sub-subdominant parts $\alpha^{\rm ssdom}(\qq,z)=-(m^3\Delta\mu/\pi\hbar^6q^3)Z/[(Z-1)^2\sqrt{1-2Z}]$ and $S_{12}^{\rm ssdom}(\qq,z)=-(m^2\mu/2\pi\hbar^4q)(1-2Z)^{-1/2}$. Furthermore, the parts of $\tilde{\chi}_{\rm coll}^{(-6)}(Z)|_{\rm 1st}$ and $\tilde{\chi}_{\rm coll}^{(-7)}(Z)|_{\rm 1st}$ regular on $\mathbb{R}^-$ are $\sim 4\Delta^2m^3\rho^{\rm sta}/\hbar^6 Z^2$ and $\sim -m^5\Delta^4/\hbar^{10} Z^3$ at infinity, so they also fall under the reasoning around (\ref{eq:schema}) and make a zero contribution to $\delta\tilde{g}_{\rm coll}(\qq)$. To see this, we expand the regular parts on $\mathbb{R}^-$, $\alpha_{\rm reg}(\qq,z)=\int_{\mathbb{R}^3}[\dd^3k/(2\pi)^3]\Delta(\eps_++\xi_++\eps_-+\xi_-)/8\eps_+\eps_-[z-(\eps_++\eps_-)]$, and $S_{12}^{\rm reg}(\qq,z)=\int_{\mathbb{R}^3}[\dd^3k/(2\pi)^3]\{(4\eps_+)^{-1}+(4\eps_-)^{-1}+(\eps_++\xi_+)(\eps_-+\xi_-)/4\eps_+\eps_-[z-(\eps_++\eps_-)]\}$ to the required order in (\ref{eq:table}). For $q\to +\infty$ at fixed $Z$, we get $S_{12}^{\rm reg}(\qq,z)=S_{12}^{\rm dom}(\qq,z)+S_{12}^{\rm sdom}(\qq,z)+S_{12}^{\rm ssdom}(\qq,z)-\rho^{\rm sta}/[E_\qq(Z-1)]+q^{-3}[\delta S_{12}^{\rm reg\ (-3)}(Z)-(m^3\mu^2/\pi\hbar^6)(1-2Z)^{-3/2}]+O(q^{-4})$ and $\alpha_{\rm reg}(\qq,z)=\alpha^{\rm dom}(\qq,z)+\alpha^{\rm sdom}(\qq,z)+\alpha^{\rm ssdom}(\qq,z)-q^{-4}(m^2\Delta/3\hbar^4)(Z-1)^{-3}[(3Z^2+5)\rho^{\rm sta}+4m\mu a_{\rm 3D}^{-1}/\pi\hbar^2]+q^{-5}[\delta\alpha_{\rm reg}^{(-5)}(Z)-(2m^4\mu^2\Delta/\pi\hbar^8) Z(Z^2-6Z+3)(Z-1)^{-4}(1-2Z)^{-3/2}]+O(q^{-6})$. The functions $\delta S_{12}^{\rm reg\ (-3)}(Z)$ and $\delta\alpha_{\rm reg}^{(-5)}(Z)$ are given by difficult integrals of the same type as (\ref{eq:asingsdom}), but here it suffices to show that $\delta S_{12}^{\rm reg\ (-3)}(Z)\sim -m^3\Delta^2/\pi\sqrt{2}\hbar^6(-Z)^{3/2}$ and $\delta \alpha_{\rm reg}^{(-5)}(Z)\sim m^4\Delta^3/4\hbar^8Z^{2}$ by considering the two scales $K_-=O(1)$ and $K_-\approx |Z|^{1/2}$.} $F(Z)=({\mathrm{i}\pi}/{2})\ln[(\sqrt{1-2Z}+\mathrm{i} Z)/(\sqrt{1-2Z}-\mathrm{i} Z)]\ \forall Z\in \mathbb{C}\setminus [1/2,+\infty[$, after showing that $\lim_{\nu\to 0^+}\mathcal{F}\left(2\mathrm{i}\nu^{1/2}+\mathrm{i}\sqrt{1+4\nu-2Z}\right)=\mathcal{F}\left(\mathrm{i}\sqrt{1-2Z}\right)=F(Z)-\pi^2/2$ by verifying in the second equality that the functions have the same first derivative with respect to $Z$, namely $\pi/[(Z-1)\sqrt{1-2Z}]$, and the same zero limit at $Z=-\infty$. For the parts that are singular on $\mathbb{R}^-$, we find the following dominant and subdominant orders:\footnote{\label{note:argsimsing} Let us explain how to find the corresponding powers $q^{-4}$ and $q^{-5}$. In expression (\ref{eq:alphaexpli}) of $\alpha(\qq,z)$, let's break down the integrand into partial fractions for the variable $z$ and keep the denominator part $z+\eps_++\eps_-$, which is the only part that can be singular on $\mathbb{R}^-$, giving $\alpha_{\rm sing}(\qq,z)=\int_{\mathbb{R}^3}[\dd^3k/(2\pi)^3](\Delta/8)[\delta\eps_+/(\eps_+\eps_-)+\delta\eps_-/(\eps_-\eps_+)](z+\eps_++\eps_-)^{-1}$ with $\delta\eps_\kk=\eps_\kk-\xi_\kk$. As $\delta\eps_\mathbf{k}\approx 1/k^2$ at infinity, the proximity zone $k_-<\eta q$ of the multiscale method gives at dominant order in $q^{-1}$ a contribution independent from the cut-off $\eta$, $\alpha_{\rm sing}^{\rm dom}(\qq,z)=2\int_{\mathbb{R}^3}[\dd^3k_-/(2\pi)^3](\Delta/8)(\delta\eps_-/\eps_-)/[E_\qq^2(Z+1)]$: we find equation (\ref{eq:asingdom}). The asymptotic zone, in which $q\to +\infty$ at fixed $\KK=\kk/q$, is obviously of dominant order $q^{-5}$ since $\dd^3k$ outputs a factor $q^3$, $\delta\eps_-/\eps_-\eps_+$ outputs a factor $q^{-6}$, and the energy denominator $z+\eps_++\eps_-$ is of order $q^2$; however, the result depends on the cutoff $\eta$ in a way that is compensated by a subdominant contribution from the proximity zone as in (\ref{eq:asingsdom}). The same reasoning on expression (\ref{eq:s12expli}) of $S_{12}(\qq,z)$ leads to a singular part on $\mathbb{R}^-$ given by $S_{12}^{\rm sing}(\qq,z)=\int_{\mathbb{R}^3}[\dd^3k/(2\pi)^3](-1/4)(\delta\eps_+\delta\eps_-/\eps_+\eps_-)(z+\eps_++\eps_-)^{-1}$ of dominant order $q^{-6}$ on the proximity zone and $q^{-7}$ on the asymptotic zone.} 
\begin{eqnarray}
\label{eq:asingdom}
\alpha_{\rm sing}^{\rm dom}(\qq,z) &=& \frac{\rho^{\rm sta}\Delta}{(2 E_\qq)^2}\frac{1}{Z+1} \\
\nonumber
\alpha_{\rm sing}^{\rm sdom}(\qq,z) &=& \lim_{\eta\to 0^+} \frac{m^4\Delta^3}{\pi^2\hbar^8q^5}\left[\frac{-1}{Z+1}\,\frac{\hbar^4q}{2m^2\Delta^2}\int_{\eta q}^{+\infty}\mathrm{d} k_-\, k_-^2\frac{2\Delta^2}{(2E_{\kk_-})^2}\right. \\
&& \left. + \int_\eta^{+\infty}\frac{\mathrm{d} K_-}{K_-^2} \int_{u_m(K_-)}^{1}\frac{\mathrm{d} u_-}{2} \frac{1+2K_-u_-+2K_-^2}{(1+2K_-u_-+K_-^2)^2}\frac{1}{Z+1+2K_-u_-+2K_-^2}\right]
\label{eq:asingsdom}
\end{eqnarray}
 where the density $\rho^{\rm sta}$ is that (\ref{eq:etat}) of the static BCS theory at zero temperature, $\eta$ is the cutoff parameter separating the proximity zone $k_-<\eta q$ from the asymptotic zone $k_->\eta q, \mathbf{k}\cdot\qq>0$ in the multiscale method of reference \cite{livre}, $1-\xi_\kk/\eps_\kk$ has been replaced by its dominant order $2\Delta^2/(2E_\kk)^2$ in the first integral, and $u_m(K)=\max(-1,-1/2K)$ has been set in the second integral as in section 4.3.5.4 of this reference ($u_-$ is the cosine of the angle between vectors $\qq$ and $\kk_-$, and we have $\KK_-=\kk_-/q$). By extending the two-dimensional trick from note \ref{note:metdev} to the three-dimensional auxiliary integral 
\begin{equation}
J=\int_{\mathbb{R}^3}\frac{\dd^3K}{(K_+^2+\veps^2)^2(K_-^2+\veps^2)^2}\frac{K_+^2+K_-^2}{Z+K_+^2+K_-^2}
\end{equation}
 (here $\KK_\pm=\mathbf{K}\pm\hat{\mathbf{q}}/2$ with $\hat{\mathbf{q}}=\qq/q$ the direction of vector $\qq$) and then performing the calculation of this integral to within $o(1)$ when ${\varepsilon}\to 0^+$ using the method in note \ref{note:beaucal}, we arrive at the explicit expression\footnote{If this trick is overlooked, we can still perform an analytical calculation for $\im[\alpha_{\rm sing}^{\rm sdom}(\qq,Z=\eps+\mathrm{i} 0^+)]$ on its support $\eps\in]-\infty,-1/2]$ using the relation $\im[1/(x+\mathrm{i} 0^+)]=-\pi\delta(x)$ under the integral sign in (\ref{eq:asingsdom}) and then taking the limit $\eta\to 0^+$ in the sense of distributions. It then remains to close the integration contour on $C_{\rm g}$ as in note \ref{note:relgen}. This leads to an equivalent but much more complicated expression of the integral (\ref{eq:defironde}) to come, $\mathcal{I}=4\phi(-1)-\int_{-\infty}^{-{3}/{2}}\mathrm{d}\eps \frac{\phi(\eps)}{(1+\eps)^2}-\int_{-1/2}^{+1/2}\frac{\dd(\delta\eps)}{2(\delta\eps)^2}[\phi(-1+\delta\eps)+\phi(-1-\delta\eps)-2\phi(-1)]+\int_{-\infty}^{-1/2}\mathrm{d}\eps\,\phi(\eps)\left[\frac{1}{2\eps^2}\ln\frac{\eps+\sqrt{-1-2\eps}}{\eps-\sqrt{-1-2\eps}}-\frac{1}{\eps(\sqrt{-1-2\eps}-\eps)}\right]$ with $\phi(\eps)=[{\pi^2}/{2}-F(\eps)]/\sqrt{1-2\eps}$.} 
\begin{equation}
\alpha_{\rm sing}^{\rm sdom}(\qq,z)= \frac{m^4\Delta^3}{\pi^2\hbar^8q^5}\left[\frac{\pi\sqrt{1+2Z}}{2Z(Z+1)^2}-\frac{F(-Z)}{2Z^2}\right]
\end{equation}
 singular on the interval $Z\in]-\infty,-1/2]$; the absence of a pole at $Z=0$ is verified by means of note \ref{note:infos}, which gives the behavior of $F(Z)$ at $Z=0$.

To summarize, at order $q^{-5}$ and remembering the curvilinear integral expression at zero temperature $\delta\tilde{g}_{\rm coll}(\qq)=\int_{C_{\rm d}} \frac{\mathrm{d} z}{(-2\mathrm{i}\pi)} \tilde{\chi}_{\rm coll}(\qq,z)$, we can identify three contributions in the Fourier transform of the collective part $\delta g_{\rm coll}(\rr,\rr')$ (\ref{eq:sepa}) of the pair distribution functions:

\noindent (i) The contribution of the dominant singular part $\alpha_{\rm sing}^{\rm dom}(\qq,z)$ in the first term of (\ref{eq:utile}) simplified as in equation (\ref{eq:substi}); since the singularity is only a simple pole at $Z=-1$, we close the integration contour $C_{\rm d}$ as in diagram (\ref{eq:schema}) to enclose the pole. The residue theorem then gives 
\begin{equation}
\label{eq:point1}
\delta\tilde{g}_{\rm coll}(\qq)|_{\rm (i)} = 2\times E_\mathbf{q} \times 
\frac{2[\alpha^{\rm dom}(\qq,z=-E_\qq)+\alpha^{\rm sdom}(\qq,z=-E_\qq)]}{S_{12}^{\rm dom}(\qq,z=-E_\qq)+S_{12}^{\rm sdom}(\qq,z=-E_\qq)}\frac{\rho^{\rm sta}\Delta}{(2E_\qq)^2}
\end{equation}
 The factor $2$ takes into account the doubling by the second term of (\ref{eq:utile}) and the factor $E_\qq$ is the Jacobian of the change of variable $z=E_\mathbf{q} Z$. The values of the functions at $Z=-1$ are easily obtained if we know that $F(-1)=\pi^2/6$ [see equation (\ref{eq:gfm})], so that 
\begin{multline}
\delta\tilde{g}_{\rm coll}(\qq)|_{\rm (i)} = -\frac{2\rho^{\rm sta}m^2\Delta^2}{\hbar^4 q^4}\frac{\frac{1}{12}-\frac{1}{4\pi a_{\rm 3D}q}+O(q^{-2})}{\frac{\sqrt{3}}{8\pi}-\frac{1}{4\pi a_{\rm 3D}q}+O(q^{-2})} \\
=-2\left(\frac{m\Delta}{\hbar^2 q^2}\right)^2\rho^{\rm sta} \left[\frac{2\pi}{3\sqrt{3}}+\left(\frac{4\pi}{9}-\frac{2\sqrt{3}}{3}\right)(q a_{\rm 3D})^{-1} + O(q^{-2})\right]
\label{eq:point1bis}
\end{multline}
 \noindent (ii) The contribution of the subdominant singular part $\alpha_{\rm sing}^{\rm dom}(\qq,z)$ in simplified form (\ref{eq:substi}); since its singularities in the complex plane are a branch cut $Z\in]-\infty,-1/2]$ and a pole at $Z=-1$, we prefer to keep the original contour $C_{\rm d}$, but flatten it over the interval $Z\in[1/2,+\infty[$ to reduce it to a real integration variable $\eps\in [1/2,+\infty[$ (we set $Z=\eps\pm\mathrm{i} 0^+$ and use the identity $\tilde{\chi}(\eps+\mathrm{i} 0^+)-\tilde{\chi}(\eps-\mathrm{i} 0^-)=2\mathrm{i}\im\tilde{\chi}(\eps+\mathrm{i} 0^+)$): 
\begin{equation}
\label{eq:point2}
\delta\tilde{g}_{\rm coll}(\qq)|_{\rm (ii)} = 2\times E_\mathbf{q} \times \int_{1/2}^{+\infty} \frac{\mathrm{d}\eps}{\pi} 
\im\left(\frac{2\alpha^{\rm dom}(\qq,Z=\eps+\mathrm{i} 0^+)}{S_{12}^{\rm dom}(\qq,Z=\eps+\mathrm{i} 0^+)}\right) \alpha_{\rm sing}^{\rm sdom}(\qq,Z=\eps)
\end{equation}
 where the factor $2\times E_\qq$ has the same origin as in (\ref{eq:point1}).

\noindent(iii)
 The contribution of the third term of (\ref{eq:utile}) simplified as in equation (\ref{eq:subs3}); we reduce to a real integration variable as in point (ii) to obtain: 
\begin{equation}
\label{eq:point3}
\delta\tilde{g}_{\rm coll}(\qq)|_{\rm (iii)} = E_\mathbf{q}\times \int_{1/2}^{+\infty} \frac{\mathrm{d}\eps}{\pi} \frac{2\alpha^{\rm dom}(\qq,Z=-\eps)}{S_{12}^{\rm dom}(\qq, Z=-\eps)} \im\left(\frac{\alpha^{\rm dom}(\qq,Z=\eps+\mathrm{i} 0^+)D_{12}^{\rm dom}(\qq,Z=\eps+\mathrm{i} 0^+)}{S_{12}^{\rm dom}(\qq,Z=\eps+\mathrm{i} 0^+)}\right)
\end{equation}
 remembering that the dominant orders of $\alpha$ and $S_{12}$ are regular functions on $\mathbb{R}^-$ and that the function $D_{12}(\qq,z)$ is related to $\alpha(\qq,z)$ by equation (\ref{eq:lien2}).

We can perform the calculation of the imaginary parts and, more generally, the integrands in (\ref{eq:point2},\ref{eq:point3}) using the following relations: 
\begin{eqnarray}
\label{eq:gfp}
F(\eps+\mathrm{i} 0^+)&\stackrel{\eps>1/2}{=}&\frac{\mathrm{i}\pi}{2}\ln\left[-\frac{\eps-\sqrt{2\eps-1}}{\eps+\sqrt{2\eps-1}}\left(1+\frac{2\mathrm{i} 0^+}{(\eps-1)\sqrt{2\eps-1}}\right)\right]=\frac{\mathrm{i}\pi}{2}\ln R(\eps) +\frac{\pi^2}{2}\sgn(\eps-1)\\
F(-\eps)&\stackrel{\eps>{-1/2}}{=}&\frac{\pi^2}{2}-\pi\arctan\frac{\sqrt{1+2\eps}}{\eps} = \pi\arctan\left(\frac{\eps}{\sqrt{1+2\eps}}\right)
\label{eq:gfm}
\end{eqnarray}
 where $\sgn$ is the sign function and we have set $R(\eps)=(\eps-\sqrt{2\eps-1})/(\eps+\sqrt{2\eps-1})\geq 0\ \forall\eps>1/2$. By collecting the three contributions (\ref{eq:point1bis},\ref{eq:point2},\ref{eq:point3}), we arrive at the announced asymptotic expansion (\ref{eq:collasymp}), with the expression of the numerical constant in the coefficient (\ref{eq:coefcoll3d}) of $q^{-5}$: 
\begin{equation}
\label{eq:dronde}
\mathcal{D}=\frac{2}{\pi^3}(\mathcal{I}+\mathcal{J})
\end{equation}
 where the integral $\mathcal{I}$ comes from point (ii) and the integral $\mathcal{J}$ comes from point (iii): 
 \begin{eqnarray} 
\label{eq:defironde}
\mathcal{I}&=& \int_{1/2}^{1} \mathrm{d}\eps \left[\frac{F(-\eps)}{2\eps^2}-\frac{\pi\sqrt{1+2\eps}}{2\eps(\eps+1)^2}\right]\frac{2\pi}{\sqrt{2\eps-1}} \\
\label{eq:defjronde}
\mathcal{J}&=&\int_{1/2}^{1}\mathrm{d}\eps\frac{[{\pi^2}/{2}-F(-\eps)][-{\pi^2}/{2}-F(-\eps)]}{\eps\sqrt{2\eps-1}\sqrt{2\eps+1}} + \int_{1/2}^{+\infty}\mathrm{d}\eps \frac{[{\pi^2}/{2}-F(-\eps)]\ln^2 R(\eps)}{4\eps\sqrt{2\eps-1}\sqrt{2\eps+1}}
\end{eqnarray}
 Here, we will simply perform a numerical calculation of these constants: 
\begin{equation}
\mathcal{I}=0.985\, 834\ldots \quad ;\quad \mathcal{J}=-9.507\, 704\ldots
\end{equation}
 However, we should mention (without going into detail) the possibility of an analytical calculation. First, we eliminate the square roots by changing the variable $\eps=(t^2+1/t^2)/4$ where $t\geq 1$; then $\sqrt{1+2\eps}=(t+1/t)/\sqrt{2}$ and $\sqrt{2\eps-1}=(t-1/t)/\sqrt{2}$. Next, we note that the numerator and denominator of the fraction under the logarithm in the definition of $F(Z)$ are squares, for example $\sqrt{1-2Z}+\mathrm{i} Z=(\sqrt{1-2Z}+\ii)^2/(2\ii)$. This allows us to write $F(-\eps)$ in the form 
\begin{equation}
F(-\eps)=\frac{\pi^2}{2} +\mathrm{i}\pi \left[\ln(t+\mathrm{i}\alpha)+\ln(t-\mathrm{i}\beta)-\ln(t-\mathrm{i}\alpha)-\ln(t+\mathrm{i}\beta)\right]\quad\mbox{where}\quad \alpha=\beta^{-1}=\frac{\sqrt{2}+\sqrt{6}}{2}
\end{equation}
 We therefore reduce (\ref{eq:defironde}) to an integral of the form $\int\mathrm{d} t R(t)\ln(t-a)$ with $\im a\neq 0$ then, by decomposing the rational fraction $R(t)$ into partial fractions on $\mathbb{C}$, to $\int\mathrm{d} t \ln(t-a)/(t-z)$ with $\im z\neq 0$ and $z\neq a$, for which a primitive (continuous on $\mathbb{R}$) is known:\footnote{If $a_{\rm I}/(a_{\rm I}-z_{\rm I})>1$, the argument of the polylogarithms crosses their branch cut $[1,+\infty[$ at $t_0=(a_{\rm I} z_{\rm R}-a_{\rm R}z_{\rm I})/(a_{\rm I}-z_{\rm I})$, but the discontinuities of $g_2$ and $g_1$ that follow compensate each other in (\ref{eq:primicont}). Here $a_{\rm R},z_{\rm R}$ and $a_{\rm I},z_{\rm I}$ are the real parts and the imaginary parts of $a,z$.

} 
\begin{equation}
\label{eq:primicont}
\int\mathrm{d} t\, \frac{\ln(t-a)}{t-z}=g_2\left(\frac{t-a}{z-a}\right)-g_1\left(\frac{t-a}{z-a}\right)\left[\ln(t-a)-\ln(z-a)\right]+\ln(z-a)\ln(t-z)
\end{equation}
 Here $g_n$ is a Bose function or polylogarithm. This suffices for $\mathcal{I}$. The same method leads, for the first integral $\mathcal{J}_1$ in (\ref{eq:defjronde}), to $\int \ln(t-a)\ln(t-b)/(t-z)$, for which a primitive (not necessarily continuous) is given by computational algebra (we do not reproduce it here). The second integral in (\ref{eq:defjronde}) is more difficult; thanks to the invariance by the transformations $t\to 1/t$ and $t\to -t$, we can write it as an integral over the entire real axis, 
\begin{equation}
\mathcal{J}_2 = \int_{-\infty}^{+\infty} \mathrm{d} t\, \frac{(-\mathrm{i}\pi t)}{1+t^4}\left\{\left[\ln(t+\mathrm{i}\alpha)-\ln(t+\mathrm{i}\beta)\right]+\left[\ln(t-\mathrm{i}\beta)-\ln(t-\mathrm{i}\alpha)\right]\right\} \ln^2\frac{|t-\alpha|\,|t+\beta|}{|t+\alpha|\,|t-\beta|}
\end{equation}
 where we have collected the logarithms whose arguments have imaginary parts of the same sign. All that remains is to apply the contour method from note \ref{note:beaucal}, after expressing $\ln|x|$ in terms of a complex logarithm 
\begin{equation}
\label{eq:decln}
\ln|x|=\ln(x+\mathrm{i} 0^+)-\mathrm{i}\pi Y(-x)=\ln(x-\mathrm{i} 0^+)+\mathrm{i} \pi Y(-x)\quad\forall x\in\mathbb{R}^*
\end{equation}
 and choosing the sign $\pm$ so that all the branch cuts of the logarithms are in the same lower or upper complex half-plane. The non-analytic part of (\ref{eq:decln}) involving the Heaviside function escapes the residue theorem, but can be reduced to an integral of a product of two logarithms over a semi-infinite interval of $\mathbb{R}$, for which the calculation can be performed as in $\mathcal{J}_1$.

\section{Short-distance behavior of certain Fourier integrals}\label{ann:courtedis}

Consider a function $\tilde{\phi}(\qq)$ from $\mathbb{R}^d$ to $\mathbb{C}$ that is isotropic, integrable in modulus, and has asymptotic behavior (\ref{eq:collasymp}). We ask how its inverse Fourier transform 
\begin{equation}
\phi(\rr)=\int_{\mathbb{R}^d}\frac{\dd^dq}{(2\pi)^d}\tilde{\phi}(\qq) \exp(\mathrm{i}\mathbf{q}\cdot\rr)
\end{equation}
 varies to order $r^2$, i.e., to within $o(r^2)$, in the vicinity of $\rr=\mathbf{0}$. In what follows, we will omit the index “coll” on the coefficients of (\ref{eq:collasymp}) to simplify and universalize.

\subsection{Two-dimensional case}

In dimension $d=2$, after angular averaging and subtraction of $\phi(\mathbf{0})$, we arrive at the following integral: 
\begin{equation}
\label{eq:int2d}
\phi(\rr)-\phi(\mathbf{0})=\int_0^{+\infty} \frac{q\mathrm{d} q}{2\pi} [J_0(qr)-1]\tilde{\phi}(\qq)
\end{equation}
 We cannot simply replace $J_0(qr)-1$ with its quadratic approximation $-(qr)^2/4$ under the integral sign without triggering a logarithmic ultraviolet divergence. After reflection, we introduce two fixed cutoff parameters $Q\gg 1$ and $0<\eta\ll 1$, we separate the integration interval into three subintervals $I_1=[0,Q], I_2=[Q,\eta/r]$ and $I_3=[\eta/r,+\infty[$, and we use three different approximations of the integrand of (\ref{eq:int2d}) adapted to each interval in the limit $r\to 0^+$. We then take the limit $Q\to +\infty$ and $\eta\to 0^+$ in the coefficients of the expansion obtained. On the interval $I_1$, we quadratize $J_0(qr)-1$; the error made is negligible $O(r^4)$. On the interval $I_2\cup I_3$, we can replace $\tilde{\phi}(\qq)$ with the dominant term $q^{-4}$ in (\ref{eq:collasymp}); since $J_0(qr)-1=O((qr)^2)$, the error made is $O(r^2/Q^2)$ (up to a factor of $\ln Q$)
 therefore affects the coefficient of $r^2$ by a negligible $O(1/Q^2)$. Still on $I_2\cup I_3$, we then perform the variable change $x=qr$, which results in an overall factor $r^2$, 
\begin{equation}
\phi(\rr)-\phi(\mathbf{0})|_{I_2\cup I_3} \simeq \frac{C r^2}{2\pi} \int_{Qr}^{+\infty} \frac{\mathrm{d} x}{x^3} \frac{\ln(A x/r)}{\ln(B x/r)} [J_0(x)-1]
\end{equation}
 In the remaining integral on $x$, we can then take the limit $r\to +\infty$, thus saying that $\ln(Ax/r)/\ln(Bx/r)\to 1$ if $x\in[\eta,+\infty[$ (the error made is $o(1)$), and approximate $J_0(x)-1$ by $-x^2/4$ if $Qr<x<\eta$ (this introduces a negligible error $O(\eta^2)$). There are two integrals left that we know how to study when $\eta\ll 1$ or $r\to 0^+$: 
\begin{multline}
\label{eq:intx1}
\int_\eta^{+\infty}\frac{\mathrm{d} x}{x^3}[J_0(x)-1]=\frac{2}{\pi}\int_0^{\pi/2} \mathrm{d}\theta\, \left[-\frac{1-\cos(\eta\cos\theta)}{2\eta^2}-\frac{1}{2\eta}\sin(\eta\cos\theta)\cos\theta+\frac{1}{2}\Ci(\eta\cos\theta)\cos^2\theta\right]\\
\underset{\eta\to 0^+}{=} \frac{1}{4}(-1-\ln 2+\gamma+\ln\eta)+O(\eta^2)
\end{multline}
 (we used the integral representation $J_0(x)=(2/\pi)\re\int_0^{\pi/2}\mathrm{d}\theta\exp(\mathrm{i} x\cos\theta)$, interchanged the integration over $x$ and over $\theta$, integrated twice by parts in the integral on $x$ to move from $x^3$ to $x$ in the denominator, recognized the cosine integral function $\Ci$, used its expansion at the origin and identity $(2/\pi)\int_0^{\pi/2}\mathrm{d}\theta\, \cos^2\theta\,\ln(\cos\theta)=(1-2\ln 2)/4$) 
\begin{multline}
\label{eq:intx2}
\int_{Qr}^{\eta}\frac{\mathrm{d} x}{x^3}\frac{\ln(A x/r)}{\ln(B x/r)}\left(-\frac{1}{4}x^2\right)=\frac{1}{4}\ln\left(\frac{Qr}{\eta}\right)-\frac{1}{4}\ln\left(\frac{A}{B}\right)\ln\left[\frac{\ln(\eta B/r)}{\ln(QB)}\right] \\
\underset{r\to 0^+}{=} \frac{1}{4}\ln\left(\frac{Qr}{\eta}\right)-\frac{1}{4}\ln\left(\frac{A}{B}\right)\left[\ln\ln\left(\frac{B}{r}\right)-\ln\ln(QB)\right]+O\left(\frac{1}{\ln r}\right)
\end{multline}
 (we used the variable change $u=\ln x$). The dependency in $\eta$ disappears in the sum of (\ref{eq:intx1}) and (\ref{eq:intx2}). To make the dependency in $Q$ disappear in the limit $Q\to +\infty$, we must return to the contribution of interval $I_1$; we transform it using the minus-plus trick as follows: 
\begin{multline}
\phi(\rr)-\phi(\mathbf{0})|_{I_1} \simeq -\frac{r^2}{8\pi} \int_0^Q\mathrm{d} q\, q^3 \tilde{\phi}(\qq) = -\frac{r^2}{8\pi}\int_0^{Q}\mathrm{d} q\, q\left[q^2\tilde{\phi}(\qq)-\frac{C}{q^2+q_0^2}\frac{\ln[A^2(q^2+q_0^2)]}{\ln[B^2(q^2+q_0^2)]}\right]\\
-\frac{Cr^2}{8\pi}\!\int_0^{Q}\!\frac{q\mathrm{d} q}{q^2+q_0^2}\frac{\ln[A^2(q^2+q_0^2)]}{\ln[B^2(q^2+q_0^2)]}\!\underset{Q\to +\infty}{=}\!\!\!\!\!-\frac{r^2}{4}\mathcal{I}_{\rm 2D}-\frac{C r^2}{8\pi}\left[\ln\left(\frac{Q}{q_{\rm dim}}\right)+\ln\left(\frac{A}{B}\right)\ln\ln(QB)\right]+O\left(\frac{{\ln^\alpha Q}}{Q^2}\right)
\end{multline}
 (the last integral in the right-hand side of the second equality is calculated using variable change $u=\ln(q^2+q_0^2)$). Here, $q_0>1/B$ is an arbitrary reference wave number, $q_{\rm dim}$ is the dimer wave number (\ref{eq:edim2d}), and we have set 
\begin{equation}
\mathcal{I}_{\rm 2D}=\int_0^{+\infty}\frac{q\mathrm{d} q}{2\pi} \left[q^2\tilde{\phi}(\qq)-\frac{C}{q^2+q_0^2}\frac{\ln[A^2(q^2+q_0^2)]}{\ln[B^2(q^2+q_0^2)]}\right] -\frac{C}{2\pi}\left[\ln\frac{q_0}{q_{\rm dim}}+\ln\left(\frac{A}{B}\right)\ln\ln (B q_0)\right] 
\end{equation}
This is a well-defined quantity given the asymptotic behavior (\ref{eq:collasymp}) of $\tilde{\phi}(\qq)$; it is also independent of $q_0$ as long as $q_0>1/B$ (it is easy to verify that its derivative with respect to $q_0$ is identically zero). Finally, after combining the contributions of the three intervals, the cutoff parameter $Q$ disappears and the desired quadratic behavior remains: 
\begin{equation}
\label{eq:theo2d}
\boxed{\phi(\rr)-\phi(\mathbf{0})\stackrel{d=2}{\underset{r\to 0^+}{=}} \frac{1}{4}r^2\left\{-\mathcal{I}_{\rm 2D}+\frac{C}{2\pi}\left[-1+\gamma+\ln(q_{\rm dim}r/2)-\ln\left(\frac{A}{B}\right)\ln(-\ln(r/B))\right]+o(1)\right\}}
\end{equation}
 We can simplify a little by replacing $\ln(-\ln(r/B))$ with $\ln(-\ln(q_{\rm dim}r))$, because this introduces a correction $O[\ln(q_{\rm dim}B)/\ln(q_{\rm dim}r)]$ that can be absorbed in $o(1)$.

\subsection{Three-dimensional case}

In dimension $d=3$, we reduce by isotropy to 
\begin{equation}
\phi(\rr)-\phi(\mathbf{0})=\int_0^{+\infty}\frac{q^2\mathrm{d} q}{2\pi^2} \left(\frac{\sin(qr)}{qr}-1\right)\tilde{\phi}(\qq)
\end{equation}
 After reflection, we find that it is sufficient to cut the integration interval into only two parts,
 using the cut $Q$. On the lower part $q\in [0,Q]$, we can replace $\sin(qr)/(qr)-1$ with its quadratic approximation $-(qr)^2/6$; the error made is a negligible $O(r^4)$. On the upper part, we can replace $\tilde{\phi}(\qq)$ with its asymptotic behavior (\ref{eq:collasymp}); as $\sin(qr)/(qr)-1=O[(qr)^2]$, the error made $\int_Q^{+\infty} O(r^2/q^2)\mathrm{d} q=O(r^2/Q)$ has no effect on the coefficient of $r^2$ when $Q\to +\infty$. On the lower part, let's immediately apply a minus-plus trick to prepare for the elimination of $Q$ in the limit where it tends towards infinity: 
\begin{multline}
\label{eq:basse3d}
\phi(\rr)-\phi(\mathbf{0})|_{q<Q}\simeq -\frac{r^2}{6} \int_0^Q\frac{q^2\mathrm{d} q}{2\pi^2} q^2\left[\tilde{\phi}(\qq)-\frac{C}{(q^2+q_0^2)^2}-\frac{D q}{(q^2+q_0^2)^3}\right]-\frac{r^2}{6} \int_0^Q\frac{q^2\mathrm{d} q}{2\pi^2}q^2\left[\frac{C}{(q^2+q_0^2)^2}\right. \\
\left. +\frac{D q}{(q^2+q_0^2)^3}\right]\underset{Q\to +\infty}{=} -\frac{r^2}{6}\left[\mathcal{I}_{\rm 3D}+\frac{QC}{2\pi^2}+\frac{D}{2\pi^2}\ln\left(\frac{Q}{k_{\Delta}}\right)-\frac{3D}{8\pi^2}+O\left(Q^{-1}\right)\right]
\end{multline}
 where we have set 
\begin{equation}
\mathcal{I}_{\rm 3D}=\int_0^{+\infty}\frac{q^2\mathrm{d} q}{2\pi^2} q^2\left[\tilde{\phi}(\qq)-\frac{C}{(q^2+q_0^2)^2}-\frac{D q}{(q^2+q_0^2)^3}\right]-\frac{3 q_0 C}{8\pi}-\frac{D}{2\pi^2} \ln\left(\frac{q_0}{k_\Delta}\right)
\end{equation}
 and $k_\Delta=(m\Delta/\hbar^2)^{1/2}$, with $q_0>0$ an arbitrary wave number whose role is to avoid infrared divergence. We verify that $\mathrm{d}\mathcal{I}_{\rm 3D}/\mathrm{d} q_0\equiv 0$, and therefore $\mathcal{I}_{\rm 3D}$, is independent of $q_0$. On the upper part $q\in [Q,+\infty[$, after changing variables $x=qr$ and passing to the limit $r\to 0^+$ to $Q$, we obtain: 
\begin{multline}
\label{eq:haute3d}
\phi(\rr)-\phi(\mathbf{0})|_{q>Q} \simeq \frac{C r}{2\pi^2} \int_{Qr}^{+\infty} \frac{\mathrm{d} x}{x^2}\left(\frac{\sin x}{x}-1\right)+\frac{Dr^2}{2\pi^2} \int_{Qr}^{+\infty} \frac{\mathrm{d} x}{x^3}\left(\frac{\sin x}{x}-1\right)\\
\underset{r\to 0^+}{=}\frac{C r}{2\pi^2}\left[-\frac{\pi}{4} +\frac{1}{6}Qr+O(r^3)\right]+\frac{Dr^2}{2\pi^2}\left[\frac{-11+6\gamma+6\ln(Qr)}{36}+O(r^2)\right]
\end{multline}
 (in the second equality, we expressed the integrals over $x$ in terms of the special functions integral sine and integral cosine, then used the known expansions of these functions at the origin). It remains to sum (\ref{eq:basse3d}) and (\ref{eq:haute3d}) to obtain the desired quadratic behavior, independent of $Q$ as it should be: 
\begin{equation}
\label{eq:theo3d}
\boxed{\phi(\rr)-\phi(\mathbf{0})\stackrel{d=3}{\underset{r\to 0^+}{=}} -\frac{C r}{8\pi} + \frac{1}{6} r^2 \left[-\mathcal{I}_{\rm 3D}+\frac{D}{2\pi^2}\left(-\frac{13}{12}+\gamma+\ln(k_\Delta r)\right)+o(1)\right]}
\end{equation}
\bibliographystyle{crunsrt}

\nocite{*}

\bibliography{english}
\end{document}